\RequirePackage{lineno}
\documentclass[aps,prd,twocolumn,showpacs]{revtex4-1}

\usepackage{preprintcover}  
\PreprintCoverPaperTitle{Combined search for the  Standard Model Higgs boson
\\ in $pp$ collisions at $\sqrt{s}=7$~TeV   with the ATLAS detector}
\PreprintIdNumber{CERN-PH-EP-2012-167}  
\PreprintCoverAbstract{
  A combined search for the Standard Model Higgs boson with the ATLAS
  detector at the LHC is presented. The datasets used correspond to
  integrated luminosities from 4.6~\infb\ to 4.9~\infb\ of
  proton-proton collisions collected at $\sqrt{s}=7$~TeV in 2011. The
  Higgs boson mass ranges of \lowerlowerObs\ to \upperlowerObs,
  \lowerIsland\ to \upperIsland, and \lowerObs\ to \upperObs\ are
  excluded at the 95\%\ confidence level, while the range
  \lowerExp\ to \upperExp\ is expected to be excluded in the absence
  of a signal. An excess of events is observed at Higgs boson mass
  hypotheses around 126~\GeV\ with a local significance of
  2.9~standard deviations ($\sigma$). The global probability for the
  background to produce an excess at least as significant anywhere in
  the entire explored Higgs boson mass range of 110--600~\GeV\ is
  estimated to be $\sim$15\%, corresponding to a significance of
  approximately $1\sigma$.
}
\PreprintJournalName{Physical Review D}

 \usepackage{subfigure}
 \usepackage{multirow}
 \usepackage{placeins}

 \usepackage{atlasphysics} 
 \usepackage{amsmath}
 \usepackage{amssymb}

\newcommand{\Pois}{{\ensuremath{\rm Pois}}}

\newcommand{\Gauss}{{\ensuremath{\rm Gauss}}}

\newcommand{\data}{{\ensuremath{\mathcal{D}}}}
\newcommand{\globs}{{\ensuremath{\mathcal{G}}}}
\newcommand{\datasim}{{\ensuremath{\mathcal{D}_{\textrm{com}}}}}
\newcommand{\F}{{\ensuremath{\mathrm{f}}}}

\newcommand{\hathatthetamu}{\ensuremath{\hat{\hat{\vec\theta}}(\mu)}}

\def\vec#1{\ifmmode
\mathchoice{\mbox{\boldmath$\displaystyle\bf#1$}}
{\mbox{\boldmath$\textstyle\bf#1$}}
{\mbox{\boldmath$\scriptstyle\bf#1$}}
{\mbox{\boldmath$\scriptscriptstyle\bf#1$}}\else
{\mbox{\boldmath$\bf#1$}}\fi}

\def\roostats{\texttt{RooStats}}

\newcommand{\bb}{\ensuremath{b\bar{b}}}

\def\mT{\ensuremath{m_{\mathrm{T}}}} 
\newcommand{\ggF}{\ensuremath{gg\to H}}
\newcommand{\VBF}{\ensuremath{qq'\to qq'H}}
\newcommand{\VH}{\ensuremath{q\bar{q}\to WH/ZH}}
\newcommand{\ttH}{\ensuremath{q\bar{q}/gg \to t\bar{t}H}}
\newcommand{\ggttH}{\ensuremath{gg \to t\bar{t}H}}

\newcommand{\ptt}{\ensuremath{p_{{\mathrm T_t}}}}
\newcommand{\hgg}{\ensuremath{H \rightarrow \gamma\gamma}}

\newcommand{\hww}{\ensuremath{H \rightarrow WW^{(*)}}}
\newcommand{\hWW}{\ensuremath{H \rightarrow WW^{(*)}}}
\newcommand{\hWWlnln}{\ensuremath{H \rightarrow WW^{(*)} \rightarrow \ell^+\nu\ell^-\overline{\nu}}}
\newcommand{\lnln}{\ensuremath{\ell^+\nu\ell^-\overline{\nu}}}
\newcommand{\hWWlnqq}{\ensuremath{H \rightarrow WW \rightarrow \ell\nu q\overline{q}'}}
\newcommand{\lnqq}{\ensuremath{\ell\nu q\overline{q}'}}
\newcommand{\hzz}{\ensuremath{H \rightarrow ZZ^{(*)}}}
\newcommand{\hZZ}{\ensuremath{H \rightarrow ZZ^{(*)}}}
\newcommand{\hZZllnn}{\ensuremath{H \rightarrow ZZ\rightarrow \ell^+\ell^-\nu\overline{\nu}}}
\newcommand{\hZZllqq}{\ensuremath{H \rightarrow ZZ\rightarrow \ell^+\ell^- q\overline{q}}}
\newcommand{\hZZllll}{\ensuremath{H \rightarrow ZZ^{(*)}\rightarrow \ell^+\ell^-\ell^+\ell^-}}
\newcommand{\htt}{\ensuremath{H \rightarrow \tau^+\tau^-}}
\newcommand{\hbb}{\ensuremath{H \rightarrow \bb}}

\newcommand{\mh}{\ensuremath{m_{H}}}

\newcommand{\lowerExp}{120~\GeV}
\newcommand{\upperExp}{560~\GeV}

\newcommand{\lowerlowerObs}{111.4~\GeV}
\newcommand{\upperlowerObs}{116.6~\GeV}

\newcommand{\lowerIsland}{119.4~\GeV}
\newcommand{\upperIsland}{122.1~\GeV}

\newcommand{\lowerObs}{129.2~\GeV}
\newcommand{\upperObs}{541~\GeV}

\newcommand{\significance}{$3.0\sigma$}
\newcommand{\significanceESS}{$2.9\sigma$}

\newcommand{\ZZbkg}{\ensuremath{ZZ^{(*)}}}

\newcommand{\httlh}{\ensuremath{H \rightarrow \tau_{\rm lep}\tau_{\rm had} }}
\newcommand{\htthh}{\ensuremath{H \rightarrow \tau_{\rm had} \tau_{\rm had}}}
\newcommand{\htthhj}{\ensuremath{H + {\rm jet} \rightarrow \tau_{\rm had} \tau_{\rm had} + {\rm jet}}}

\newcommand{\httll}{\ensuremath{H\rightarrow \tau_{\rm lep}\tau_{\rm lep} }}
\newcommand{\lh}{\ensuremath{\tau_{\rm lep} \tau_{\rm had}}}
\newcommand{\hh}{\ensuremath{\tau_{\rm had}\tau_{\rm had}}}
\newcommand{\tauleptaulep}{\ensuremath{\tau_{\rm lep} \tau_{\rm lep}}}

\newcommand{\hWlvbb}{\ensuremath{WH \rightarrow \ell\nu b\overline{b}}}
\newcommand{\hZllbb}{\ensuremath{ZH \rightarrow \ell^+\ell^- b\overline{b}}}
\newcommand{\hZvvbb}{\ensuremath{ ZH \rightarrow \nu \overline{\nu} b\overline{b} }}

\newcommand{\infb}{fb$^{-1}$}

\usepackage{hypernat}

\usepackage{graphicx}
\usepackage{dcolumn}
\usepackage{bm}
\usepackage[hyperindex,breaklinks]{hyperref}

\newcommand{\mycomment}[1]{}

\def\ptt{\ensuremath{p_{\mathrm{Tt}}}}

\begin{document}

\title{Combined search for the  Standard Model Higgs boson
in $pp$ collisions \\ at $\sqrt{s}=7$~TeV   with the ATLAS detector}

\author{The ATLAS Collaboration}

\begin{abstract}

  A combined search for the Standard Model Higgs boson with the ATLAS
  detector at the LHC is presented. The datasets used correspond to
  integrated luminosities from 4.6~\infb\ to 4.9~\infb\ of
  proton-proton collisions collected at $\sqrt{s}=7$~TeV in 2011. The
  Higgs boson mass ranges of \lowerlowerObs\ to \upperlowerObs,
  \lowerIsland\ to \upperIsland, and \lowerObs\ to \upperObs\ are
  excluded at the 95\%\ confidence level, while the range
  \lowerExp\ to \upperExp\ is expected to be excluded in the absence
  of a signal. An excess of events is observed at Higgs boson mass
  hypotheses around 126~\GeV\ with a local significance of
  2.9~standard deviations ($\sigma$). The global probability for the
  background to produce an excess at least as significant anywhere in
  the entire explored Higgs boson mass range of 110--600~\GeV\ is
  estimated to be $\sim$15\%, corresponding to a significance of
  approximately $1\sigma$.
\end{abstract}

\maketitle

\clearpage

\section{Introduction}

Probing the mechanism for electroweak symmetry breaking (EWSB) is one
of the prime objectives of the Large Hadron Collider (LHC).  In the
Standard Model (SM)~\cite{Glashow:1961tr,Weinberg:1967tq,sm_salam},
the electroweak interaction is described by a local gauge field theory
with an $SU(2)_L \otimes U(1)_Y$ symmetry, and EWSB is achieved via
the Higgs mechanism with a single $SU(2)_L$ doublet of complex scalar
fields~\cite{Englert:1964et,Higgs:1964ia,Higgs:1964pj,Guralnik:1964eu,
  Higgs:1966ev,Kibble:1967sv}.  After EWSB the electroweak sector has
massive $W^\pm$ and $Z$ bosons, a massless photon, and a massive
CP-even, scalar boson, referred to as the Higgs boson. Fermion masses
are generated from Yukawa interactions with couplings proportional to
the masses of fermions. The mass of the Higgs boson, \mH, is a free
parameter in the SM.  However, for a given \mH\ hypothesis the cross
sections of the various Higgs boson production processes and the branching
fractions of the decay modes can be predicted, allowing a combined
search with data from several search channels.

Combined searches at the CERN LEP $e^+e^{-}$ collider excluded the 
production of a SM Higgs boson with mass below 114.4~\GeV\ at 95\% confidence level (CL)~\cite{LEP}.
The combined searches at the Fermilab Tevatron $p\overline{p}$
collider excluded the production of a SM Higgs boson with a mass between
147~\GeV\ and 179~\GeV, and between 100~\GeV\ and 106~\GeV\ at 95\%\
CL~\cite{tevatronHiggs}.  Precision
electroweak measurements are sensitive to \mh\ via radiative
corrections and indirectly constrain the SM Higgs boson mass to be
less than 158~\GeV\ \cite{Grunewald:1313716} at 95\% CL.

In 2011, the LHC delivered an integrated luminosity of 5.6~\infb\ of
proton-proton ($pp$) collisions at a center-of-mass energy of 7~TeV to
the ATLAS detector~\cite{atlas-det}. Of the 4.9~\infb\ collected, the
integrated luminosity used in the individual Higgs search channels is
between 4.6~\infb\ and 4.9~\infb, depending on the data quality
requirements specific to each channel.

This paper presents a combined search for the SM Higgs boson in the
decay modes \hgg, \hZZ, \hWW, \htt, and \hbb, with subsequent decays
of the $W$, $Z$, and $\tau$ leading to different final states. Some
searches are designed to exploit the features of the production modes
$pp\rightarrow H$ (gluon fusion), $pp \rightarrow qqH$ (vector boson
fusion) and $pp\rightarrow VH$ with $V = W^{\pm}$ or $Z$ (associated
production with a gauge boson).  In order to enhance the search
sensitivity, the various decay modes are further subdivided into
sub-channels with different signal and background contributions and
different sensitivities to systematic uncertainties.  While the
selection requirements for individual search channels are disjoint,
each selection is, in general, populated by more than one combination 
of Higgs boson production and decay.  For instance, Higgs boson 
production initiated by vector
boson fusion (VBF) can contribute significantly to a search channel
optimized for gluon fusion production.

The ATLAS collaboration has previously published a similar but less
extensive combined search for the Higgs
boson~\cite{PreMoriondCombPaper} in data taken at the LHC in 2011.
The CMS collaboration has also performed a combined analysis of Higgs
searches with data collected in 2011 and have obtained similar
results~\cite{CMScombination}.  In comparison to the analysis of
Ref.~\cite{PreMoriondCombPaper}, the \htt\ and \hbb\ channels have
been added, the \hWWlnln\ analysis has been updated and extended to
cover the mass range range of 110--600~\GeV, and the \hWWlnqq\, 
\hZZllnn, and \hZZllqq\ analyses have been
updated to use the full 2011 dataset.  Both the \hWWlnln\ and
\hWWlnqq\ analyses include a specific treatment of the 2-jets final
state targeted at the VBF production process.

The different channels entering the combination are summarized in 
Table~\ref{tab:channels}.  After describing the general approach to statistical
modeling in Section~\ref{S:Modeling}, the individual channels and the
specific systematic uncertainties are described in
Section~\ref{S:Inputs} and Section~\ref{sec:syst}, respectively.  The
statistical procedure is described in
Section~\ref{S:StatisticalProcedure} and the resulting exclusion
limits and compatibility with the background-only hypothesis are
presented in Section~\ref{S:Exclusion} and Section~\ref{S:p0},
respectively.

\begin{table*}[htb]
\center
\caption{Summary of the individual channels entering  the  combination. 
The transition points between separately optimized  \mh\ regions are indicated when applicable.   The symbols $\otimes$ and $\oplus$ represent direct products or sums over sets of selection requirements. The details of the sub-channels are given in Section~\ref{S:Inputs}.
  \label{tab:channels}} 
\vspace{1em}
\begin{tabular}{c|ccccc}\hline\hline
\multirow{2}{*}{Higgs Decay}	& Subsequent  &
\multirow{2}{*}{Sub-Channels}	&$\mh$ Range  & {$\int$ L $dt$ } & \multirow{2}{*}{Ref.} \\ 
		 				&  Decay 		&   &  [\GeV] & [\infb] &  \\ \hline\hline
\multirow{1}{*}{$H\to\gamma\gamma$} & -- & 9 sub-channels \{\ptt\ $\otimes$ $\eta_\gamma$ $\otimes$ $\rm conversion$\}  & 110--150 & 4.9  & \cite{ggPaper} \\\hline
\multirow{3}{*}{$H\to ZZ^{(*)}$} 	& $\ell\ell\ell'\ell'$ & $\{4e,2e2\mu,2\mu2e,4 \mu\}$ & 110--600 & 4.8 & \cite{4lPaper}  \\
						& $\ell\ell\nu\bar{\nu}$ & $\{ee,\mu\mu\}$ $\otimes$ \{low, high pile-up periods\} & 200--280--600 & 4.7 & \cite{may_llvv}  \\
						& $\ell\ell q\bar{q}$ &  \{$b$-tagged, untagged\} & 200--300--600 & 4.7 & \cite{may_llqq}   \\ \hline
\multirow{2}{*}{$H\to WW^{(*)}$} 	& $\ell\nu\ell\nu$ &  $\{ee,e\mu,\mu\mu\}$ $\otimes$ \{0-jets, 1-jet, 2-jets\} $\otimes$  \{low, high pile-up periods\}  & 110--200--300--600 & 4.7 & \cite{may_lvlv} \\
						& $\ell\nu q\overline{q}'$ & $\{e,\mu\}$ $\otimes$ \{0-jets, 1-jet, 2-jets\} & 300--600 & 4.7 & \cite{may_lvqq}  \\ \hline
\multirow{4}{*}{$H\to \tau^+\tau^-$} 	& $\tau_{\rm lep}\tau_{\rm lep}$ &  $\{e\mu\} \otimes \{$0-jets$\}$  $\oplus$ \{$\ell\ell$\} $\otimes$ \{1-jet, 2-jets, $VH\}$ & 110--150 & 4.7 & \\
      & \multirow{2}{*}{$\tau_{\rm lep} \tau_{\rm had}$} & $\{e,\mu \}$ $\otimes$ \{0-jets\} $\otimes$  $\{ E_{\rm T}^{\rm miss}  < 20~\textrm{\GeV}, E_{\rm T}^{\rm miss}  \ge 20~\textrm{\GeV}\} $  &  \multirow{2}{*}{110--150} &  \multirow{2}{*}{4.7} &  \cite{may_tautau}   \\
						& & $\oplus$  $\{e,\mu \}$ $\otimes$ \{1-jet\} $\oplus$ \{$\ell$\} $\otimes$ \{2-jets\}  &  &  &  \\
						& $\tau_{\rm had}\tau_{\rm had}$ & \{1-jet\} & 110--150 & 4.7 &  \\ \hline
\multirow{3}{*}{$VH\to b\overline{b}$} & $Z\to\nu \overline{\nu}$ & $ E_{\rm T}^{\rm miss}\in \{120-160, 160-200, \ge 200$ \GeV \}   & 110-130 & 4.6  & \\
						& $W\to\ell\nu$ & $p_{\rm T}^W\in \{< 50, 50-100, 100-200, \ge 200$  \GeV \} & 110-130 & 4.7 & \cite{may_bb}\\
						& $Z\to \ell\ell$ &   $p_{\rm T}^Z\in \{< 50, 50-100, 100-200, \ge 200$  \GeV \}  & 110-130 & 4.7 &\\ \hline \hline

\end{tabular}
\end{table*}


\section{Statistical Modeling}
\label{S:Modeling}

In this combined analysis, a given search channel, indexed by $c$, is
defined by its associated event selection criteria, which may select
events from various physical processes.  In addition to the number of
selected events, $n$, each channel may make use of an invariant or
transverse mass distribution of the Higgs boson candidates.  
The discriminating variable is denoted $x$ and its 
probability density function (pdf) is written as
$f(x|\vec\alpha)$, where $\vec\alpha$ represents both theoretical
parameters such as \mh\ and nuisance parameters associated with
various systematic effects.  These distributions are normalized to
unit probability.  The predicted number of events satisfying the
selection requirements is parameterized as $\nu(\vec\alpha)$.  For a
channel with $n$ selected events, the data consist of the values of
the discriminating variables for each event $\data =
\{x_1,\dots,x_{n}\}$.  The probability model for this type of data is
referred to as an unbinned extended likelihood or marked Poisson model
$\F$, given by
\begin{linenomath}
\begin{equation}
\label{Eq:markedPoisson}
\F(\data|\vec\alpha) = \Pois(n|\nu(\vec\alpha)) \prod_{e=1}^n f(x_e|\vec\alpha) \; .
\end{equation}
\end{linenomath}

For each channel several signal and background scattering processes
contribute to the total rate $\nu$ and the overall pdf
$f(x|\vec\alpha)$.  Here, the term \textit{process} is used for any
set of scattering processes that can be added incoherently. The total
rate is the sum of the individual rates
\begin{linenomath}
\begin{equation}
\nu(\vec\alpha) = \sum_{k\in\textrm{processes}} \nu_k(\vec\alpha)
\end{equation}
\end{linenomath}
and the total pdf is the weighted sum
\begin{linenomath}
\begin{equation}
f(x|\vec\alpha) = \frac{1}{\nu(\vec\alpha)} \sum_{k\in\textrm{processes}} \nu_k(\vec\alpha) f_k(x|\vec\alpha)\;.
\end{equation}
\end{linenomath}

Using $e$ as the index over the $n_c$ events in the $c^{\rm th}$
channel, $x_{ce}$ is the value of the observable $x$ for the $e^{\rm
  th}$ event in channels 1 to $c^{\rm max}$.  The total data are a
collection of data from individual channels:
\mbox{$\datasim=\{\data_1, \dots, \data_{c_{\rm max}}\}$}.
 The combined model can then be
written as follows
\begin{linenomath}
\begin{equation}
\label{Eq:simultaneous}
\F_{\textrm{com}}(\datasim|\vec\alpha) = \prod_{c = 1}^{c_{\rm max}} \left[ \Pois(n_c|\nu_c(\vec\alpha)) \prod_{e=1}^{n_c} f_c(x_{ce}|\vec\alpha) \right] .
\end{equation}
\end{linenomath}

\subsection{Parameterization of the Model}
\label{S:parameters}

The parameter of interest is the overall signal strength factor $\mu$,
which acts as a scale factor to the total rate of signal events.  This
global factor is used for all pairings of production cross sections
and branching ratios. The signal strength is defined such that $\mu=0$
corresponds to the background-only model and $\mu=1$ corresponds to the SM Higgs
boson signal.  It is convenient to separate the full list of
parameters $\vec\alpha$ into the parameter of interest $\mu$, the
Higgs boson mass \mh, and the nuisance parameters $\vec\theta$, i.e.
$\vec\alpha=(\mu,\mh,\vec\theta)$.

Each channel in the combined model uses either the reconstructed
transverse mass or the invariant mass of the Higgs candidate as a
discriminating variable. Two approaches are adopted to model the
signal pdfs at intermediate values of \mh\ where full simulation has
not been performed. The first, used in the \hgg\ channel, is based on
a continuous parameterization of the signal as a function of \mh\
using an analytical expression for the pdf validated with simulated
Monte Carlo (MC) samples. The second, used in channels where pdfs are
modeled with histograms, is based on an interpolation procedure using
the algorithm of Ref.~\cite{Read:1999kh}.

\subsection{Auxiliary Measurements}\label{S:AuxMeas}

The nuisance parameters  represent uncertain 
aspects of the model, such as  the background normalization, 
reconstruction efficiencies, energy scale and resolution, 
luminosity, and theoretical predictions. These nuisance
parameters are often estimated from auxiliary measurements, such as 
control regions, sidebands, or dedicated calibration measurements.
 A detailed account of these measurements is beyond the scope of this paper and is given
in the references for the individual
channels~\cite{ggPaper,4lPaper,may_lvlv,may_lvqq,may_llvv,may_llqq,may_tautau,may_bb}.

Each parameter $\alpha_p$ with a dedicated auxiliary measurement
$\F_{\rm aux}(\data_{\rm aux} | \alpha_p, \vec{\alpha}_{\rm other})$
provides a maximum likelihood estimate for $\alpha_p$, $a_p$,
and a standard error $\sigma_p$.  Thus, the detailed probability model
for an auxiliary measurement is approximated as
 \begin{linenomath}
\begin{equation}
\F_{\rm aux}(\data_{\rm aux} | \alpha_p, \vec{\alpha}_{\rm other})
\rightarrow f_p(a_p | \alpha_p,\sigma_p) \;.
\end{equation}
\end{linenomath}
The $f_p(a_p | \alpha_p, \sigma_p)$ are referred to as constraint
terms. 

The fully frequentist procedure applied for the present analysis
includes randomizing the $a_p$ when constructing the ensemble of
possible experiment outcomes.  In the hybrid frequentist-Bayesian
procedures used at LEP and the Tevatron, the $a_p$ are held constant
and the nuisance parameters $\alpha_p$ are randomized according to the
prior probability density
 \begin{linenomath}
\begin{equation}
 \label{eq:urprior}
 \pi(\alpha_p | a_p) \propto f(a_p|\alpha_p,\sigma_p) \eta(\alpha_p)\; ,
 \end{equation}
\end{linenomath}
where $\eta(\alpha_p)$ is an original prior, usually taken to be
constant.

The set of nuisance parameters constrained by auxiliary measurements
is denoted $\mathbb{S}$ and the set of estimates of those parameters,
also referred to as global observables which augments $\datasim$, is
denoted $\mathcal{G}=\{a_p\}$ with $p\in\mathbb{S}$.  Including the
constraint terms explicitly, the model can be rewritten to

\begin{linenomath}
\begin{center}
\begin{eqnarray}
\label{Eq:ftot}
\F_{\textrm{tot}}(\datasim, \globs|\vec\alpha) & = & \prod_{c=\rm 1}^{c_{\rm max}} \left[ \Pois(n_c|\nu_c(\vec\alpha)) \prod_{e=1}^{n_c} f_c(x_{ce}|\vec\alpha) \right] \nonumber
\\
&  & \cdot \prod_{p \in \mathbb{S}} f_p(a_p | \alpha_p,\sigma_p)\; .
\end{eqnarray}
\end{center}
\end{linenomath}

The use of a Gaussian constraint term $f_p(a_p|\alpha_p, \sigma_p) =
\Gauss(a_p|\alpha_p,\sigma_p)$ is problematic if the parameter is
intrinsically non-negative, as is the case for event yields and energy scale
uncertainties.  This is particularly important
when the relative uncertainty is large.  An alternative constraint
term defined only for positive parameter values is the log-normal
distribution, which is given by
\begin{linenomath}
\begin{equation}
\label{eq:lognormal}
f_p(a_p|\alpha_p) = \frac{1}{\sqrt{2 \pi} \ln \kappa} \frac{1}{a_p}
\exp \left[ - \frac{ ( \ln (a_p/\alpha_p ))^{2} } 
{2 (\ln\kappa)^2 } \right] \;.
\end{equation}
\end{linenomath}

The conventional choice $\kappa = 1+\sigma_{\rm rel}$ is made, where
$\sigma_{\rm rel}$ is the relative uncertainty $\sigma_{p}/a_p$ from
the observed auxiliary measurement~\cite{LHC-HCG}.

Using the log-normal distribution for $a_p$ is equivalent to having a
Gaussian constraint for the transformed parameter \mbox{$a_p' =\ln
  a_p$} and \mbox{$\alpha_p'=\ln \alpha_p$}.  

For channels that use histograms based on simulated MC samples, the parametric pdf
$f(x|\vec\alpha)$ is formed by interpolating between histogram
variations evaluated at $\alpha_p=a_p\pm \sigma_p$.  Since the variations
need not be symmetric, the function is treated in a piecewise way 
using a sixth-order  polynomial to interpolate in the range
$\alpha_p \in [a_p-\sigma_p,a_p+\sigma_p]$ with coefficients chosen to
match the first and second derivatives~\cite{Moneta:2010pm}.
Henceforth, the prime will be suppressed and $\alpha_p$ will refer to
the transformed nuisance parameter.
 
Not all systematic uncertainties have an associated auxiliary
measurement.  For example, uncertainties associated with the choice of
renormalization and factorization scales and missing higher-order
corrections in a theoretical calculation are not statistical in
nature.  In these cases, the frequentist form of the constraint term
is derived  assuming, by convention, a log-normal prior probability density on
these parameters and inverting Eq.~(\ref{eq:urprior}).

\section{Individual Search Channels}
\label{S:Inputs}

All the channels combined to search for the SM Higgs boson use the
complete 2011 dataset passing the relevant quality requirements.  The
Higgs boson decays considered are \hgg, \hww, \hzz, \htt, and \hbb. In
modes with a $W$ or $Z$ boson, an electron or muon is required for
triggering. In the \htt\ channel, almost all combinations of
subsequent $\tau$ decays are considered. 
The results in
the $\gamma\gamma$ and $\ell^+\ell^-\ell^+\ell^-$ modes are the same
as in the previously published combination~\cite{PreMoriondCombPaper},
but all other channels have been updated.  A summary of the individual channels 
contributing to this combination is given in Table~\ref{tab:channels}.

The invariant and transverse mass distributions for the individual
channels are shown in Figs.~\ref{fig:massplots1} and
\ref{fig:massplots2}, with several sub-channels merged.

\begin{itemize}

\item \hgg: This analysis is unchanged with respect to the previous
  combined search~\cite{PreMoriondCombPaper,ggPaper} and is carried
  out for \mH\ hypotheses between 110~\GeV\ and 150~\GeV.  Events are
  separated into nine independent categories of varying
  sensitivity. The categorization is based on the pseudorapidity of
  each photon, whether it was reconstructed as a converted or
  unconverted photon, and the momentum component of the diphoton
  system transverse to the diphoton thrust axis (\ptt).  The mass
  resolution is approximately 1.7\%\ for \mh$\sim$120~\GeV.
  \end{itemize}
  
  \begingroup
   \begin{figure*}[htb]
    \begin{center}
      \subfigure[]{    \includegraphics[width=.3\textwidth]{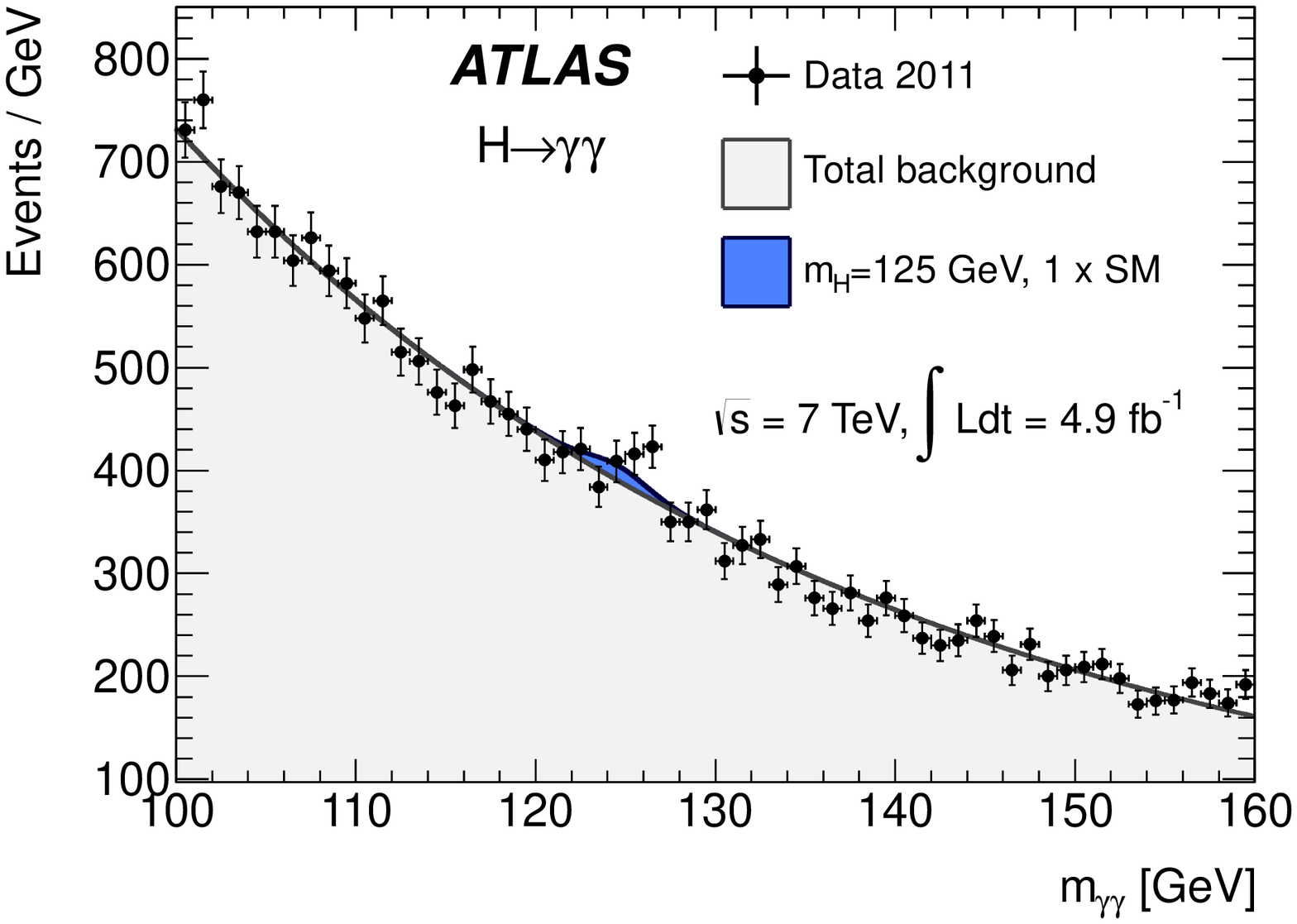}}
      \subfigure[]{    \includegraphics[width=.3\textwidth]{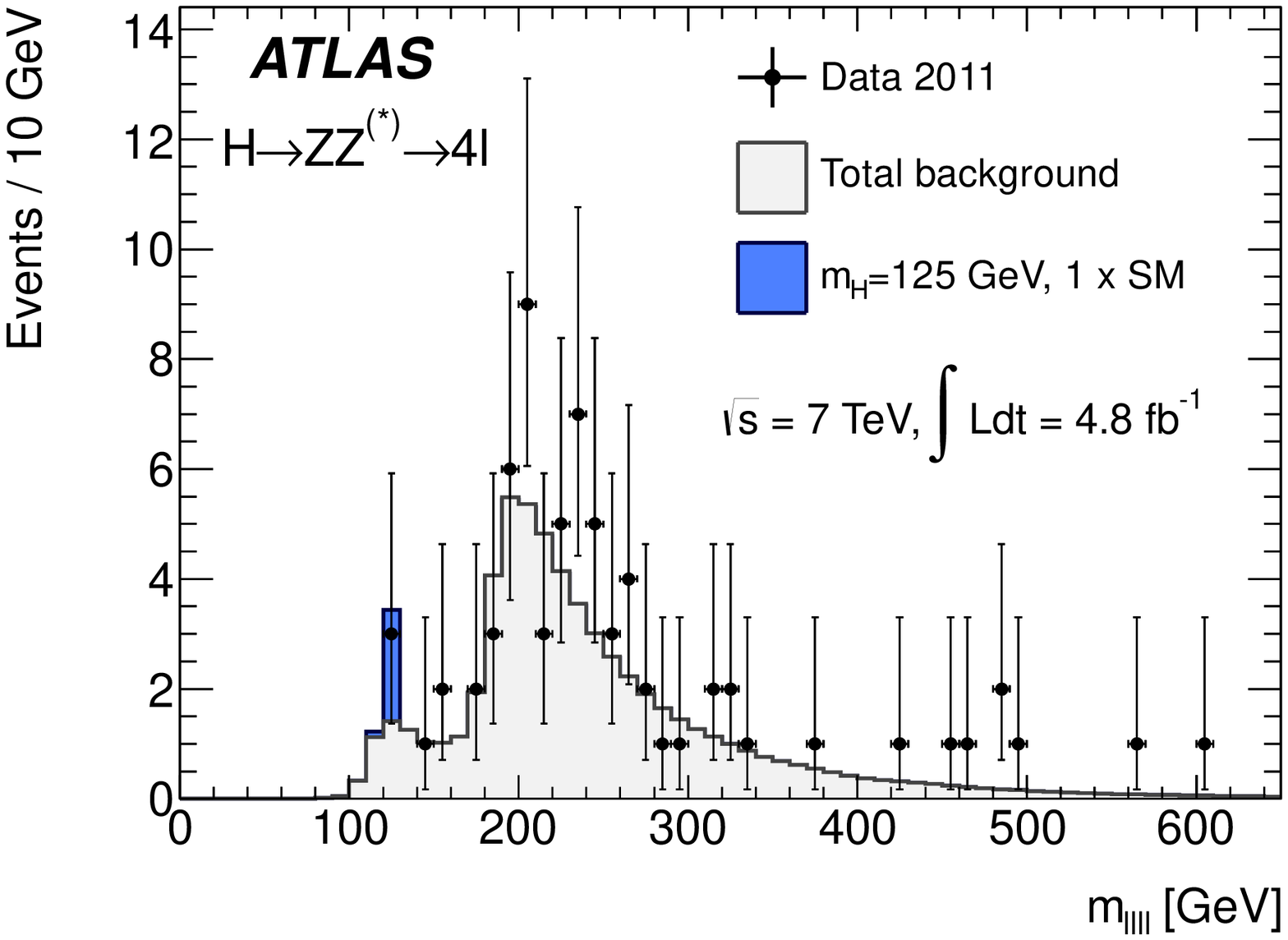}}
      \subfigure[]{    \includegraphics[width=.3\textwidth]{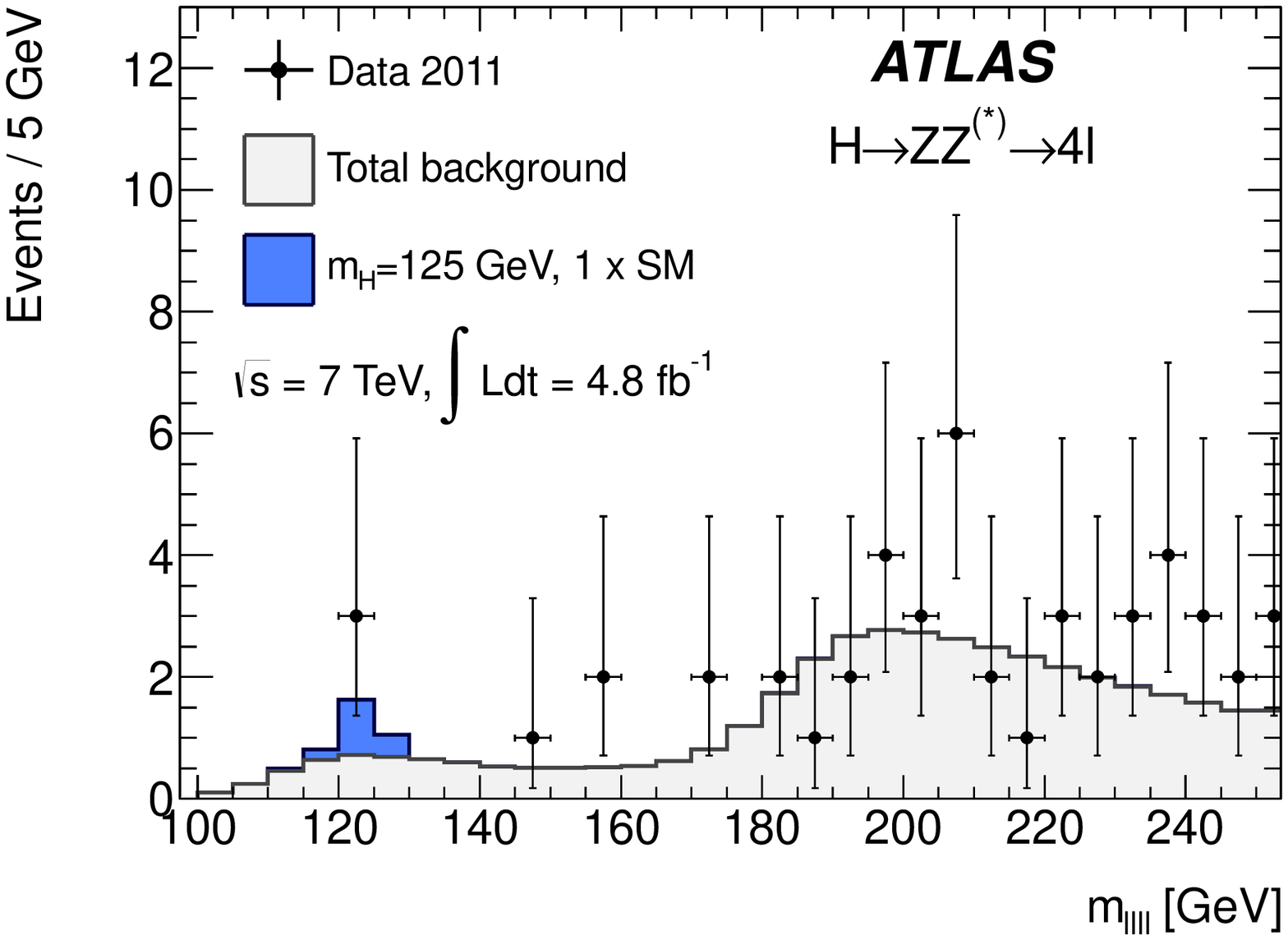}}
      \subfigure[]{    \includegraphics[width=.3\textwidth]{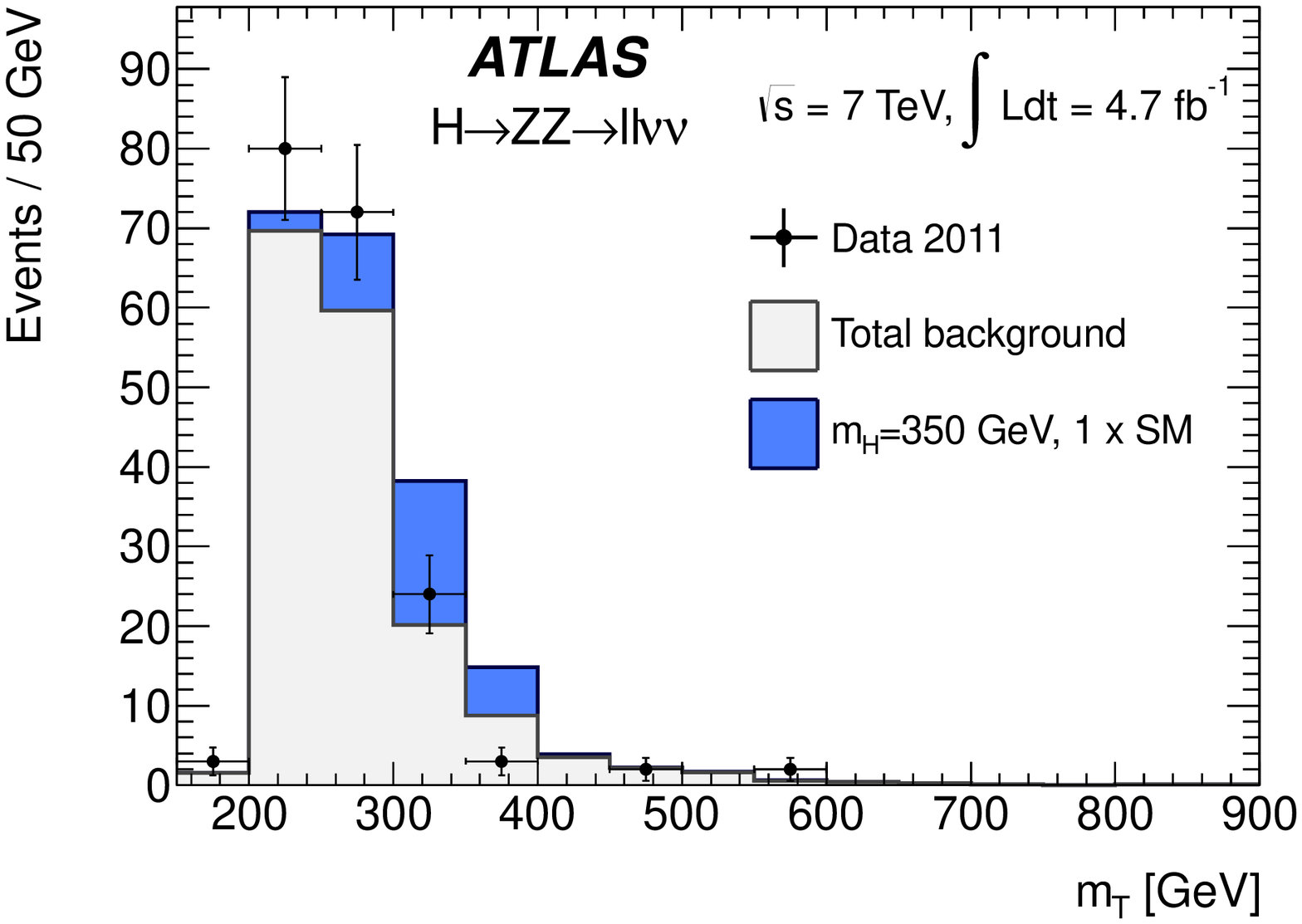}}
      \subfigure[]{    \includegraphics[width=.3\textwidth]{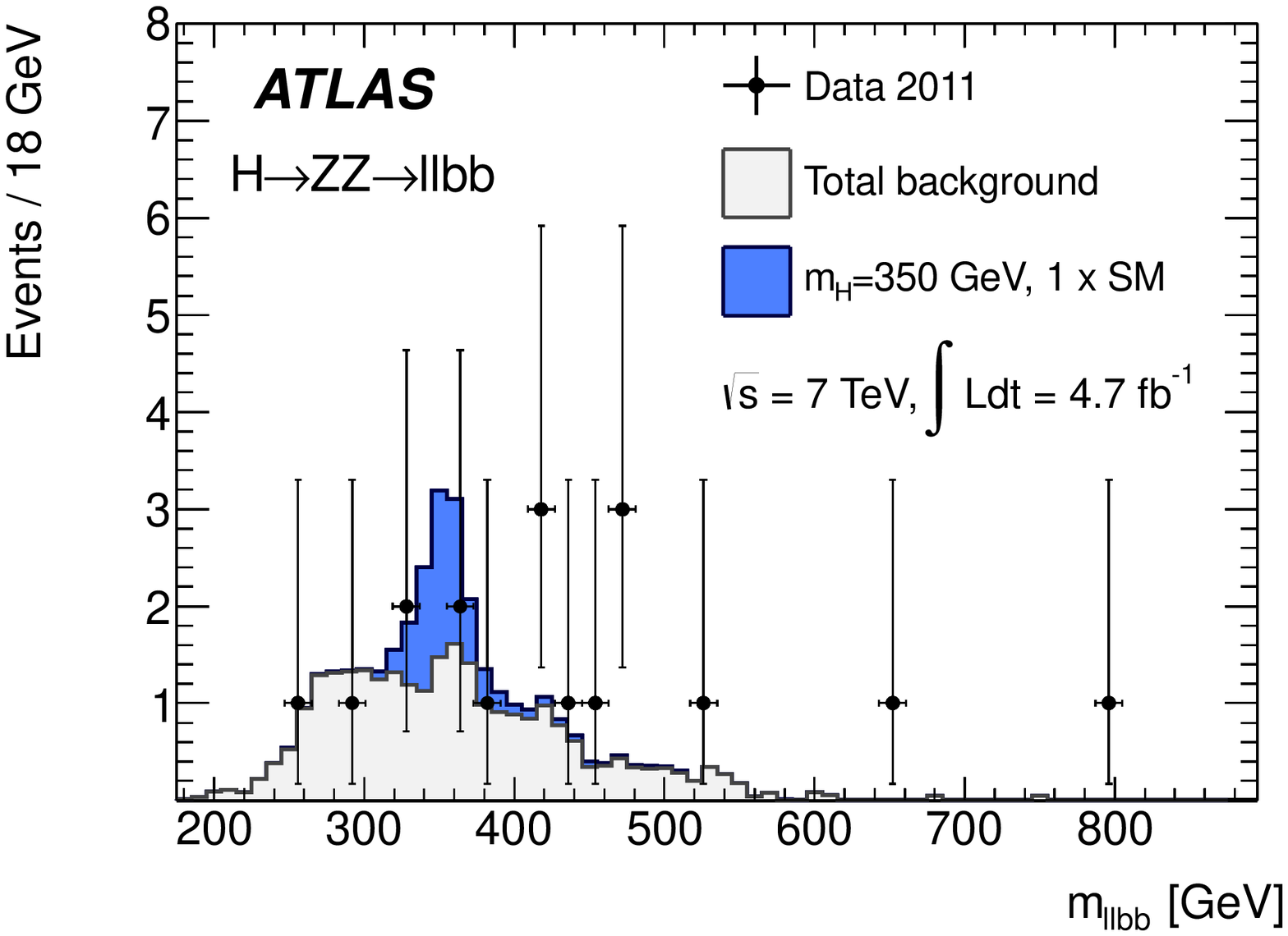}}
      \subfigure[]{    \includegraphics[width=.3\textwidth]{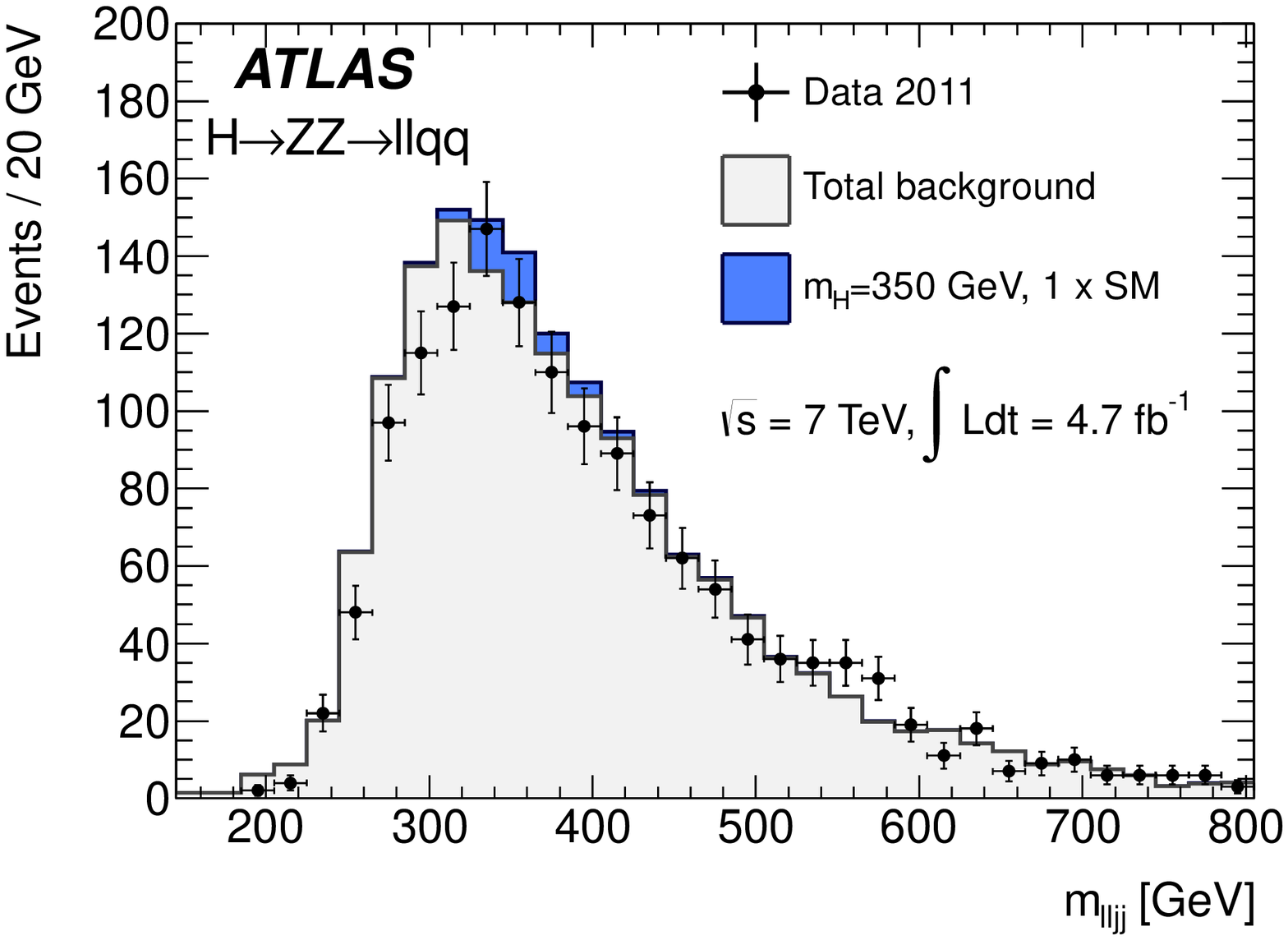}}
      \subfigure[]{    \includegraphics[width=.3\textwidth]{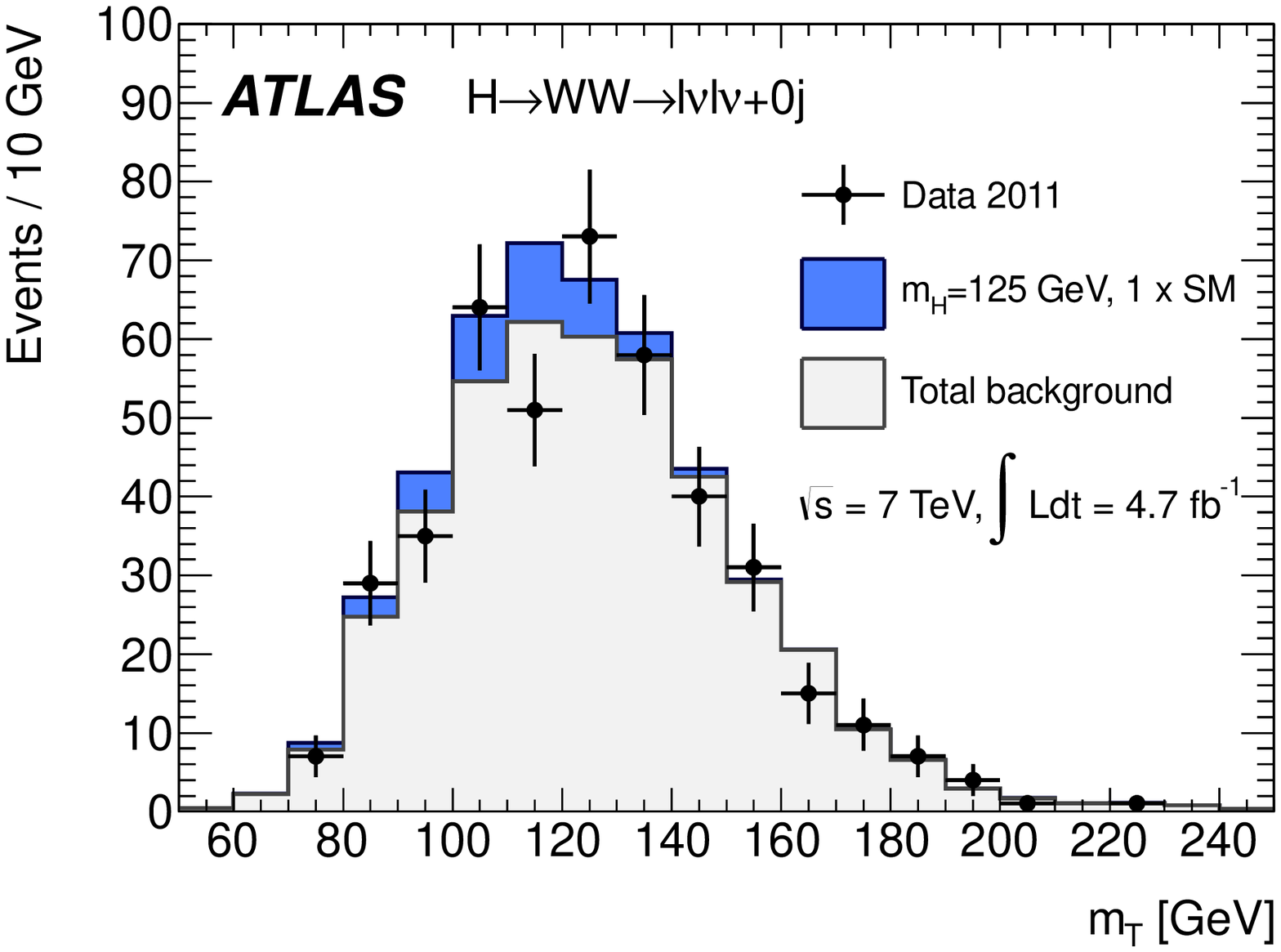}}
      \subfigure[]{    \includegraphics[width=.3\textwidth]{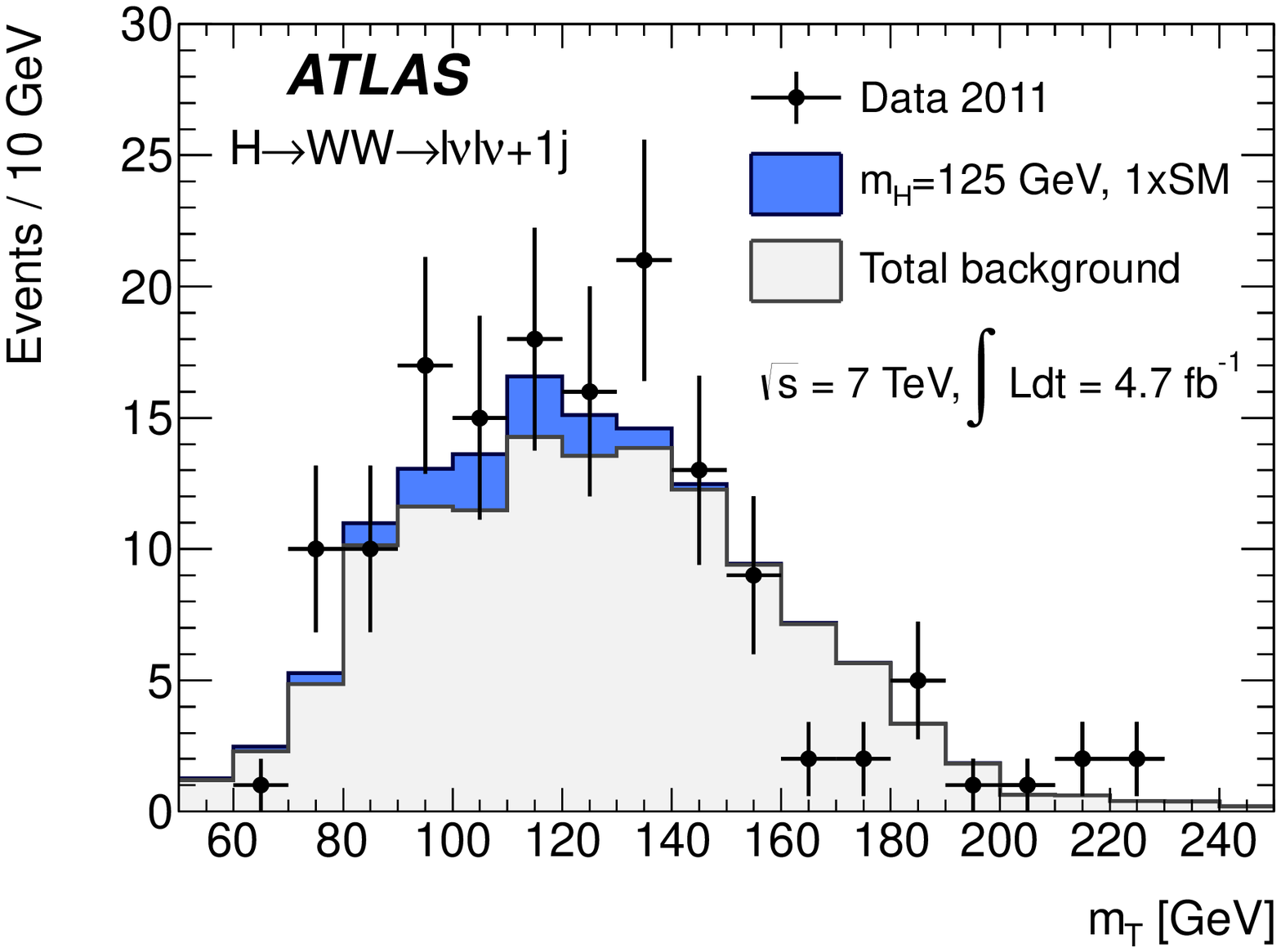}}
      \subfigure[]{    \includegraphics[width=.3\textwidth]{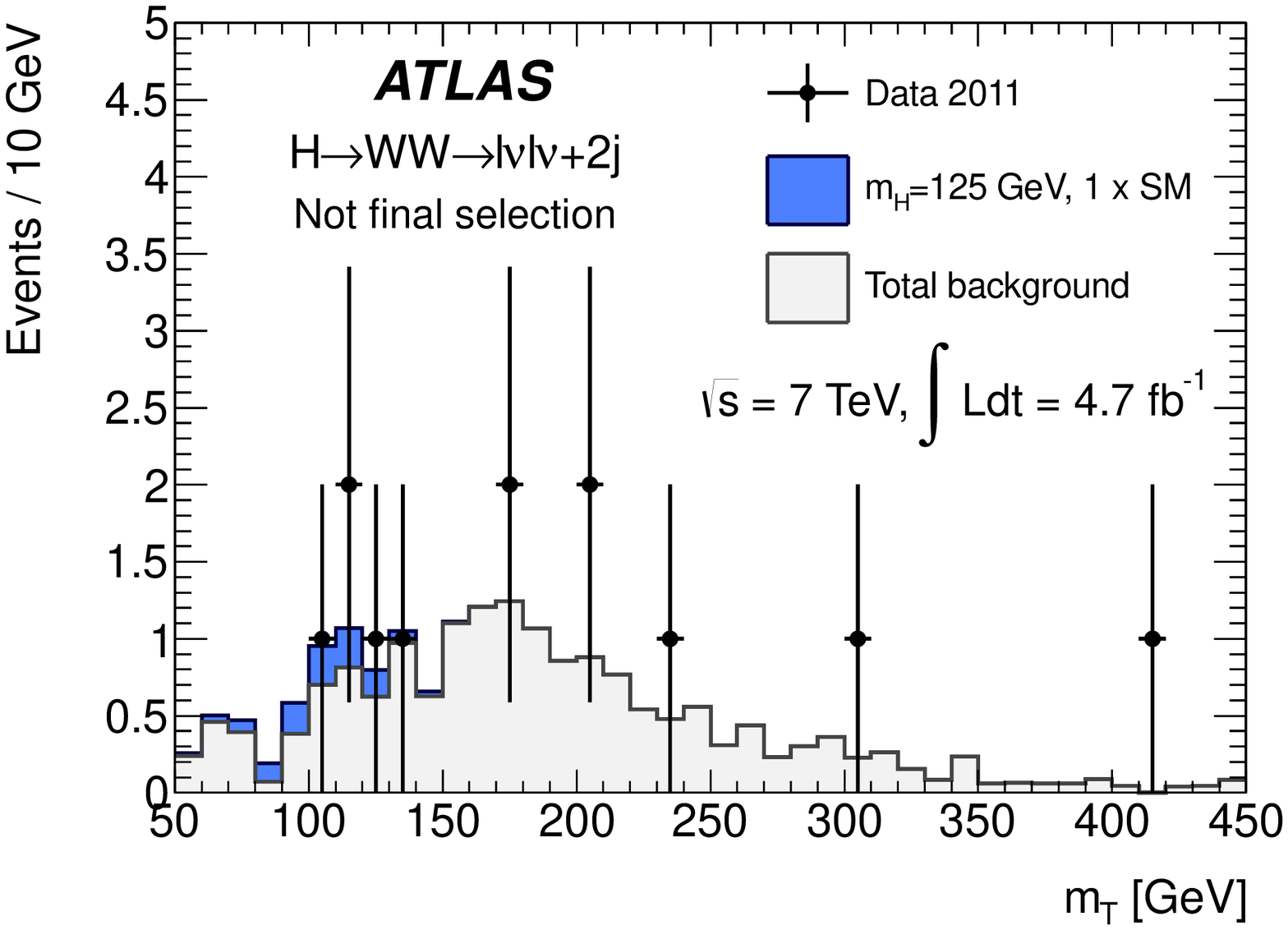}}
      \subfigure[]{    \includegraphics[width=.3\textwidth]{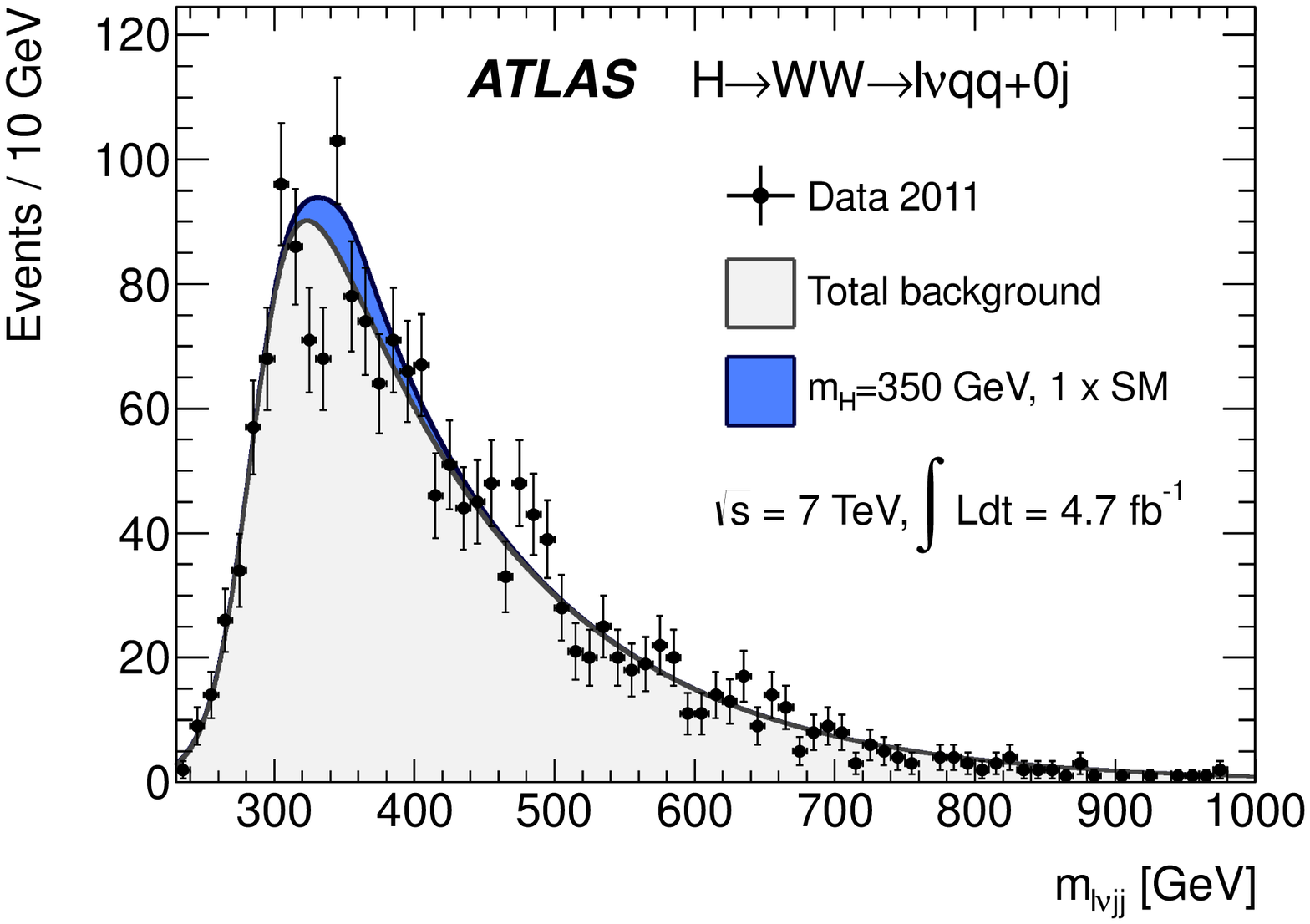}}
      \subfigure[]{    \includegraphics[width=.3\textwidth]{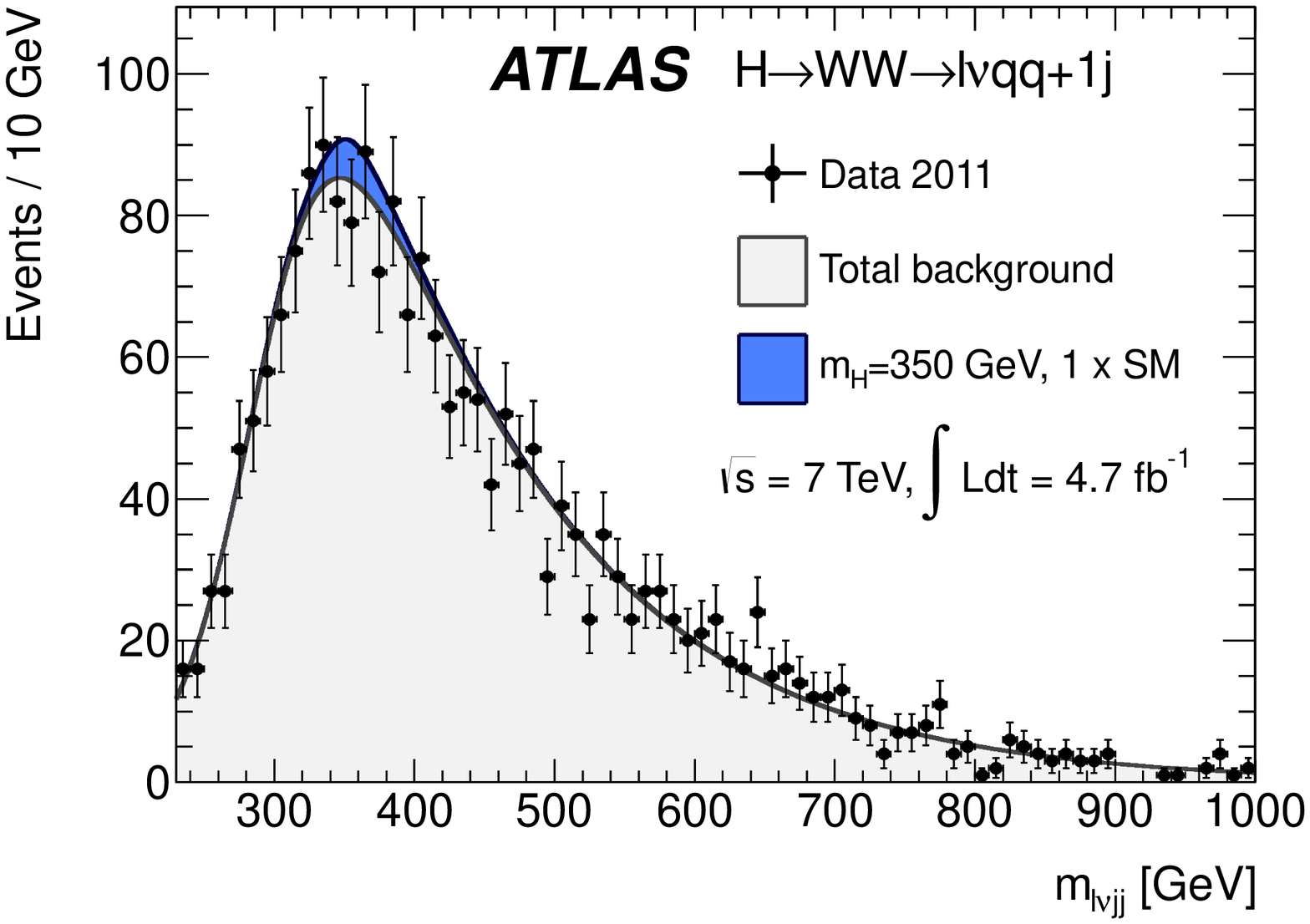}}
      \subfigure[]{    \includegraphics[width=.3\textwidth]{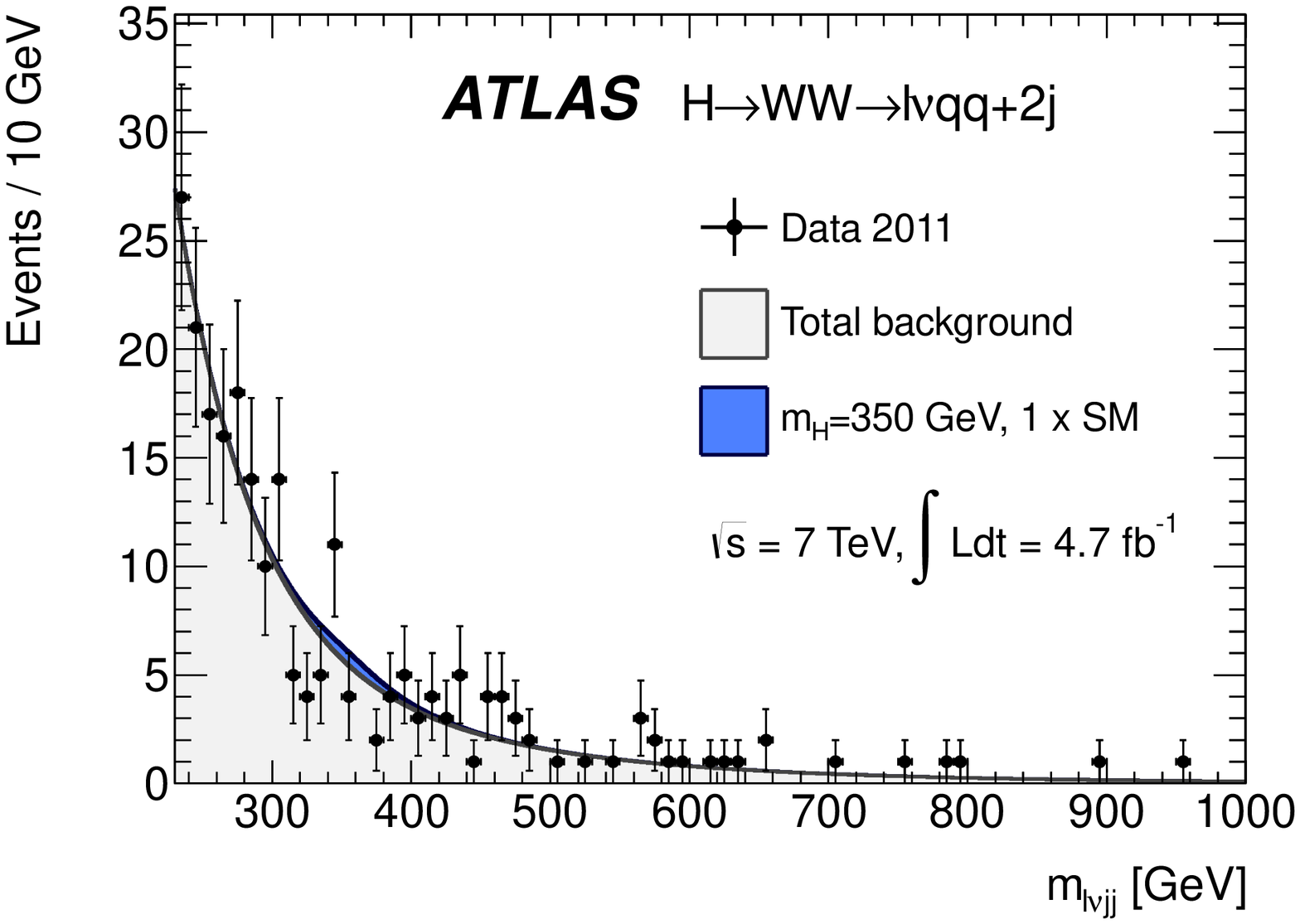}}
      \caption{Invariant or transverse mass distributions for the
        selected candidate events, the total background and the signal
        expected in the following channels: (a) \hgg, (b) \hZZllll\ in
        the entire mass range, (c) \hZZllll\ in the low mass range,
        (d) \hZZllnn, (e) {\it b-tagged} selection and (f) {\it
          untagged} selection for \hZZllqq, (g) \hWWlnln+0-jets, (h)
        \hWWlnln+1-jet, (i) \hWWlnln+2-jets, (j) \hWWlnqq+0-jets, (k)
        \hWWlnqq+1-jet and (l) \hWWlnqq+2-jets. The  \hWWlnln+2-jets distribution is shown before the final selection requirements are applied.}
       \label{fig:massplots1}	
    \end{center}
  \end{figure*}

   \begin{figure*}[h]
    \begin{center}
      \subfigure[]{ \includegraphics[width=.3\textwidth]{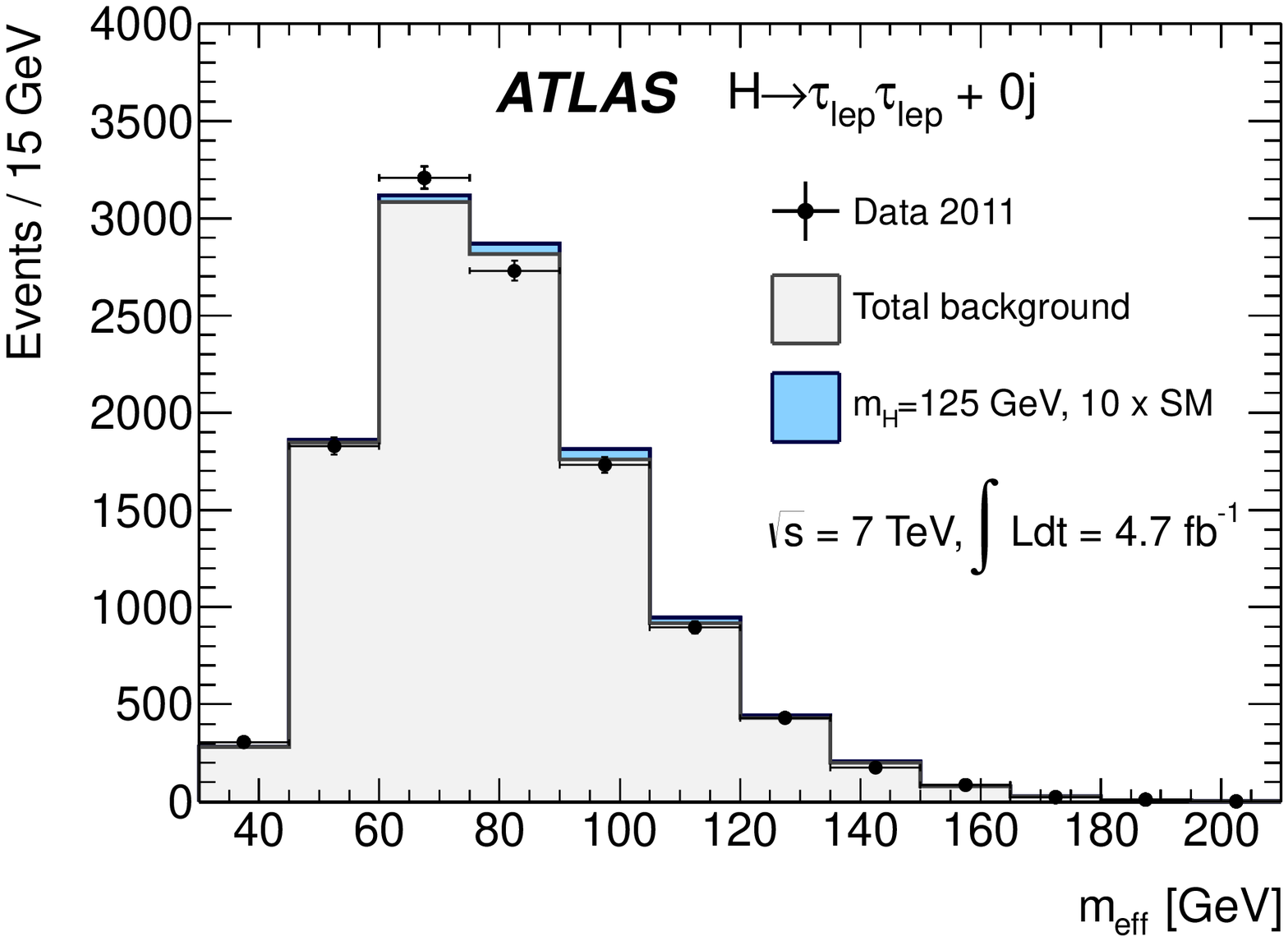}}
      \subfigure[]{ \includegraphics[width=.3\textwidth]{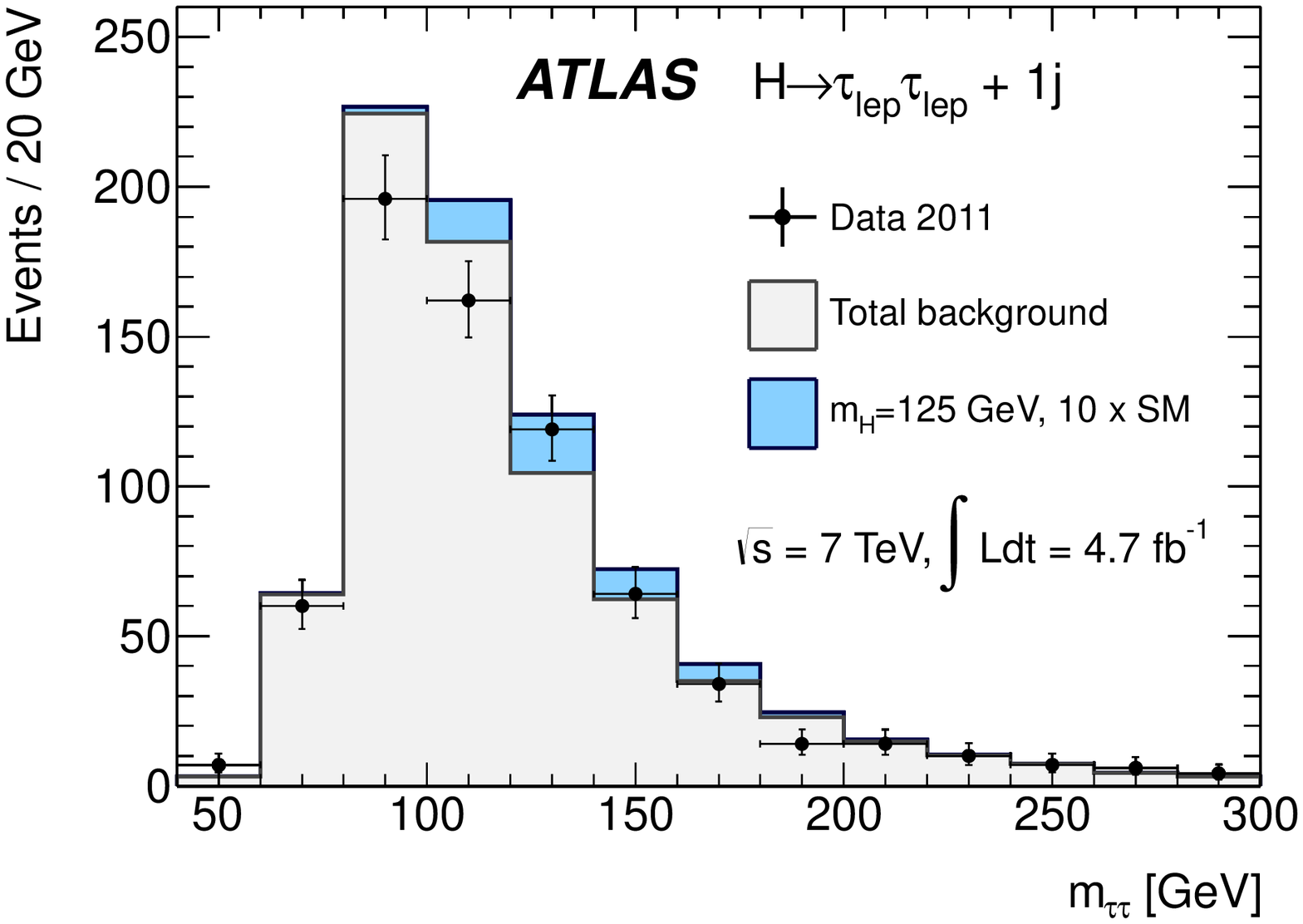}}
      \subfigure[]{ \includegraphics[width=.3\textwidth]{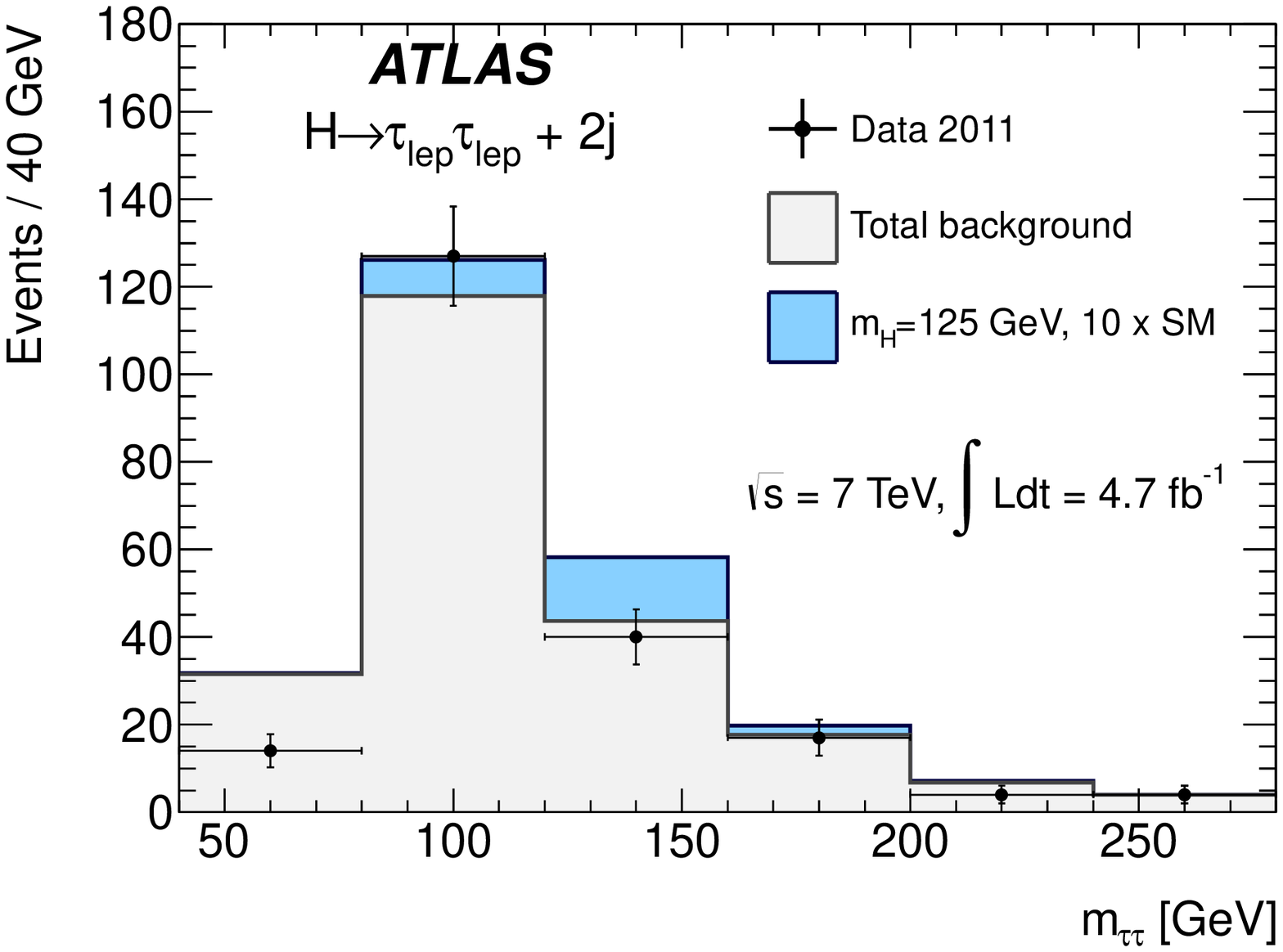}}
      \subfigure[]{ \includegraphics[width=.3\textwidth]{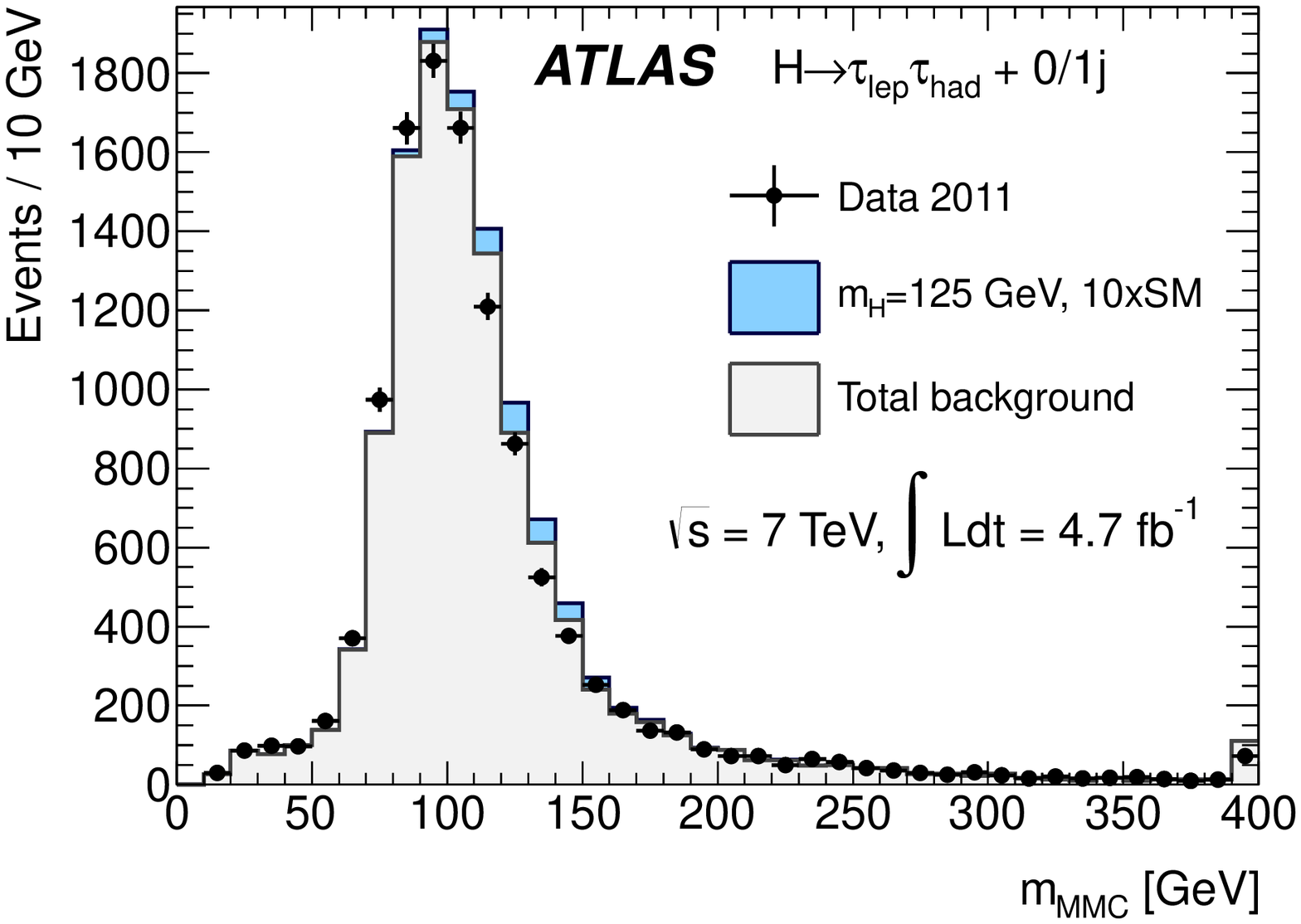}}
      \subfigure[]{ \includegraphics[width=.3\textwidth]{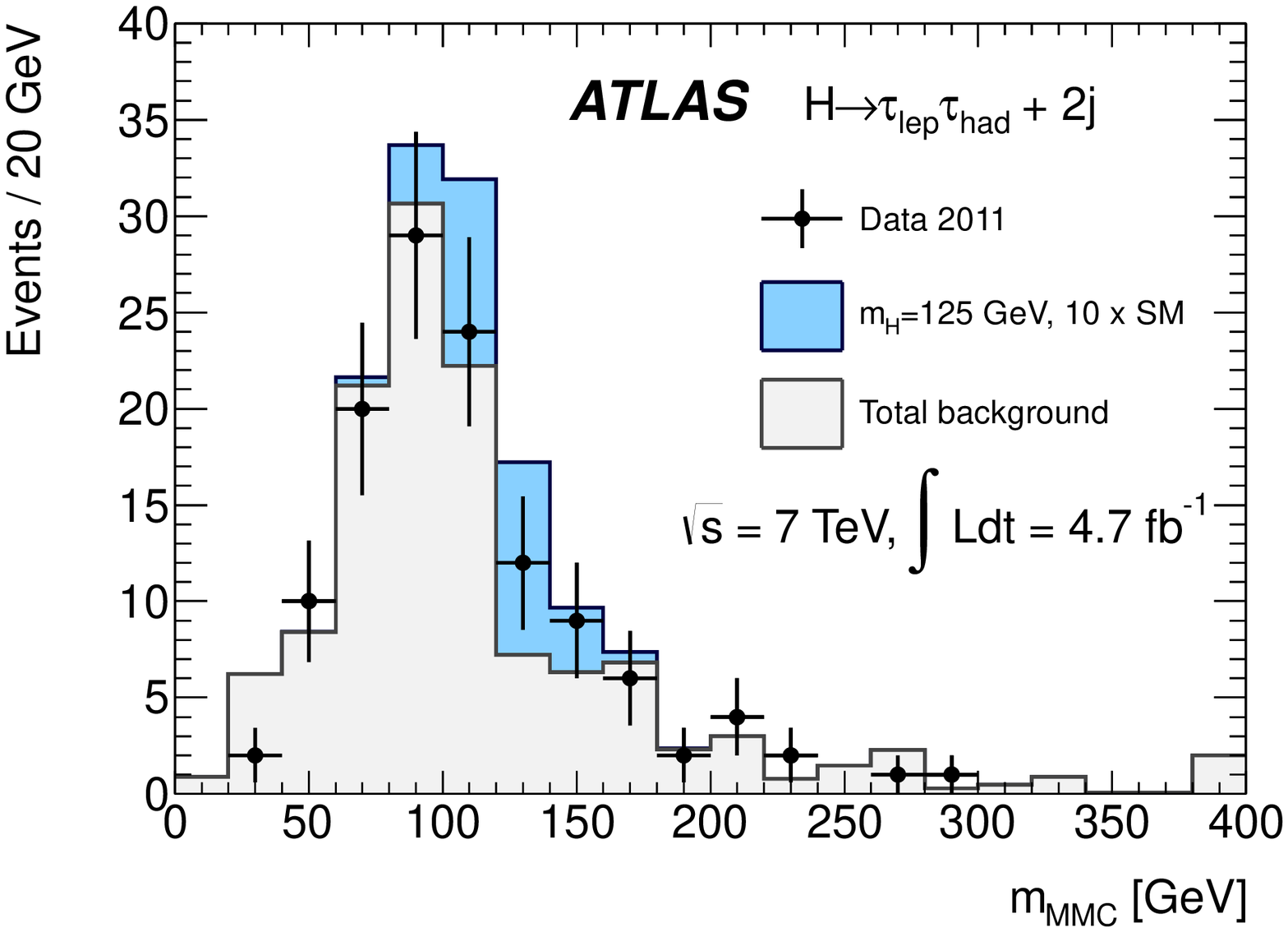}}
      \subfigure[]{ \includegraphics[width=.3\textwidth]{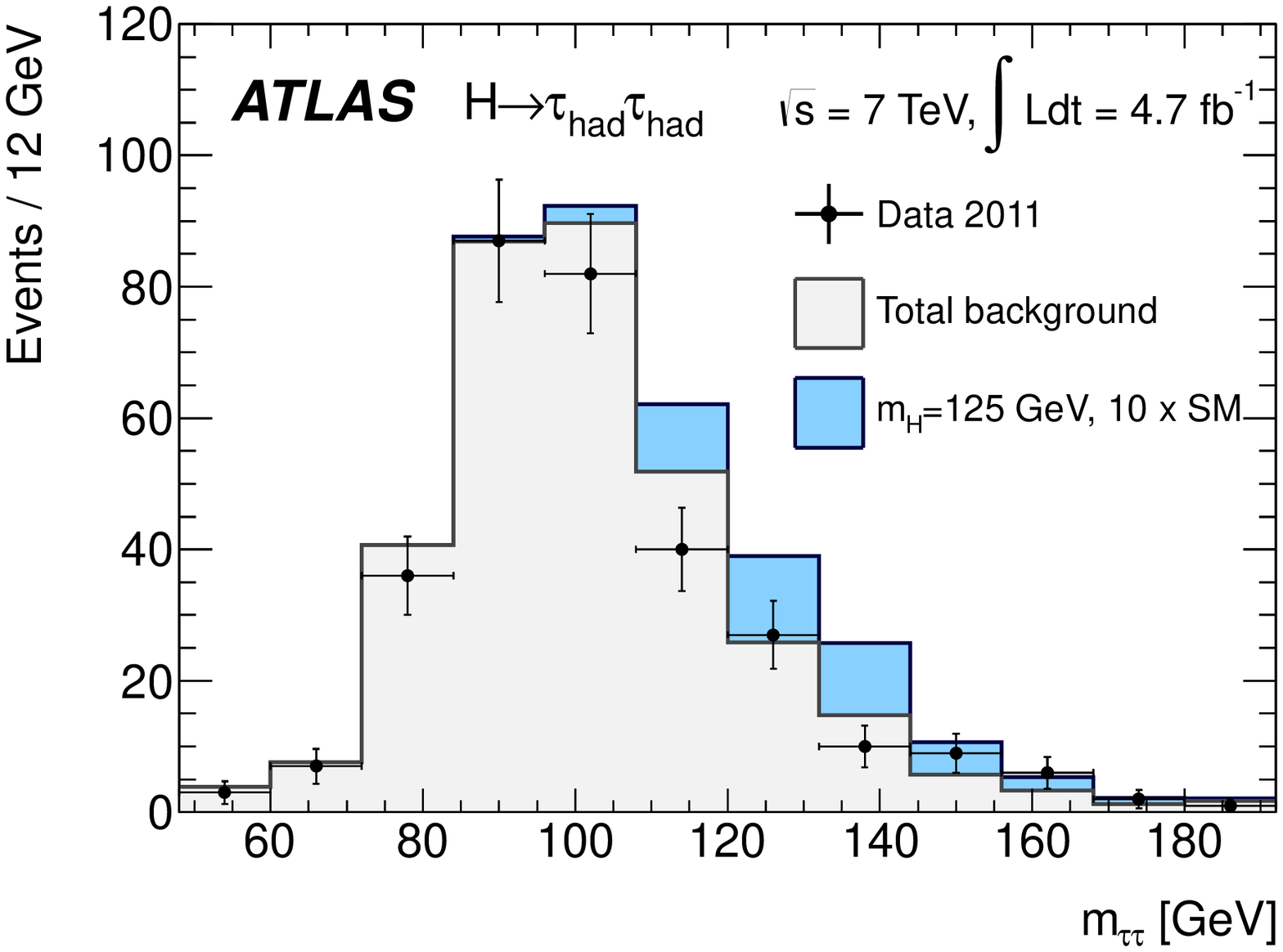}}
      \subfigure[]{ \includegraphics[width=.3\textwidth]{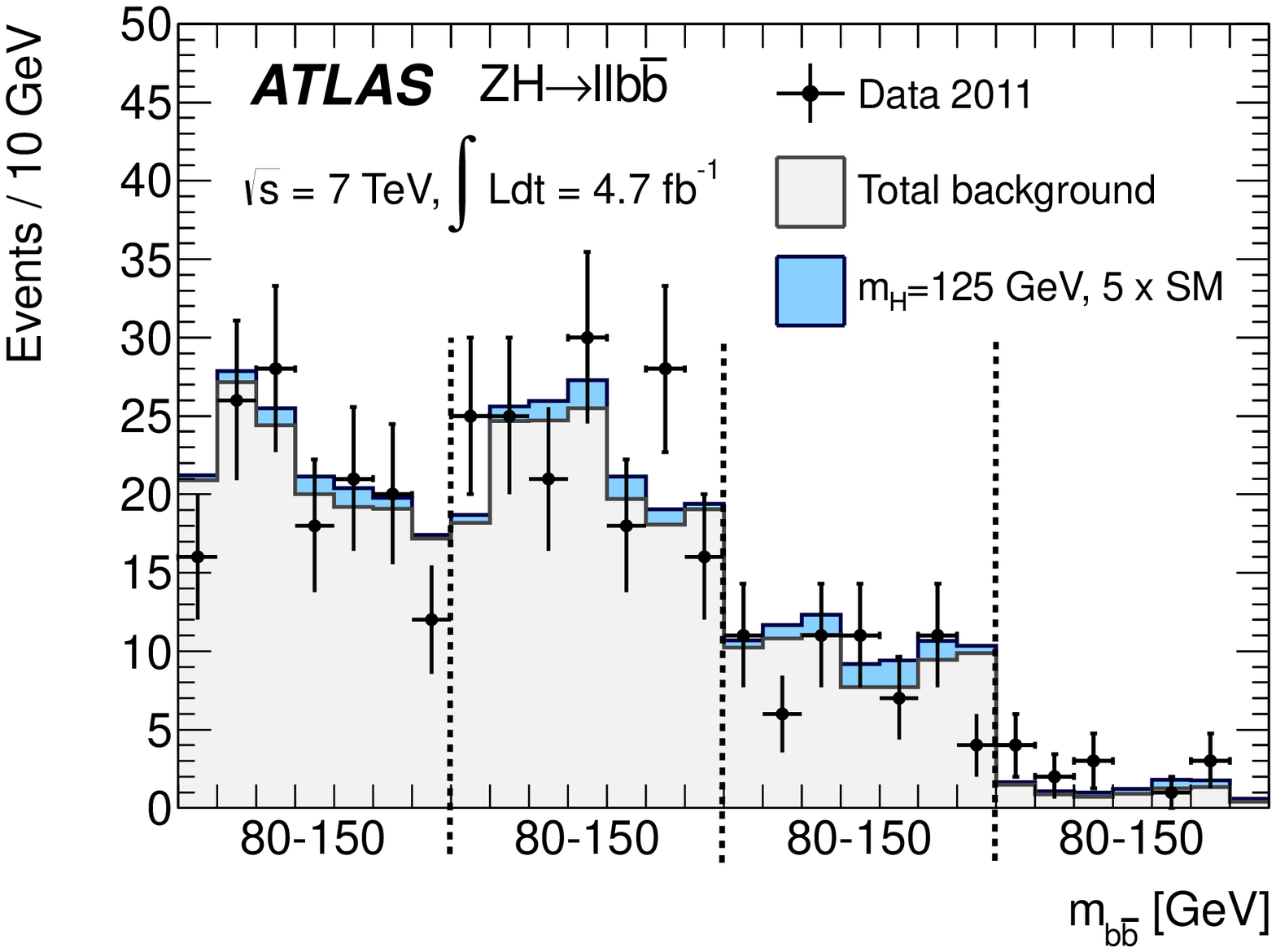}}
      \subfigure[]{ \includegraphics[width=.3\textwidth]{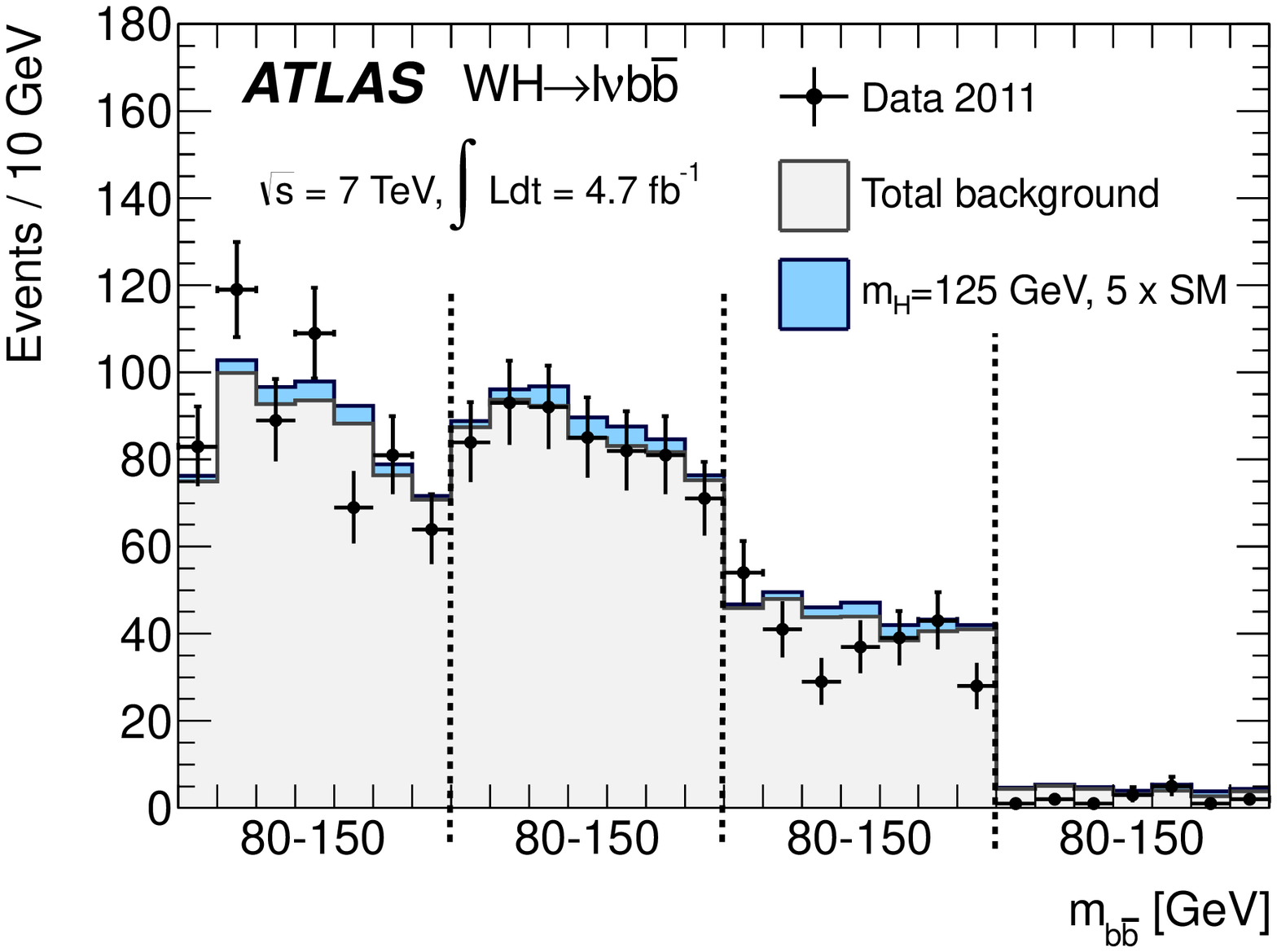}}
      \subfigure[]{ \includegraphics[width=.3\textwidth]{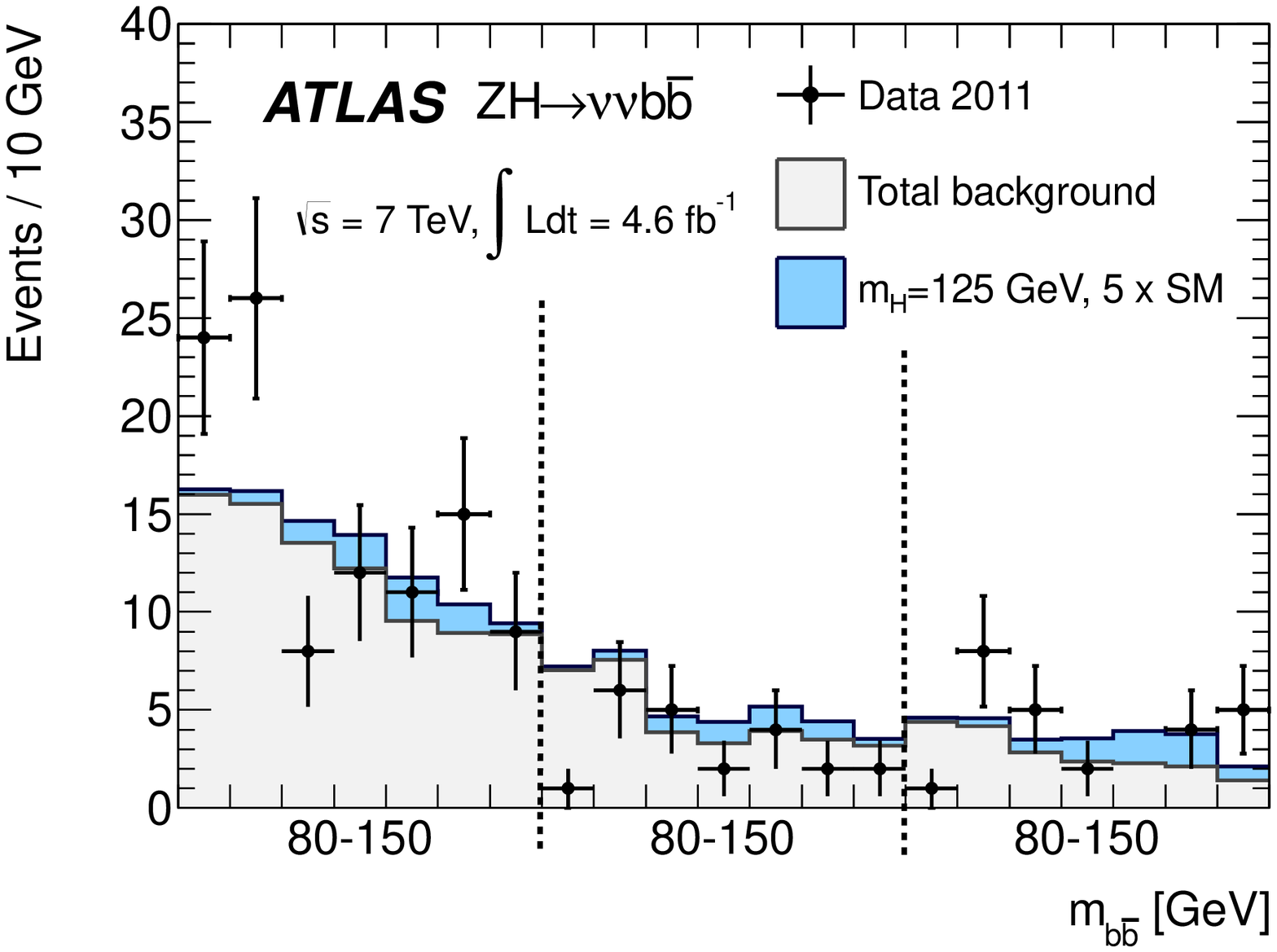}}
      \caption{Invariant or transverse mass distributions for the
        selected candidate events, the total background and the signal
        expected in the following channels: (a) \httll+0-jets, (b)
        \httll\ 1-jet, (c) \httll+2-jets, (d) \httlh+0-jets and
        1-jet, (e) \httlh+2-jets, (f) \htthh. The $b\overline{b}$
        invariant mass for (g) the $ZH \rightarrow \ell^+\ell^-
        b\bar{b}$, (h) the $WH \rightarrow \ell\nu b\bar{b}$ and (i)
        the $ZH \rightarrow \nu\overline{\nu} b\bar{b}$ channels. The
        vertical dashed lines illustrate the separation between the
        mass spectra of the subcategories in $p_{\rm T}^Z$, $p_{\rm
          T}^W$, and $E_{\rm T}^{\rm miss}$, respectively. The signal distributions
        are lightly shaded where they have been scaled by a factor of
        five or ten for illustration purposes. }
       \label{fig:massplots2}	
    \end{center}
  \end{figure*}
\endgroup

\clearpage

\begin{itemize}

\item \hzz: In the $ZZ^{(*)}$ decay mode at least one $Z$ is
  required to decay to charged leptons, while the other decays to
  either leptons, neutrinos, or jets.
  
  - \hZZllll: This analysis, described in Ref.~\cite{4lPaper}, is
  performed for \mH\ hypotheses in the 110~\GeV\ to 600~\GeV\ mass
  range and is unchanged with respect to the previous combined
  search~\cite{PreMoriondCombPaper}.  The main irreducible \ZZbkg\
  background is estimated using a combination of Monte Carlo
  simulation and the observed data. The reducible $Z$+jets background,
  which mostly impacts the low four-lepton invariant mass region, is
  estimated from control regions in the data. The top-quark (\ttbar)
  background normalization is validated using a dedicated control
  sample. Four categories of events are defined by the lepton flavor
  combinations and the four-lepton invariant mass is used as a
  discriminating variable.  The mass resolution is approximately
  1.5\%\ in the four-muon channel and 2\%\ in the four-electron
  channel for \mh$\sim$120~\GeV.

 - \hZZllnn: This analysis~\cite{may_llvv} is split into two regimes
 according to the level of pile-up, i.e. the average number of $pp$
  collisions per bunch crossing.  The first 2.3~\infb\ of data
 had an average of about six pile-up collisions per event and the
 subsequent 2.4~\infb\ had an average of about twelve.  The search is
 performed for \mH\ hypotheses ranging from 200~\GeV\ to 600~\GeV. The
 analysis is further categorized by the flavor of the leptons from the
 $Z$ decay. The selection is optimized separately for Higgs boson
 masses above and below 280~\GeV.  The $\ell^+\ell^{-}$ invariant mass
 is required to be within 15~\GeV\ of the $Z$ boson mass. The inverted
 requirement is applied to same-flavor leptons in the \hWWlnln\
 channel to avoid overlap in the selection. The transverse mass
 (\mT), computed from the dilepton transverse momentum and the
 missing transverse momentum, is used as a discriminating variable.

 - \hZZllqq: This search is performed for \mH\ hypotheses ranging from
 200~\GeV\ to 600~\GeV\ and is separated into search regions above and
 below \mH=300~\GeV, for which the event selections are independently
 optimized. The dominant background arises from $Z$+jets production,
 which is estimated using sidebands of the dijet invariant mass
 distribution in data.  To profit from the relatively large rate of
 $b$-jets from $Z$ boson decays present in the signal compared to the
 rate of $b$-jets found in the $Z$+jets background, the analysis is
 divided into two categories.  The first category contains events in
 which the two jets are $b$-tagged and the second uses events with
 less than two $b$-tags. The analysis~\cite{may_llqq} takes advantage
 of a highly efficient $b$-tagging algorithm~\cite{BTagging} and the
 sideband to constrain the background yield.  Using the $Z$
 boson mass constraint improves the mass resolution of the
 $\ell^+{\ell^-}q\overline{q}$ system by more than a factor of
 two. The invariant mass of the $\ell^+{\ell^-}q\overline{q}$ system
 is used as a discriminating variable.

\item \hww: Two sets of channels are devoted to the decay of the Higgs
  boson into a pair of $W$ bosons, namely the \lnln\ and \lnqq\
  channels.

  - \hWWlnln: The updated analysis~\cite{may_lvlv} is performed for
  \mH\ values from 110~\GeV\ up to 600~\GeV. Events with two leptons
  are classified by the number of associated jets (0, 1 or 2), where
  the two-jet category has selection criteria designed to enhance
  sensitivity to the VBF production process. The events are further
  divided by the flavors of the charged leptons, $ee$, $e\mu$ and
  $\mu\mu$ where the mixed mode ($e\mu$) has a much smaller background
  from the Drell-Yan process. As in the case of \hZZllnn, the samples
  are split according to the pile-up conditions and analyzed
  separately.  Each sub-channel uses the $WW$ transverse mass
  distribution, except for the 2-jets category, which does not use 
  a discriminating variable.

  - \hWWlnqq: This analysis is performed for \mH\ hypotheses ranging
  from 300~\GeV\ to 600~\GeV. A leptonically decaying $W$ boson is
  tagged with an isolated lepton and missing transverse momentum
  ($E_{\rm T}^{\rm miss}$).  Additionally, two jets with an invariant
  mass compatible with a second $W$ boson~\cite{may_lvqq} are required.  The $W$
  boson mass constraint allows the reconstruction of the Higgs boson
  candidate mass on an event-by-event basis by using a quadratic
  equation to solve for the component of the neutrino momentum along
  the beam axis. Events where this equation has imaginary solutions
  are discarded in order to reduce tails in the mass distribution.
  The analysis searches for a peak in the reconstructed $\ell \nu
  q\overline{q}'$ mass distribution.  The background is modeled with a
  smooth function.  The analysis is further divided by lepton flavor
  and by the number of additional jets (0, 1 or 2), where the two jet
  channel is optimized for the VBF production process.

\item \htt: The analyses~\cite{may_tautau} are categorized by the
  decay modes of the two $\tau$ leptons, for \mH\ hypotheses ranging
  from 110~\GeV\ to 150~\GeV\ (the leptonically decaying $\tau$
  leptons are denoted $\tau_{\rm lep}$ and the 
  hadronically decaying $\tau$ leptons are denoted $\tau_{\rm
    had}$). Most of these sub-channels are triggered using leptons,
  except for the fully hadronic channel $H\rightarrow \tau_{\rm had} \tau_{\rm had}$, which is triggered with
  specific double hadronic $\tau$ decay selections.  All the searches
  using $\tau$ decay modes have a significant background from $Z \ra
  \tau^+\tau^-$ decays, which are modeled using an embedding technique
  where $Z \ra \mu^+\mu^-$ candidates selected in the data have the
  muons replaced by simulated $\tau$ decays~\cite{may_tautau}. These
  embedded events are used to describe this background process.
  \vspace{1em}

  - \httll: In this channel events are separately analyzed in four
  disjoint categories based on the number of reconstructed jets in the
  event~\cite{may_tautau}.  There are two categories specifically
  aimed towards the gluon fusion production process, with or without a
  jet, one for the VBF production process and one for the Higgs boson
  production in association with a hadronically decaying vector
  boson. Each jet category requires at least one jet with \pt\ above
  40 \GeV.
  The collinear approximation~\cite{Ellis:1987xu} is used to
  reconstruct the $\tau\tau$ invariant mass, which is used as the
  discriminating variable.  All three combinations of $e$ and $\mu$
  are used, except in the 0-jets category, which uses only the $e\mu$
  candidate events where the effective mass is used as a
  discriminating variable.

  - \httlh: There are seven separate categories in this
  sub-channel. The selection of VBF-like events requires two jets with
  oppositely signed pseudorapidities $\eta$, $|\Delta \eta_{jj}|>3.0$
  and a dijet invariant mass larger than 300 \GeV, in which events
  with electrons and muons are combined due to the limited number of
  candidates. In the other sub-channels, electron and muon final
  states are considered separately. The remaining candidate events are
  categorized according to the number of jets with transverse momenta
  in excess of 25~\GeV, the 0-jets category being further subdivided
  based on whether the $E_{\rm T}^{\rm miss}$ exceeds 20~\GeV\ or
  not. The Missing Mass Calculator (MMC)
  technique~\cite{Elagin:2010aw} is used to estimate the $\tau\tau$
  invariant mass, which is used as a discriminating variable.

  - \htthhj: Events are triggered using a selection of two
  hadronically decaying $\tau$ leptons with transverse energy
  thresholds varying according to the running
  conditions~\cite{may_tautau}. Two oppositely-charged hadronically
  decaying $\tau$ candidates are required along with one jet with
  transverse momentum larger than 40~\GeV, $E_{\rm T}^{\rm
    miss}>20$~\GeV\ and a reconstructed invariant mass of the two
  $\tau$ leptons and the jet greater than 225~\GeV.  In addition to
  the $Z$ background there is a significant multijet background which
  is estimated using data-driven methods.  The $\tau\tau$ invariant
  mass is estimated via the collinear approximation and is used as a
  discriminating variable after further selections on the momentum
  fractions carried away by visible $\tau$ decay products.

\item \hbb: The \hZllbb, \hZvvbb, and \hWlvbb\ analyses~\cite{may_bb} are
  performed for \mH\ ranging from 110~\GeV\ to 130~\GeV.  All three
  analyses require two $b$-tagged jets (one with $p_{\rm T}>45$~\GeV\
  and the other with $p_{\rm T}>25$~\GeV) and the invariant mass of
  the two $b$-jets, $m_{bb}$, is used as a discriminating variable.
  The \hZllbb\ analysis requires a dilepton invariant mass in the
  range 83~\GeV$< m_{\ell\ell} < 99$~\GeV\ and $E_{\rm T}^{\rm
    miss}<50$~\GeV\ to suppress the \ttbar\ background.  The \hWlvbb\
  analysis requires $E_T^{\rm miss}>25$~\GeV, the transverse mass of
  the lepton-$E_{\rm T}^{\rm miss}$ system to be in excess of
  $40$~\GeV, and no additional leptons with $p_{\rm T}>20$~\GeV. The
  \hZvvbb\ analysis requires $E_{\rm T}^{\rm miss}>120$~\GeV, as well
  as $p_{\rm T}^{\rm miss}>30$~\GeV, where $p_{\rm T}^{\rm miss}$ is
  the missing transverse momentum determined from the tracks
  associated with the primary vertex.  To increase the sensitivity of
  the search, the $m_{bb}$ distribution is examined in sub-channels
  with different signal-to-background ratios. In the searches with one
  or two charged leptons, the division is made according to four bins
  in transverse momentum $p_{\rm T}^V$ of the reconstructed vector
  boson $V$: $p_{\rm T}^V <50$~\GeV, 50~\GeV~$\le p_{\rm T}^V
  <100$~\GeV, 100~\GeV~$\le p_{\rm T}^V < 200$~\GeV\ and $p_{\rm T}^V
  \ge 200$~\GeV. In the $ZH\to\nu\bar{\nu}b\overline{b}$ search the
  $E_{\rm T}^{\rm miss}$ is used to define three sub-channels
  corresponding to 120~\GeV~$< E_{\rm T}^{\rm miss} < 160$~\GeV,
  160~\GeV~$\le E_{\rm T}^{\rm miss} < 200$~\GeV, and $E_{\rm T}^{\rm
    miss} \ge 200$~\GeV.  No categorization is made based on lepton
  flavor.
\end{itemize}

\section{Systematic Uncertainties}
\label{sec:syst}

The sources of systematic uncertainties and their effects on the
signal and background rates $\nu_k(\vec\alpha)$ and discriminating
variable distributions $f_k(x|\vec\alpha)$ are described in detail for
each channel in
Refs.~\cite{ggPaper,4lPaper,may_lvlv,may_lvqq,may_llvv,may_llqq,may_tautau,may_bb}.
The sources of systematic uncertainty are decomposed into uncorrelated
components, such that the constraint terms factorize as in
Eq.~(\ref{Eq:ftot}).  The main focus of the combination of channels is
the correlated effect of given sources of uncertainties across
channels.  Typically, the correlated effects arise from the
ingredients common to several channels, for example the simulation,
the lepton and photon identification, and the integrated luminosity. The sources
of systematic uncertainty affecting the signal model are frequently
different from those affecting the backgrounds, which are often
estimated from control regions in the data.  The dominant
uncertainties giving rise to correlated effects are those associated
with theoretical predictions for the signal production cross sections
and decay branching fractions, as well as those related to detector
response affecting the reconstruction of electrons, photons, muons,
jets, $E_{\rm T}^{\rm miss}$ and $b$-tagging. The log-normal
constraint terms are used for uncertainties in the signal and
background normalizations, while Gaussian constraints are used for
uncertainties affecting the shapes of the pdfs.

\subsection{Theoretical  Uncertainties Affecting the Signal}

The Higgs boson production cross sections are computed up to
next-to-next-to-leading order (NNLO)
\cite{Djouadi1991,Dawson:1990zj,Spira:1995rr,Harlander:2002wh,Anastasiou:2002yz,Ravindran:2003um}
in QCD for the gluon fusion (\ggF) process, including soft-gluon
resummation up to next-to-next-to-leading log
(NNLL)~\cite{Catani:2003zt} and next-to-leading-order (NLO)
electroweak corrections~\cite{Aglietti:2004nj,Actis:2008ug}.  These
predictions are compiled in
Refs.~\cite{Anastasiou:2008tj,deFlorian:2009hc,Baglio:2010ae}.  The
cross section for the VBF process is estimated at
NLO~\cite{Ciccolini:2007jr,Ciccolini:2007ec,Arnold:2008rz} and
approximate NNLO QCD~\cite{Bolzoni:2010xr}.
The cross sections for the associated production processes (\VH) are
computed at NLO~\cite{Han:1991ia,Ciccolini:2003jy},
NNLO~\cite{Brein:2003wg} QCD and NLO
electroweak~\cite{Ciccolini:2003jy}.
The cross section for the associated production with a \ttbar\ pair
(\ttH) are estimated at
NLO~\cite{Beenakker:2001rj,Beenakker:2002nc,Reina:2001sf,Dawson:2002tg,Dawson:2003zu}.
The Higgs boson production cross sections and decay branching
ratios~\cite{Djouadi:1997yw,hdecay2,Bredenstein:2006rh,Bredenstein:2006ha,Actis:2008ts},
as well as their related uncertainties, are compiled in
Ref.~\cite{LHCHiggsCrossSectionWorkingGroup:2011ti}.  The QCD scale
uncertainties for \mH=120~\GeV\ amount to $^{+12}_{-8}$\%\ for the
\ggF\ process, $\pm1$\%\ for the \VBF\ and associated $WH/ZH$
processes, and $^{+3}_{-9}$\%\ for the \ttH\ process. The
uncertainties related to the parton distribution functions (PDF)
amount to $\pm 8$\%\ for the predominantly gluon-initiated processes
\ggF\ and \ggttH, and $\pm 4$\%\ for the predominantly quark-initiated
processes \VBF\ and $WH/ZH$ processes~\cite{Botje:2011sn}. The
theoretical uncertainty associated with the exclusive Higgs boson
production process with one additional jet in the \hWWlnln\ channel
amounts to $\pm 20$\%\ and is treated according to the prescription of
Refs.~\cite{LHC-HCG,Dittmaier:2012vm}.  An additional theoretical
uncertainty on the signal normalization, to account for effects
related to off-shell Higgs boson production and interference with
other SM processes, is assigned at high Higgs boson masses ($\mH>
300$~\GeV) and estimated as
$\pm$150\%$\times$(\mH/TeV)$^3$~\cite{lhcCombination,Passarino:2010qk,Anastasiou:2011pi,Dittmaier:2012vm}.

\subsection{Theoretical Uncertainties Affecting the Background}

In the \hgg\ and \hWWlnqq\ channels the backgrounds are estimated from
a fit to the data. This removes almost all sensitivity to the
corresponding theoretical uncertainties. In the case of the \hgg\
analysis an additional uncertainty is assigned to take into account
possible inadequacies of the analytical background model chosen.  
Theoretical uncertainties enter in all other channels where theoretical 
calculations are used for background estimates. In particular,
both signal and background processes are sensitive to the parton
distribution functions, the underlying event simulation, and the
parton shower model.

The $ZZ^{(*)}$ continuum process is the main background for the
\hZZllll\ and \hZZllnn\ analyses and is also part of the backgrounds
in the \hZZllqq\ channel. An NLO prediction~\cite{mcfm6} is used for
the normalization. The QCD scale uncertainty has a $\pm5$\%\ effect on
the expected $ZZ^{(*)}$ background, and the effect due to the PDF and
$\alpha_{\rm S}$ uncertainties is $\pm 4$\%\ and $\pm$8\% for
quark-initiated and gluon-initiated processes respectively. An
additional theoretical uncertainty of $\pm$10\%\ on the inclusive
$ZZ^{(*)}$ cross section is conservatively included due to the missing
higher-order QCD corrections for the gluon-initiated process. This
theoretical uncertainty is treated as uncorrelated for the different
channels due to the different acceptance in the \hZZllll\ and
\hZZllnn\ channels and because its contribution to the \hZZllqq\
channel is small.

In most other channels the overall normalization of the main
backgrounds is not estimated from theoretical predictions; however,
simulations are used to model the pdfs $f(x|\vec\alpha)$ or the scale
factors used to extrapolate from the control regions to the signal
regions.  For example, in the \hWWlnln\ channel the main backgrounds are continuum
$WW^{(*)}$ and $t\overline{t}$ production. Their normalizations are
estimated in control regions; however, the factors used to extrapolate
to the signal region are estimated with the NLO
simulation~\cite{mcatnlo}.

\subsection{Experimental Uncertainties}

The uncertainty on the integrated luminosity is considered as being
fully correlated among channels and amounts to $\pm$3.9\%~\cite{Luminosity,
  LuminosityCONF}.

The detector-related sources of systematic uncertainty can affect
various aspects of the analysis: (a) the overall normalization of the
signal or background, (b) the migration of events between categories
and (c) the shape of the discriminating variable distributions
$f(x|\vec\alpha)$.  Similarly to the theoretical uncertainties,
experimental uncertainties on the event yields (a) are treated using a
log-normal $f_p(a_p|\alpha_p)$ constraint pdf. In cases (b) and (c) a
Gaussian constraint is applied.

The experimental sources of systematic uncertainty are modeled using
the classification detailed below. Their effect on the signal and
background yields in each channel separately is reported in
Table~\ref{tab:sys}.   The various sources of systematic uncertainty
have in some cases been grouped for a concise presentation (e.g. the
jet energy scale and $b$-tagging efficiencies), while the full
statistical model of the data provides a more detailed account of the
various systematic effects including the effect on the pdfs
$f(x|\vec\alpha)$. The assumptions made in the treatment of
systematics are outlined below.

\begingroup
\begin{table*}[htb]
  \begin{center}
    \vspace{-0.5cm}
    \caption{ A summary of the main correlated experimental systematic
      uncertainties. The uncorrelated systematic uncertainties are
      summarized in a single combined number. The uncertainties
      indicate the $\pm 1\sigma$ relative variation in the signal and
      background yields in (\%). The signal corresponds to a Higgs
      boson mass hypothesis of 125~\GeV\ except for \hZZllqq,
      \hZZllnn\ and \hWWlnqq, which are quoted at 350~\GeV.}
    \label{tab:sys}
    \vspace{0.5cm}
    \begin{tabular}{lccccccccccc}
      \hline \hline
      & \multicolumn{1}{c}{\hgg} & \multicolumn{3}{c}{\hzz}  & \multicolumn{2}{c}{\hww}  & \multicolumn{3}{c}{\htt} & \multicolumn{2}{c}{\hbb} \\
      \vspace{0.04cm}
      &   & \multicolumn{1}{c}{$\ell\ell\ell\ell$}  & \multicolumn{1}{c}{$\ell\ell\nu\nu$}  & \multicolumn{1}{c}{$\ell\ell qq$}  & \multicolumn{1}{c}{$\ell\nu \ell\nu$}  & \multicolumn{1}{c}{$\ell\nu qq$}   &\multicolumn{1}{c}{\lh}   &\multicolumn{1}{c}{\tauleptaulep}   &\multicolumn{1}{c}{\hh}   &\multicolumn{1}{c}{$ZH$}   &\multicolumn{1}{c}{$WH$}   \vspace{0.04cm} \\ 
      \hline
      \hline
      \multicolumn{12}{c}{Relative Uncertainty on Signal Yields}\\
      \hline\hline
      \vphantom{\rule{0cm}{0.4cm}} Luminosity    & $\pm 3.9$  &  $\pm 3.9$  &  $\pm 3.9$  &  $\pm 3.9$   & $\pm 3.9$  & $\pm 3.9$  & $\pm 3.9$  &  $\pm 3.9$  & $\pm 3.9$  & $\pm 3.9$  & $\pm 3.9$   \\
      \hline\hline
      \vspace{0.04cm}
      \vphantom{\rule{0cm}{0.4cm}} e/$\gamma$ efficiency   & ~$^{+13.5}_{-11.9}$   & $\pm3.2$  & ~$\pm1.3$  & - & $\pm1.5$  & ~$\pm0.9$ & ~$\pm 2.9$  &  ~$\pm 2.0$ & -  & ~$\pm1.2$  & - \\
      \vspace{0.04cm}
      \vphantom{\rule{0cm}{0.4cm}} e/$\gamma$ energy scale & -  & -   &$\pm0.4$  & - & $\pm0.7$  &  -  &  ~$^{+1.4}_{+0.3}$  & ~$\pm0.3$  & -  & ~$^{+0.3}_{-0.4}$  & $\pm0.2$ \\
      \vspace{0.04cm}
      \vphantom{\rule{0cm}{0.4cm}} e/$\gamma$ resolution  & -     &  -   & $\pm 0.1$   &  - & $\pm 0.1$  & -   & - & ~$^{+0.2}_{-0.5}$   & -  & - & ~$^{-0.2}_{-0.1}$  \\ 
      \hline
      \vspace{0.04cm}
      \vphantom{\rule{0cm}{0.4cm}}  $\mu$ efficiency      & -   &$\pm0.2$    & $\pm 0.4$ & -  & $\pm0.1$   & ~$\pm0.3$  & $\pm 1.0$  & $\pm 2.0$   & -  & $\pm 0.4$  &  -  \\
      \vspace{0.04cm}
      \vphantom{\rule{0cm}{0.4cm}} $\mu$ resolution (ID) & -  &  - & $\pm 0.1$ &  -    & $\pm 0.1$    & -  & - &  ~$^{+0.2}_{-0.5}$  & -  & $\pm 0.1$ &  -   \\
      \vspace{0.04cm}
      \vphantom{\rule{0cm}{0.4cm}} $\mu$ resolution (MS) & -  &  - &  $\pm 0.1$  &  -  & $\pm 0.1$   &  -  & -  & - & -  & ~$^{+0.2}_{-0.1}$  & -  \\ 
      \hline
      \vspace{0.04cm}
      \vphantom{\rule{0cm}{0.4cm}} Jet/$E_{\rm T}^{\rm miss}$ energy scale  & -   &  - &$\pm 2.5$ & ~$^{+3.5}_{-3.4}$   & ~$^{+2.2}_{-3.4}$  & ~$^{+7.6}_{-7.0}$  &  ~$^{+1.3}_{-1.8}$  &  $\pm 0.9$  & ~$^{+13.7}_{-16.5}$   & ~$^{+4.1}_{-5.0}$  & ~$^{+2.7}_{-5.1}$  \\
      \vspace{0.04cm}
      \vphantom{\rule{0cm}{0.4cm}} Jet energy resolution   & -   & -  & $\pm 1.1$   &  $\pm 4.2$   & $\pm 1.1$    & ~$^{+8.4}_{-7.8}$   & -   & $\pm 0.3$  & $\pm 2.4$  & $\pm2.9$  & $\pm 2.6$  \\
      \hline
      \vspace{0.04cm}
      \vphantom{\rule{0cm}{0.4cm}} $b$-tag efficiency      & -   & - & ~$\pm0.9$   & $\pm0.02$ & ~$\pm0.02$   & ~$^{+6.1}_{-5.7}$   & -  &  -  & -   & $\pm9.5$  & $\pm9.4$  \\ \hline
      \vspace{0.04cm}
      \vphantom{\rule{0cm}{0.4cm}} $\tau$ efficiency       & - & -  & - & -  & -  & -  & ~$\pm 4.2$  & -  & ~$\pm 8.0$  & -  & - \\ \hline \hline
      \vspace{0.04cm}
      \vphantom{\rule{0cm}{0.4cm}} Uncorrelated Uncertainties & $\pm 0.2$  & $\pm 5.0$ & ~$^{+7.8}_{-7.2}$  & ~$^{+12.0}_{-10.7}$  & $\pm 3.4$  & $\pm 1.5$  & -  & ~$^{+1.1}_{-2.0}$  & ~$^{+3.7}_{-4.2}$ & $\pm 1.7$  & $\pm 2.8$ \\
      \hline\hline
      \multicolumn{12}{c}{Relative Uncertainty on Background Yields}\\
      \hline\hline
      \vphantom{\rule{0cm}{0.4cm}} Luminosity     & -   & ~$^{+3.7}_{-3.5}$   &  ~$^{+2.8}_{-2.7}$   & ~$\pm 0.2$  & ~$\pm 0.5$  & -   & ~$^{+2.7}_{-2.6}$  & ~$^{+3.5}_{-3.4}$  & -  & -   &  -  \\
      \hline\hline
      \vspace{0.04cm}
      \vphantom{\rule{0cm}{0.4cm}} e/$\gamma$ efficiency   & -   & $\pm1.8$   & $\pm0.9$  & -  & $\pm 1.4$  & $\pm0.9$ & $\pm2.0$ & $^{+0.5}_{-1.4}$ & -   & -  & - \\
      \vspace{0.04cm}
      \vphantom{\rule{0cm}{0.4cm}} e/$\gamma$ energy scale  & -   &  -   &  $^{+3.1}_{-2.2}$ & -   & $^{+0.5}_{-0.4}$   & -  &  ~$^{+0.8}_{-0.5}$  & $\pm0.7$  & -  & $\pm0.1$  & $\pm 0.3$ \\
      \vspace{0.04cm}
      \vphantom{\rule{0cm}{0.4cm}} e/$\gamma$ resolution  & -   &  -  & $^{+1.1}_{-0.8}$  & -   & $\pm 0.2$  & - & - & ~$^{+1.6}_{-1.7}$   & - &  ~$^{+0.6}_{-0.2}$  & $\pm 0.1$ \\ 
      \hline
      \vspace{0.04cm}
      \vphantom{\rule{0cm}{0.4cm}}  $\mu$ efficiency        & -    &$\pm0.1$  & ~$\pm0.3$  & - & $\pm0.12$   & ~$\pm0.3$  & $\pm 0.7$ &  $^{+0.5}_{-1.5}$  &  -    & -   & - \\
      \vspace{0.04cm}
      \vphantom{\rule{0cm}{0.4cm}} $\mu$ resolution (ID) & -  &  -  & ~$\pm 0.2$  & -    & $\pm 0.2$    &  -  & - & ~$^{+1.6}_{-1.8}$    &  - & $\pm 0.1$  & -  \\
      \vspace{0.04cm}
      \vphantom{\rule{0cm}{0.4cm}} $\mu$ resolution (MS)  & -   &  -  &  $\pm 0.2$ & -   & $\pm0.2$   & -  & -  & -  & -  & $\pm 0.1$ &  -  \\ 
      \hline
      \vspace{0.04cm}
      \vphantom{\rule{0cm}{0.4cm}} Jet/$E_{\rm T}^{\rm miss}$  energy scale  & -  & -  & ~$^{+6.1}_{-4.6}$   & $\pm 0.4$  & ~$^{+4.0}_{-5.6}$  & - & - & -  & -   & ~$^{+0.5}_{-0.0}$ & ~$^{+2.2}_{-1.6}$ \\
      \vspace{0.04cm}
      \vphantom{\rule{0cm}{0.4cm}} Jet energy resolution    & -  & -   & $\pm 1.7$ & $\pm 0.1$   & $\pm 1.2$  & -  & -   & -   & -  & $\pm 0.3$  & $\pm 0.9$  \\
      \hline
      \vspace{0.04cm}
      \vphantom{\rule{0cm}{0.4cm}} $b$-tag efficiency     & -  &  -  & ~$^{+5.2}_{-4.4}$  & -  & ~$^{+1.4}_{-1.1}$  & -  & - & ~$\pm 0.1$  & - &  $\pm 1$ & $\pm 1$ \\ \hline
      \vspace{0.04cm}
      \vphantom{\rule{0cm}{0.4cm}} $\tau$ efficiency        & -  & - & -  & - & -  & -  & ~$\pm 3.0$  & - & -  & -  & -\\ \hline \hline
      \vspace{0.04cm}
      \vphantom{\rule{0cm}{0.4cm}} Uncorrelated Uncertainties & -  & ~$\pm 10.0$  & ~$\pm 4.9$ & ~$^{+2.3}_{-2.1}$ & ~$\pm 12.0$  & -  & ~$\pm10.2$  & ~$^{+5.5}_{-6.3}$  & ~$\pm10.3$ & ~$\pm 5.5$  & ~$^{+2.8}_{-2.9}$ \\
      \hline\hline
    \end{tabular}
  \end{center}
\end{table*} 
\endgroup

\begin{itemize}

\item The uncertainty in the trigger and identification efficiencies are
 treated as fully correlated for electrons and photons. 
 The energy scale and resolution
  for photons and electrons are treated as uncorrelated sources of uncertainty.

\item 
  The uncertainties affecting muons are separated 
  into those related to the inner detector (ID) and
  the muon spectrometer (MS)  in order to provide a
  better description of the correlated effect among channels using
  different muon identification criteria and different ranges of muon
  transverse momenta.

\item The Jet Energy Scale  and Jet Energy Resolution
  are sensitive to a number of uncertain quantities, 
  which depend on $p_T$, $\eta$, and flavor of the jet. 
  Measurements of the JES and JER result in complicated correlations 
  among these components.  Building a  complete model of the response 
  to these correlated sources of uncertainty is intricate.  Here,
  a simplified scheme is used in which independent JES and JER nuisance 
  parameters  are associated to channels with significantly different
  kinematic requirements and scattering processes with different
  kinematic distributions or flavor composition.  This scheme includes
  a specific treatment for $b$-jets. The sensitivity of the results to
  various assumptions in the correlation between these sources of
  uncertainty has been found to be negligible. Furthermore, an
  additional component to the uncertainty in $E_{\rm T}^{\rm miss}$,
  which is uncorrelated with the JES uncertainty, is included.

\item While the $\tau$ energy scale uncertainty is expected to be
  partially correlated with the JES, here it is treated as an
  uncorrelated source of uncertainty.  This choice is based on the
  largely degenerate effect due to the uncertainty associated with the
  embedding procedure, in which the simulated detector response to
  hadronic $\tau$ decays is merged with a sample of $Z\to\mu^+\mu^-$
  data events.  Furthermore, the uncertainty of this embedding
  procedure is treated separately for signal and background processes, 
  which is a conservative approach given that $Z\to\tau^+\tau^-$ sideband 
  effectively constrains this nuisance parameter.

\item The $b$-tagging systematic uncertainty is decomposed into five
  fundamental sources in the \hbb\ channels, while a simplified
  model with a single source is used in the \hZZllqq\ channel. 
  The uncertainty in the $b$-veto is considered uncorrelated with the uncertainty in the 
  $b$-tagging efficiency.

\end{itemize}

The effect of these systematic uncertainties depends on the final
state, but is typically small compared to the theoretical uncertainty
of the production cross section.

The electron and muon energy scales are directly constrained by
$Z\rightarrow e^{+}e^{-}$ and $Z\rightarrow \mu^{+}\mu^{-}$ events;
the impact of the resulting systematic uncertainty on the four-lepton
invariant mass is of the order of $\pm$0.5\% for electrons and
negligible for muons. The impact of the photon energy scale systematic
uncertainty on the diphoton invariant mass is approximately $\pm$0.6\%.

\subsection{Background Measurement Uncertainty}

The estimates of background normalizations and model parameters 
from control regions or sidebands are the main remaining source of uncertainty. 
Because of the differences in control regions these uncertainties are not correlated
across channels.

In the case of the \hbb\ channels the background normalizations are
constrained both from sideband fits and from auxiliary measurements based
on the MC prediction of the main background processes ($Z$+jets,
$W$+jets and \ttbar).

The uncorrelated sources of systematic uncertainties are summarized as
a single combined number in Table~\ref{tab:sys} for each channel.

\subsection{Summary of the Combined Model}

To cover the search range efficiently between \mh=110~\GeV\ and
\mh=600~\GeV, the signal and backgrounds are modeled and the
combination performed in \mh\ steps that reflect the interplay between
the invariant mass resolution and the natural width of the Higgs
boson. In the low mass range, where the high mass-resolution \hgg\ and
\hZZllll\ channels dominate, the signal is modeled in steps from
500~MeV to 2~\GeV.  For higher masses the combination is performed
with step sizes ranging from 2~GeV to 20~GeV. The \mh\
step sizes are given in Table~\ref{tab:StepSizes}.

\begin{table}[h]
  \caption{Step sizes in Higgs boson mass hypotheses at which the signal and backgrounds are modeled.}
\label{tab:StepSizes}
  \vspace{1em}
\begin{tabular}{cc}
  \hline\hline
  \mH\ [\GeV] &  Step size  \\ \hline\hline
  110--120 &  1~GeV \\
  120--130 &  0.5~GeV \\
  130--150 &  1~GeV \\
  150--290 &  2~GeV \\
  290--350 &  5~GeV \\
  350--400 &  10~GeV \\
  400--600 &  20~GeV  \\ \hline\hline
\end{tabular}
\end{table}

The combined model and statistical procedure are implemented within
the \texttt{RooFit}, \texttt{HistFactory}, and \texttt{RooStats}
software framework~\cite{ROOFIT,HistFactory,Moneta:2010pm}.  As shown in
Table~\ref{tab:channels}, the number of channels included in the
combination depends on the hypothesized value of \mh.  The details for
the number of channels, nuisance parameters, and constraint terms for
various \mh\ ranges are shown in Table~\ref{tab:modelSummary}.  For
\mh$ = 125$~\GeV\ there are 70 channels included in the combined
statistical model and the associated dataset is comprised of more than
22,000 unbinned events and 8,000 bins.  For \mh$ = 350$~\GeV\ there are
46 channels included in the combined statistical model and the
associated dataset is comprised only of binned distributions, for
which there are more than 4,000 bins. Due to the limited size of the MC
samples, additional nuisance parameters and constraint terms are
included in the model to account for the statistical uncertainty in the MC
templates.  The difference
between the number of nuisance parameters and the number of
constraints reported in Table~\ref{tab:modelSummary} corresponds to
the number of nuisance parameters without external constraints, which
are estimated in data control regions or sidebands.

\begin{table}[h]
  \caption{Details of the combined model for different \mh\ ranges. The table shows the number of channels, nuisance parameters associated with systematic uncertainties, number of constraint terms, and number of additional nuisance parameters and constraints associated to limited MC sample sizes.}
  \vspace{1em}
  \label{tab:modelSummary}
  \begin{tabular}{ccccc}
    \hline\hline
    \mH\ [\GeV] &  Channels & Nuisance  & Constraints & MC stat     \\ \hline\hline
    110--130 & 70  & 287 & 159 & 180   \\ 
    131--150 & 67  & 210 & 119 & 140   \\
    152--198 & 46  & 83  & 80  & 78    \\
    200--278 & 52  & 199 & 119 & 120   \\
    280--295 & 52  & 208 & 118 & 119   \\
    300     & 58  & 281 & 120 & 121   \\
    305--400 & 46  & 269 & 111 & 112   \\
    420--480 & 46  & 238 & 111 & 112   \\
    500--600 & 46  & 201 & 111 & 112   \\ \hline\hline
  \end{tabular}
\end{table}

\section{Statistical Procedures}
\label{S:StatisticalProcedure}

The procedures for computing frequentist $p$-values used for
quantifying the agreement of the data with the background-only
hypothesis and for determining exclusion limits are based on the
profile likelihood ratio test statistic.

For a given dataset $\datasim$ and values for the global observables
$\globs$ there is an associated likelihood function of $\mu$ and
${\vec \theta}$ derived from the combined model over all channels
including all constraint terms in Eq.~(\ref{Eq:ftot})
\begin{linenomath}
\begin{equation}
  L(\mu,\vec\theta;\mh,\datasim,\globs) = \F_{\rm tot}(\datasim,\globs|\mu,\mh,\vec\theta) \;.
\end{equation}
\end{linenomath}
The notation $L(\mu,\vec\theta)$ leaves the dependence on the data
implicit.

\subsection{The Test Statistics and Estimators of $\mu$ and  $\vec\theta$}

The statistics used to test different values of the strength parameter
$\mu$ are defined in terms of a likelihood function $L(\mu,\vec\theta)$.  
  The maximum likelihood estimates (MLEs)
$\hat\mu$ and $\hat{\vec\theta}$ are the values of the parameters that
maximize the likelihood function $L(\mu,\vec\theta)$.  The conditional
maximum likelihood estimate (CMLE) $\hathatthetamu$ is the value of
$\vec\theta$ that maximizes the likelihood function with $\mu$ fixed.
The tests are based on the profile likelihood ratio $\lambda(\mu)$,
which reflects the level of compatibility between the data and $\mu$.
It is defined as
\begin{linenomath}
\begin{equation}
  \label{eq:lambda}  {\lambda}({\mu}) =  \frac{ L(\mu, \hat{\hat{\vec{\theta}}}(\mu)) }
  {L(\hat{\mu}, \hat{\vec{\theta}}) } \;.
\end{equation}
\end{linenomath}

Physically, the rate of signal events is non-negative, thus $\mu\ge
0$.  However, it is convenient to define the estimator $\hat\mu$ as the 
value of $\mu$ that  maximizes the likelihood, even if is negative (as long as the pdf
$f_c(x_c | \mu,\vec\theta)\ge 0$ everywhere).  In particular,
$\hat\mu<0$ indicates a deficit of events with respect to the
background in the signal region.  Following Ref.~\cite{Cowan:2010st} a
treatment equivalent to requiring $\mu\ge 0$ is to allow $\mu< 0$ and
impose the constraint in the test statistic itself, i.e.
\begin{linenomath}
\begin{equation}
\label{eq:lambdatilde}  \tilde{\lambda}({\mu}) =  \left\{ 
\begin{array}{ll} \frac{ L(\mu, \hat{\hat{\vec{\theta}}}(\mu)) }
{L(\hat{\mu}, \hat{\vec{\theta}}) } & \hat{\mu} \ge 0  , \\*[0.3 cm]
\frac{ L(\mu, \hat{\hat{\vec{\theta}}}(\mu)) }  {L(0,
\hat{\hat{\vec{\theta}}}(0)) } & \hat{\mu} < 0  .
              \end{array} \right. 
\end{equation}
\end{linenomath}

To quantify the significance of an excess, the test statistic
$\tilde{q}_0$ is used to test the background-only hypothesis
$\mu=0$ against the alternative hypothesis $\mu>0$.  It is defined as
\begin{linenomath}
\begin{equation}
\tilde{q}_{0} =  \left\{ \!
\! \begin{array}{ll} - 2 \ln {\lambda}(0)  & \hat{\mu} > 0 ,
\\*[0.2 cm] + 2 \ln {\lambda}(0) & \hat{\mu} \le 0 .
              \end{array} \right.
\end{equation}
\end{linenomath}
Instead of defining the test statistic to be identically zero for
$\hat{\mu}\le0$ as in Ref.~\cite{Cowan:2010st}, this sign change is
introduced in order to probe $p$-values larger than 50\%.

For  setting an upper limit on $\mu$, the test statistic $\tilde{q}_{\mu}$ is used to
test the hypothesis of signal events being produced at a rate $\mu$
against the alternative hypothesis of signal events being produced at a
lesser rate $\mu'<\mu$:
\begin{linenomath}
\begin{equation}
\label{eq:qmutilde}  
\tilde{q}_{\mu} =  \left\{ \!
\! \begin{array}{ll} - 2 \ln \tilde{\lambda}(\mu)  & \hat{\mu} \le \mu ,
\\*[0.2 cm] + 2 \ln \tilde{\lambda}(\mu) & \hat{\mu} > \mu .
              \end{array} \right.  \quad 
\end{equation}
\end{linenomath}

Again,  a sign change is introduced in order to probe $p$-values larger than 50\%.  
The test statistic $ - 2 \ln {\lambda}(\mu)$ is used to differentiate signal events
being produced at a rate $\mu$ from the alternative hypothesis of
signal events being produced at a different rate $\mu'
\ne\mu$.

Tests of $\mu$ are carried out with the Higgs mass \mH\ fixed
to a particular value, and the entire procedure is repeated for
values of \mH\ spaced in small steps.

\subsection{The Distribution of the Test Statistic and $p$-values}

When calculating upper limits, a range of values of $\mu$ is explored
using the test statistic $\tilde{q}_\mu$. The value of the test
statistic for the observed data is denoted $\tilde{q}_{\mu,\rm obs}$.
One can construct the distribution of $\tilde{q}_\mu$ assuming a
different value of the signal strength $\mu'$, which is denoted
\begin{linenomath}
\begin{equation}
f(\tilde{q}_\mu | \mu', \mh, \vec\theta) \;.
\end{equation}
\end{linenomath}

The distribution depends explicitly on \mh\ and $\vec\theta$.  The
$p$-value is given by the tail probability of this distribution, and
thus the $p$-value will also depend on \mh\ and $\vec\theta$.  The
reason for choosing the test statistic based on the profile likelihood
ratio is that, with sufficiently large numbers of events, the
distribution of the profile likelihood ratio with $\mu=\mu'$ is
independent of the values of the nuisance parameters and, thus also
the associated $p$-values.

In practice, there is generally some residual dependence of the
$p$-values on the value of $\vec\theta$. The values of the nuisance
parameters that maximize the $p$-value are therefore sought.
Following Refs.~\cite{Cranmer,Hybrid,Bodhi,paper2010,LHC-HCG}, the
$p$-values for testing a particular value of $\mu$ are based on the
distribution constructed at
$\hat{\hat{\vec{\theta}}}(\mu,\textrm{obs})$, the CMLE estimated with
the observed data, as follows:
\begin{linenomath}
\begin{equation}
p_{\mu}=\int_{\tilde q_{\mu,\rm obs}}^{\infty} f(\tilde q_\mu|\mu,\mh,\hat{\hat{\vec{\theta}}}(\mu,\textrm{obs})) \,d\tilde q_\mu \; .
\end{equation}
\end{linenomath}
The ensemble includes randomizing both $\data$ and $\globs$.

Here the distribution of $\tilde{q}_\mu$ assumes that the
data $\data$ as well as the global observables  $\globs$ are treated
as measured quantities, i.e., they fluctuate upon repetition of the
experiment according to the model $\F_{\rm tot}(\datasim,\globs|\mu,\mh,\vec\theta)$.

Upper limits for the strength parameter $\mu$ are calculated using
the $CL_s$ procedure~\cite{Read:2002hq}.  To calculate this limit,
the quantity $CL_s$ is defined as the ratio
\begin{linenomath}
\begin{equation}
CL_s(\mu)=\frac{p_\mu}{1-p_b} \; ,
\end{equation}
\end{linenomath}
where $p_b$ is the $p$-value derived from the same test statistic
under the background-only hypothesis
\begin{linenomath}
\begin{equation}
\label{eq:pb}
p_b=1-\int_{\tilde q_{\mu,obs}}^\infty f(\tilde q_\mu|0,\mh,\hat{\hat{\vec{\theta}}}(\mu=0,\textrm{obs}))d\tilde q_\mu \;.
\end{equation}
\end{linenomath}

The $CL_s$ upper limit on $\mu$ is denoted $\mu_{\rm up}$ and obtained
by solving for $CL_s(\mu_{\rm up})=5\%$.  A value of $\mu$ is regarded
as excluded at the 95\% confidence level if $\mu>\mu_{\rm up}$.

The significance of an excess is based on the compatibility of the
data with the background-only hypothesis.  
This compatibility is quantified by the following $p$-value:
\begin{linenomath}
\begin{equation}
\label{eq:p0}
  p_0=\int_{\tilde q_{0,obs}}^\infty f(\tilde q_0|0,\mh,\hat{\hat{\vec{\theta}}}(\mu=0,\textrm{obs}))d\tilde q_0 \;.
\end{equation}
\end{linenomath}
Note that $p_0$ and $p_b$ are both $p$-values of the
background-only hypothesis, but the test statistic $\tilde q_0$ in
Eq.~(\ref{eq:p0}) is optimized for discovery while the test statistic
$\tilde{q}_\mu$ in Eq.~(\ref{eq:pb}) is optimized for upper limits.

It is customary to convert the background-only $p$-value into an equivalent Gaussian 
significance $Z$ (often written $Z\sigma$).   The conversion is defined as
\begin{linenomath}
\begin{equation}
Z = \Phi^{-1}(1-p_0) \; ,
\end{equation}
\end{linenomath}
\noindent
where $\Phi^{-1}$ is the inverse of the cumulative distribution for a
standard Gaussian.  

\subsection{Experimental Sensitivity and Bands}

It is useful to quantify the experimental sensitivity by means of the
significance one would expect to find if a given signal hypothesis were
true.  Similarly, the expected upper limit is the median upper limit one would
expect to find if the background-only hypothesis were true.  
Although these are useful quantities, they are subject to a
certain degree of ambiguity because the median values depend on the
assumed values of all of the  parameters of the model, including the nuisance parameters.

Here, the expected upper limit is defined as the median of the distribution $f(\mu_{\rm
  up}|0,\mh,\hat{\hat{\vec{\theta}}}(\mu=0,\textrm{obs}))$ and the
expected significance is based on the median of the distribution
$f(p_0 | 1,\mh,\hat{\hat{\vec\theta}}(\mu=1, \rm obs))$.
The expected limit and significance thus have a small residual
dependence on the observed data through $\hat{\hat{\vec\theta}}(\mu,
\rm obs)$.

These distributions are also used to define bands around the median
upper limit.  The standard presentation of upper limits includes the
observed limit, the expected limit, a $\pm 1\sigma$ and a $\pm 2\sigma$ band.  More precisely, 
the edges of these bands, denoted $\mu_{{\rm up}\pm 1}$ and  $\mu_{{\rm up} \pm 2}$, are
defined by
\begin{linenomath}
\begin{equation}
  \int_{0}^{\mu_{{\rm up}\pm N}}  f(\mu_{\rm up}|0,\mh,\hat{\hat{\vec{\theta}}}(\mu=0,\textrm{obs})) d\mu_{\rm up} = \Phi(\pm N) 
\end{equation}
\end{linenomath}

 \subsection{Asymptotic Formalism }
\label{S:asymptotic}

For large data samples, the asymptotic distributions \mbox{$f(\tilde q_\mu | \mu', \mh,
  \vec\theta)$} and \mbox{$f(\tilde q_0 | \mu', \mh, \vec\theta)$} are
known and described in Ref.~\cite{Cowan:2010st}.  These formulae
require  the variance of the maximum likelihood estimate of
$\mu$ given $\mu'$ is the true value:
 \begin{linenomath}
\begin{equation}
 \sigma_{\mu'}=\sqrt{\textrm{var}[\hat\mu | \mu']}\;.
 \end{equation}
\end{linenomath}

One result of Ref.~\cite{Cowan:2010st} is that $\sigma_{\mu'}$ can be
estimated with an artificial dataset referred to as the {\it Asimov}
dataset. This dataset is defined as a binned dataset, where the
number of events in bin $b$ is exactly the number of events expected
in bin $b$.  
The value of the test statistic
evaluated on the Asimov data is denoted $\tilde
q_{\mu,A_{\vec\alpha}}$, where the subscript $A_{\vec\alpha}$ denotes 
that this is the Asimov data associated with $\vec\alpha$. 
 A convenient way to estimate the variance of $\hat\mu$ is
\begin{linenomath}
\begin{equation}
\label{eq:sigmaofmu}
\sigma_{\mu'} \approx \frac{\mu-\mu'}{\sqrt{\tilde q_{\mu,A_{\mu'}}}} \;.
\end{equation}
\end{linenomath}

In the asymptotic limit, $\tilde
q_\mu$ is parabolic, thus $\sigma_{\mu'}$ is independent of $\mu$. In
Ref.~\cite{Cowan:2010st}, the bands around the expected limit are
given by
\begin{linenomath}
\begin{equation}
\label{eq:originalBands}
\mu_{{\rm up}+N}=\sigma(\Phi^{-1}(1-0.05 \,\Phi(N))+N) \;.
\end{equation}   
\end{linenomath}
\noindent
An improved procedure is used here in order to capture the leading
deviations of $\tilde q_\mu$ from a parabola, corresponding to
departures in the distribution of $\hat\mu$ from a Gaussian
distribution centered at $\mu'$ with variance $\sigma_{\mu'}^2$.
Finding the upper limit $\mu_{\rm up}$ (still using the formulae in
Ref.~\cite{Cowan:2010st}) for an Asimov dataset constructed with
$\vec\alpha_N=(\mu_N,\mh,\hat{\hat{\vec\theta}}(\mu_N,\rm obs))$,
where $\mu_N$ is the value of $\mu$ corresponding to the edges of the
$\mu_{{\rm up}\pm N}$ band.  It is found to be more accurate than
Eq.~(\ref{eq:originalBands}) in reproducing the bands obtained with
ensembles of pseudo-experiments.  The value of $\mu_N$ used for the
$+N\sigma$ band is the value of $\mu$ that satisfies $\sqrt{\tilde
  q_{{\mu_N,A_0}}}=N$.  The choice of
$\hat{\hat{\vec\theta}}(\mu_N,\rm obs)$ is more indicative of
${\hat{\vec\theta}}$ for the pseudo-experiments that have $\mu_{\rm
  up}$ near the corresponding band.

\subsection{Bayesian Methods}
\label{S:BayesianMethods}

A posterior distribution for the signal strength parameter $\mu$ can
be obtained, via Bayes's theorem, with $\F_{\rm
  tot}(\data,\globs|\vec\alpha)$ and a prior $\eta(\vec\alpha)$.  The
information about the nuisance parameters from auxiliary measurements
is incorporated into $\F_{\rm tot}(\data,\globs|\vec\alpha)$ through
the constraint terms $f_p(a_p|\alpha_p)$.  As in
Eq.~(\ref{eq:urprior}), the prior on $\eta(\alpha_p)$ is taken as a
uniform distribution.  The upper limits are calculated for a given
value of \mh, so no prior on \mh\ is needed.  The prior on the signal
strength parameter $\mu$ is taken to be uniform as this choice leads
to one-sided Bayesian credible intervals that correspond numerically
with the $CL_s$ upper limits in the frequentist formalism for the
simple Gaussian and Poisson cases, and have been observed to coincide
in more complex situations.  The one-sided, 95\% Bayesian credible
region is defined as
\begin{linenomath}
\begin{equation}
\int_{0}^{\mu_{\rm up}} \F_{\rm tot}(\data_{\rm sim},\globs|\mu,\mh,\vec\theta)   \,\eta(\mu) \eta(\vec\theta) \, d \mu \, d\vec\theta = 0.95\;.
\end{equation}
\end{linenomath}
The integration over nuisance parameters is carried out with Markov
Chain Monte Carlo using the Metropolis-Hastings algorithm in the
\roostats\ package~\cite{Moneta:2010pm}.

\subsection{Correction for the Look Elsewhere Effect}
\label{S:LEE}

By scanning over \mh\ and repeatedly testing the background-only hypothesis,
the procedure is subject to effects of multiple testing referred to as the look-elsewhere effect.
In principle, the confidence intervals can be constructed in the
$\mu-\mh$ plane using the profile likelihood ratio
$\lambda(\mu,\mh)$. For $\mu=0$ there is no signal present, thus the
model is independent of \mh.  This leads to a background-only
distribution for $-2\ln \lambda(0,\mh)$ that departs from a chi-square
distribution, thus complicating the calculation of a global $p_0$ when
\mh\ is not specified. The global test statistic is the supremum of
$q_0(m_H)$ with respect to $m_H$
\begin{linenomath}
\begin{equation}
q_0(\hat{m}_H) = \sup_{\mh} q_0(m_H) \, .
\end{equation}
\end{linenomath}

In the asymptotic regime and for very small $p$-values, a
procedure~\cite{LEE}, based on the result of Ref.~\cite{Davies}, exists to
estimate the tail probability for $q_0(\hat{m}_H)$.  The procedure
requires an estimate for the average number of up-crossings of
$q_0(m_H)$ above some threshold.  Due to the \mh-dependence in the
model, changes in cuts, and the list of channels included, it is
technically difficult to estimate this quantity with ensembles of
pseudo-experiments.  Instead, a simple alternative method is used in
which the average number of up-crossings is estimated by counting the
number of up-crossings with the observed data. When the trials factor
is large, the number of up-crossings at low thresholds is also large
and thus a satisfactory estimate of the average~\cite{LHC-HCG}. 
This procedure has been checked using a large
number of pseudo-experiments in a simplified test case, where it provided a
good estimate of the trials factor.

\section{Exclusion Limits}
\label{S:Exclusion}

The model discussed in Section~\ref{S:Modeling} is used for all
channels described in Section~\ref{S:Inputs} and the systematic
uncertainties summarized in Section~\ref{sec:syst}.  The statistical
methods described in Section~\ref{S:StatisticalProcedure} are used to
set limits on the signal strength as a function of \mh.

\begin{figure}[!htb]
  \begin{center}
	\subfigure[]{ \includegraphics[width=.46\textwidth]{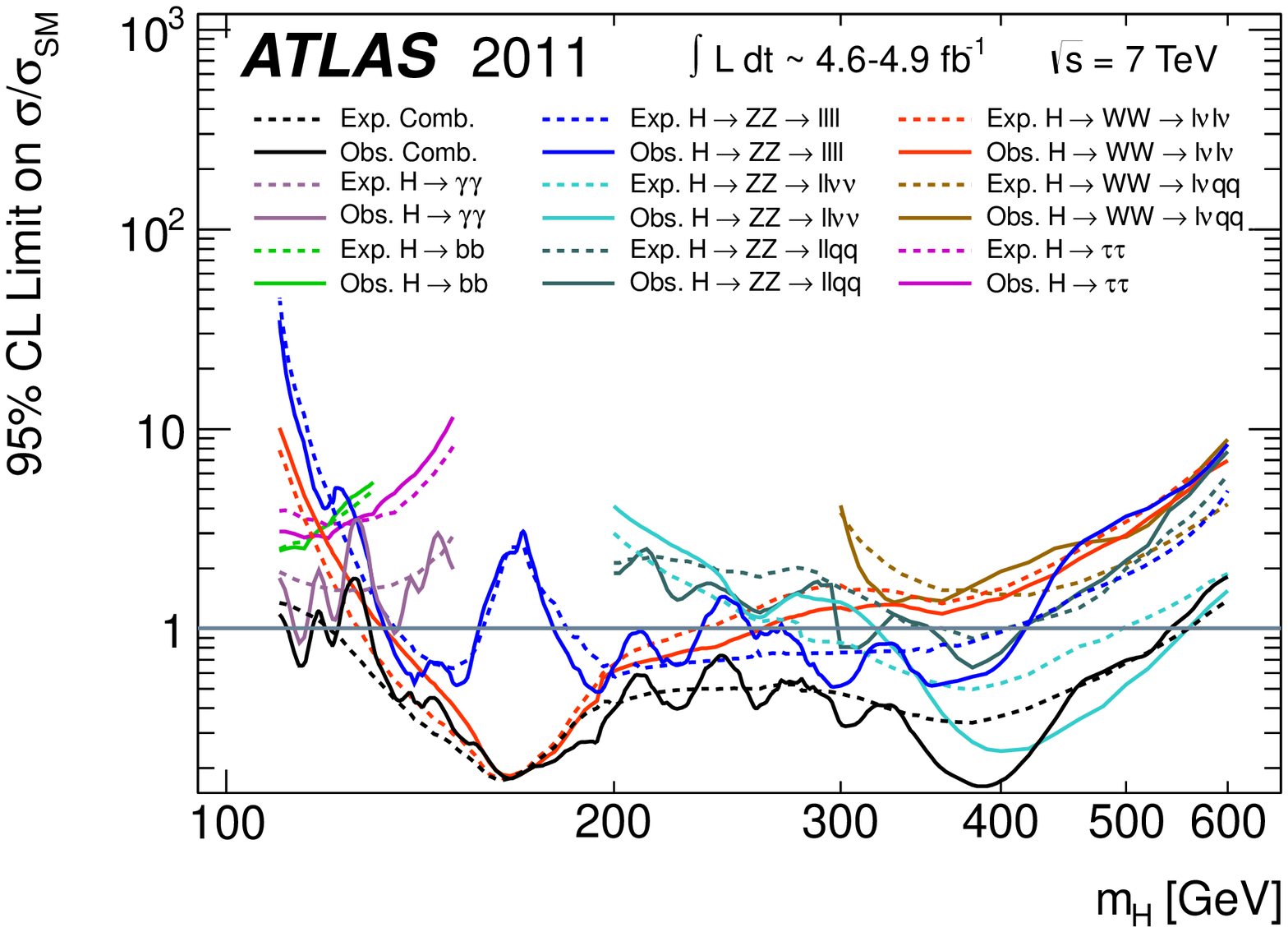} }
	\subfigure[]{   \includegraphics[width=.46\textwidth]{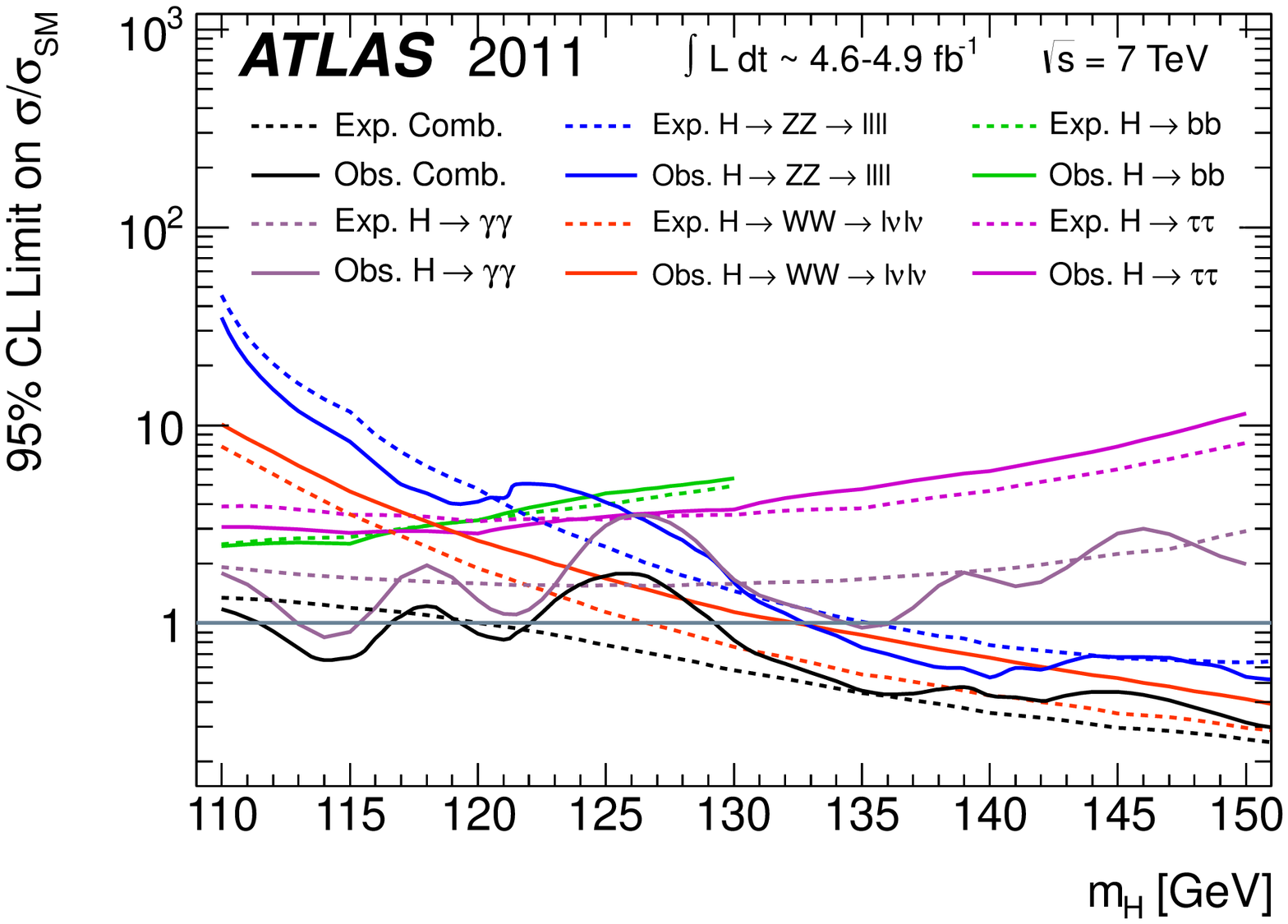}}
  \end{center}
\vspace*{-0.5cm}
\caption{ The observed (solid) and expected (dashed) 95\%~CL cross
  section upper limits for the individual search channels and the
  combination, normalized to the SM Higgs boson production cross
  section, as a function of the Higgs boson mass; (a) for the full
  Higgs boson mass hypotheses range and (b) in the low mass range. The
  expected limits are those for the background-only hypothesis i.e. in
  the absence of a Higgs boson signal. }
  \label{fig:inputs}
\end{figure}

\begin{figure}[!htb]
  \begin{center}
	\subfigure[]{  \includegraphics[width=.45\textwidth]{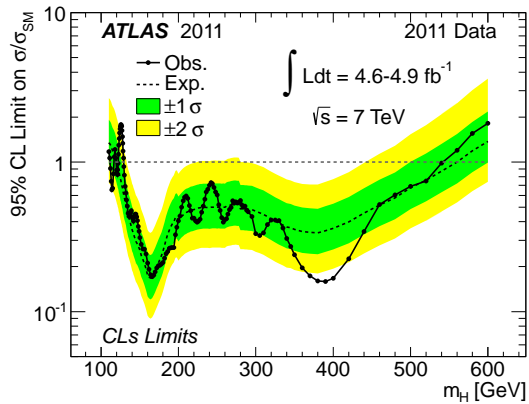} }
	\subfigure[]{    \includegraphics[width=.45\textwidth]{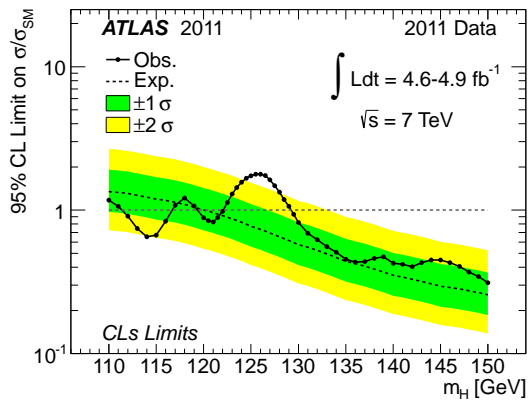} }
  \end{center}
  \caption{The observed (full line) and expected (dashed line) 95\%~CL
    combined upper limits on the SM Higgs boson production cross
    section divided by the SM expectation as a function of \mh, (a) in
    the full mass range considered in this analysis and (b) in the low
    mass range.  The dotted curves show the median expected limit
    in the absence of a signal and the green and yellow bands indicate
    the corresponding $\pm 1\sigma$ and $\pm 2\sigma$\ intervals.}
  \label{fig:CLs}
\end{figure}

The expected and observed limits from the individual channels entering
this combination are shown in Fig.~\ref{fig:inputs}. The combined
95\%~CL exclusion limits on $\mu$ are shown in Fig.~\ref{fig:CLs} as a
function of \mh.  These results are based on the asymptotic
approximation. The $\pm 1\sigma$ and $\pm 2 \sigma$ variation bands
around the median background expectation are calculated using the
improved procedure described in Section~\ref{S:asymptotic}, which
yields slightly larger bands compared to those in
Ref.~\cite{PreMoriondCombPaper}. Typically the increase in the bands
is of the order of $\sim 5$\%\ for the $\pm 1 \sigma$ band and
10--15\%\ for the $\pm 2 \sigma$ band. This procedure has been
validated using ensemble tests and the Bayesian calculation of the
exclusion limits with a uniform prior on the signal strength described in 
Section~\ref{S:BayesianMethods}. These
approaches yield limits on $\mu$ which typically agree with the
asymptotic results within a few percent.

The expected 95\%~CL exclusion region for the SM ($\mu=1$) hypothesis
covers the \mH\ range from \lowerExp\ to \upperExp.  The addition of
the \htt\ and $H\to b\bar{b}$ channels as well as the update of the
\hWWlnln\ channel bring a significant gain in sensitivity in the
low-mass region with respect to the previous combined search. For
Higgs boson mass hypotheses below approximately 122~GeV the dominant
channel is  \hgg. For mass hypotheses larger than 122~GeV but
smaller than 200~GeV the \hWWlnln\ channel is the most sensitive.  
In the mass range between $\sim$200~GeV and
$\sim$300~GeV the \hZZllll\ dominates. For higher mass hypotheses the
\hZZllnn\ channel leads the search sensitivity.  The updates of the
\hWWlnln, \hWWlnqq, \hZZllnn, and \hZZllqq\ channels improve the
sensitivity in the high-mass region.

The observed exclusion regions range from \lowerlowerObs\ to \upperlowerObs,
from \lowerIsland\ to \upperIsland, and from \lowerObs\ to \upperObs\
at 95\%~CL under the SM ($\mu=1$) hypothesis. The mass range
\upperIsland\ to \lowerObs\ is not excluded due to the observation of
an excess of events above the expected background. This excess and its
significance are discussed in detail in Section~\ref{S:p0}.

Two mass regions where the observed exclusion is stronger than
expected can be seen in Fig.~\ref{fig:CLs}.  In
the low mass range, Higgs mass hypotheses in the \lowerlowerObs\ to
\upperlowerObs\ range are excluded due mainly to a local deficit of
events in the diphoton channel with respect to the expected
background.  A similar deficit is observed in the high mass region in
the range 360~GeV to 420~GeV, resulting from deficits in the \hZZllll\
and \hZZllnn\ channels.  Both fluctuations correspond to approximately
two standard deviations in the distribution of upper limits expected
from background only.

\begin{figure}[htb]
  \begin{center}
	\subfigure[]{    \includegraphics[width=.45\textwidth]{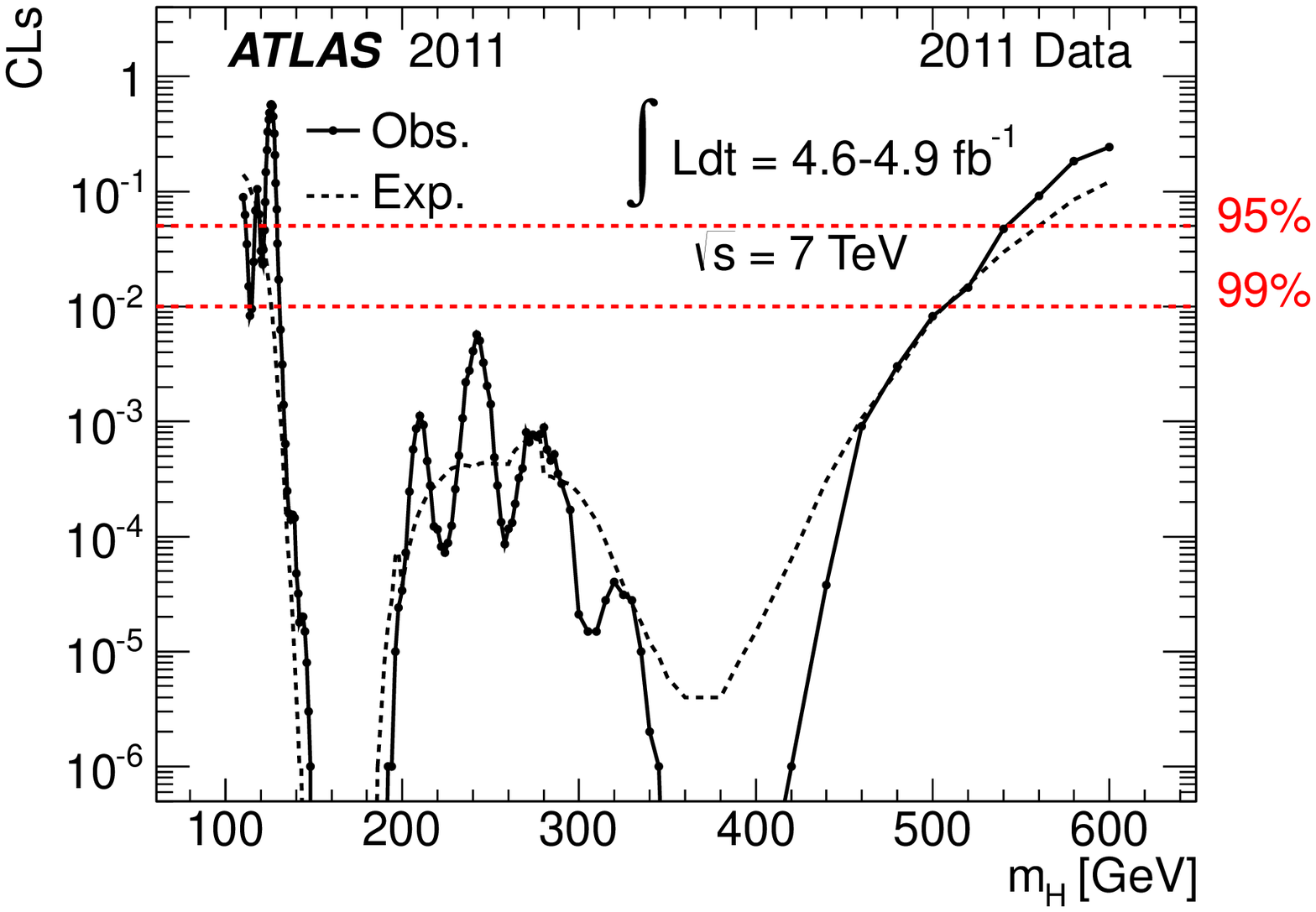} }
	\subfigure[]{    \includegraphics[width=.45\textwidth]{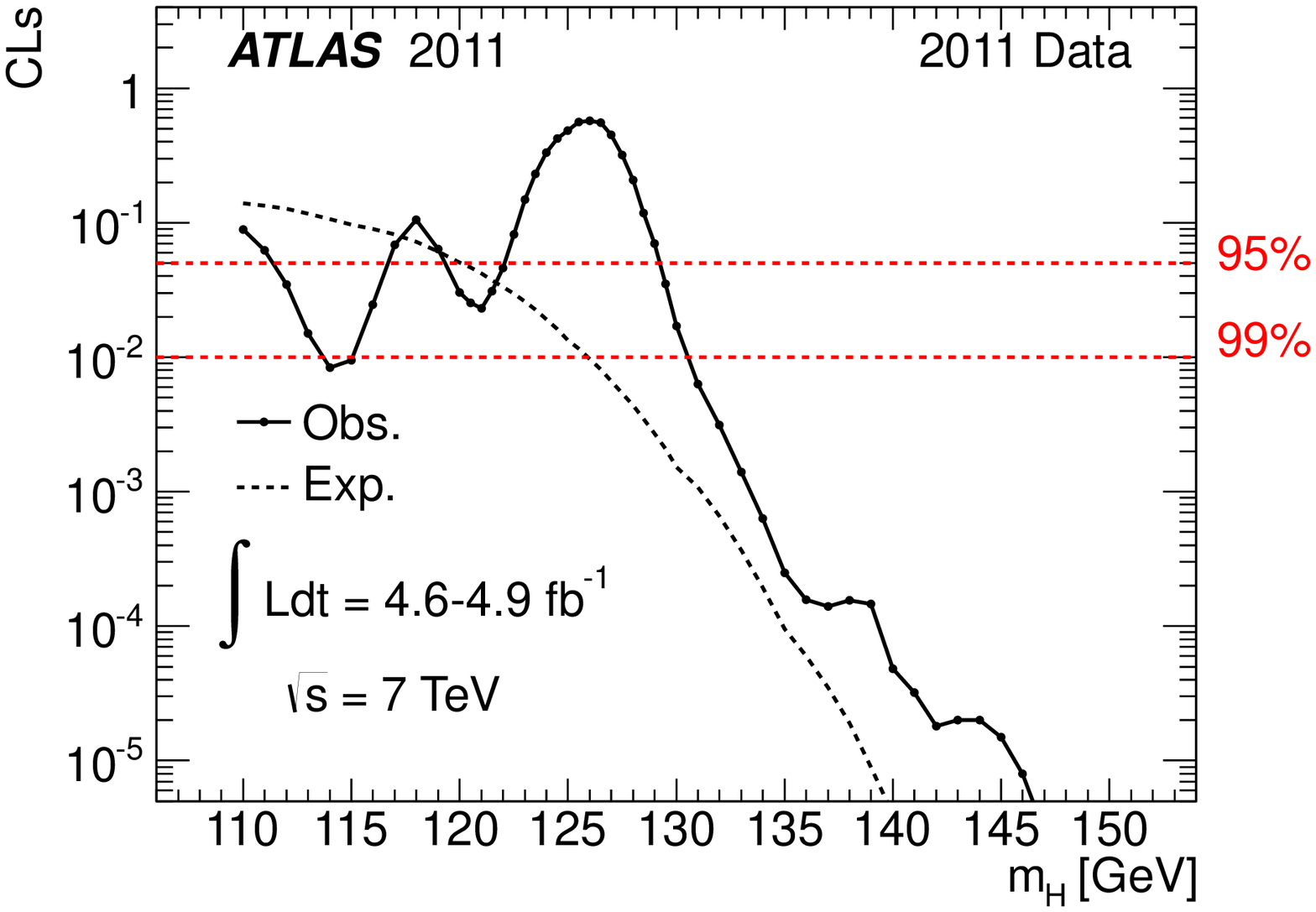} }
  \end{center}
  \caption{The value of the combined $CL_s$ for $\mu=1$ (testing the
    SM Higgs boson hypothesis) as a function of \mh, (a) in the full
    mass range of this analysis and (b) in the low mass range. The
    regions with $CL_s < \alpha$ are excluded at the ($1-\alpha$) CL.}
  \label{fig:CLsMass}
\end{figure}

A small mass region near $\mh\sim245~$\GeV\, was not excluded at the
95\%~CL in the combined search of Ref.~\cite{PreMoriondCombPaper},
mainly due to a slight excess in the \hZZllll\ channel. This mass
region is now excluded.  The $CL_s$ values for $\mu=1$ as a function
of the Higgs boson are shown in Fig.~\ref{fig:CLsMass}, where it can
also be seen that the region between 130.7~\GeV\ and 506~\GeV\ is
excluded at the 99\%~CL.  The observed exclusion covers a large part
of the expected exclusion range.


\begin{figure}[htb]
  \begin{center}
    \subfigure[]{ \includegraphics[width=.45\textwidth]{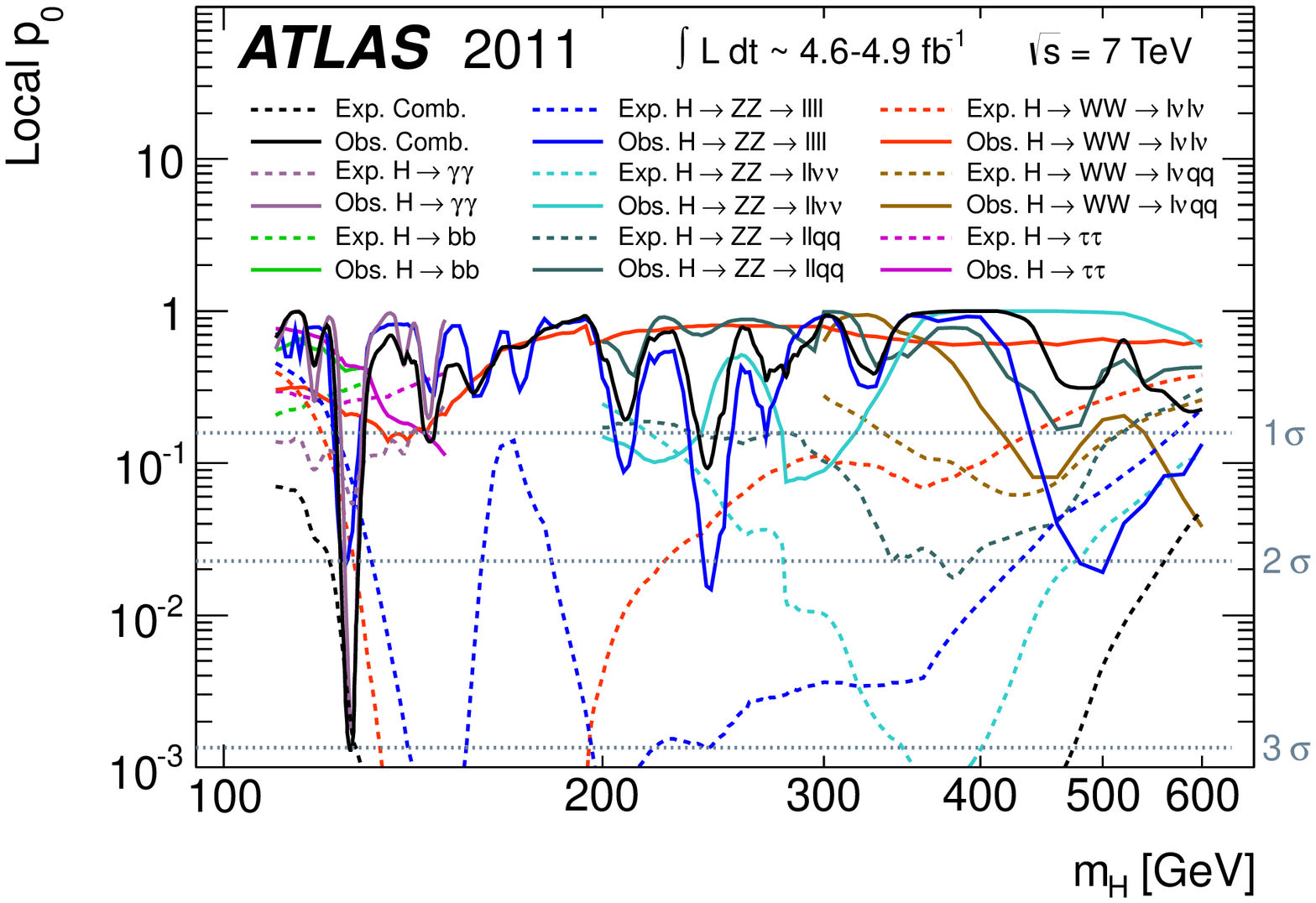}}
    \subfigure[]{ \includegraphics[width=.45\textwidth]{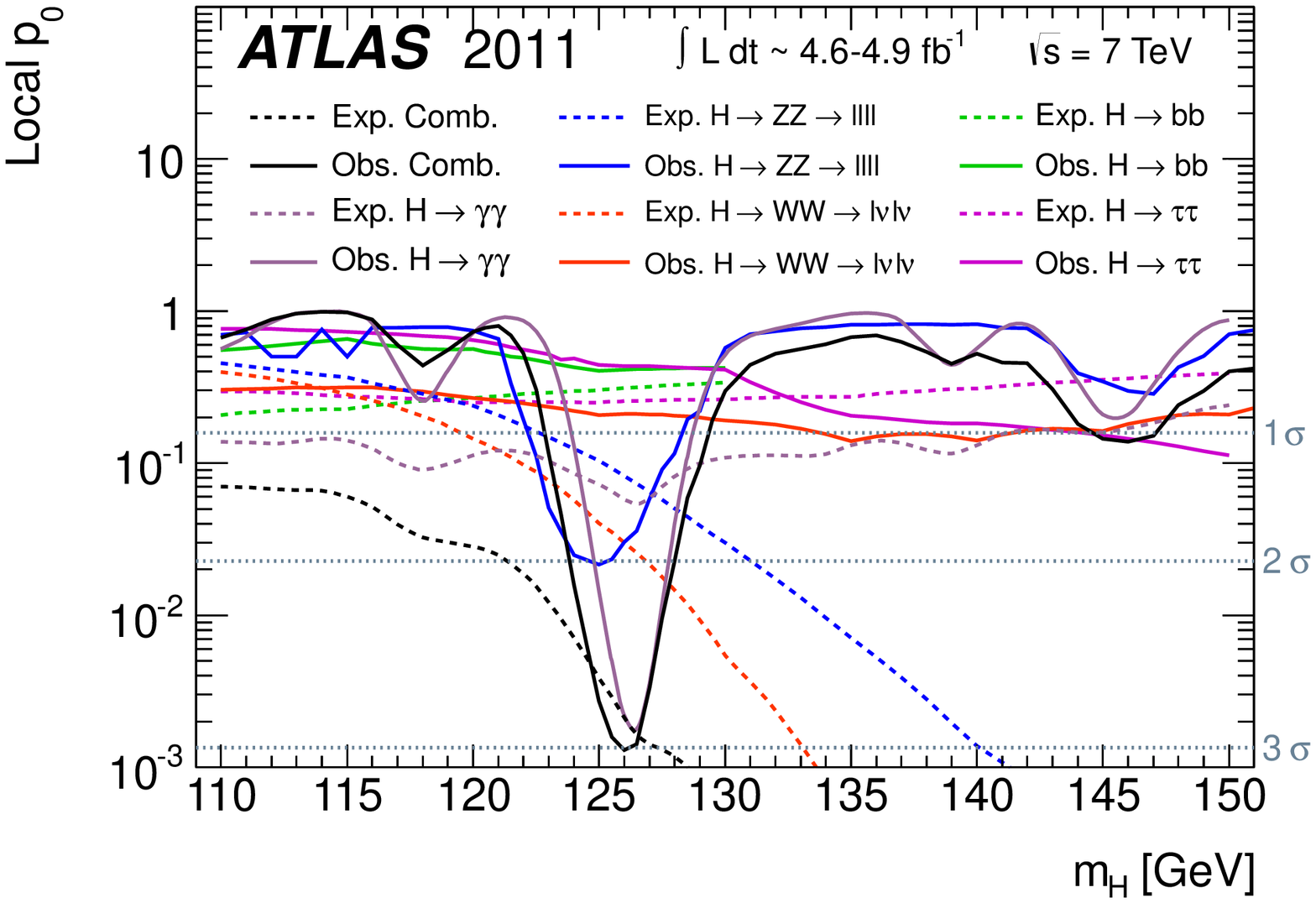}}
    \subfigure[]{\hspace{.7cm} \includegraphics[width=.45\textwidth]{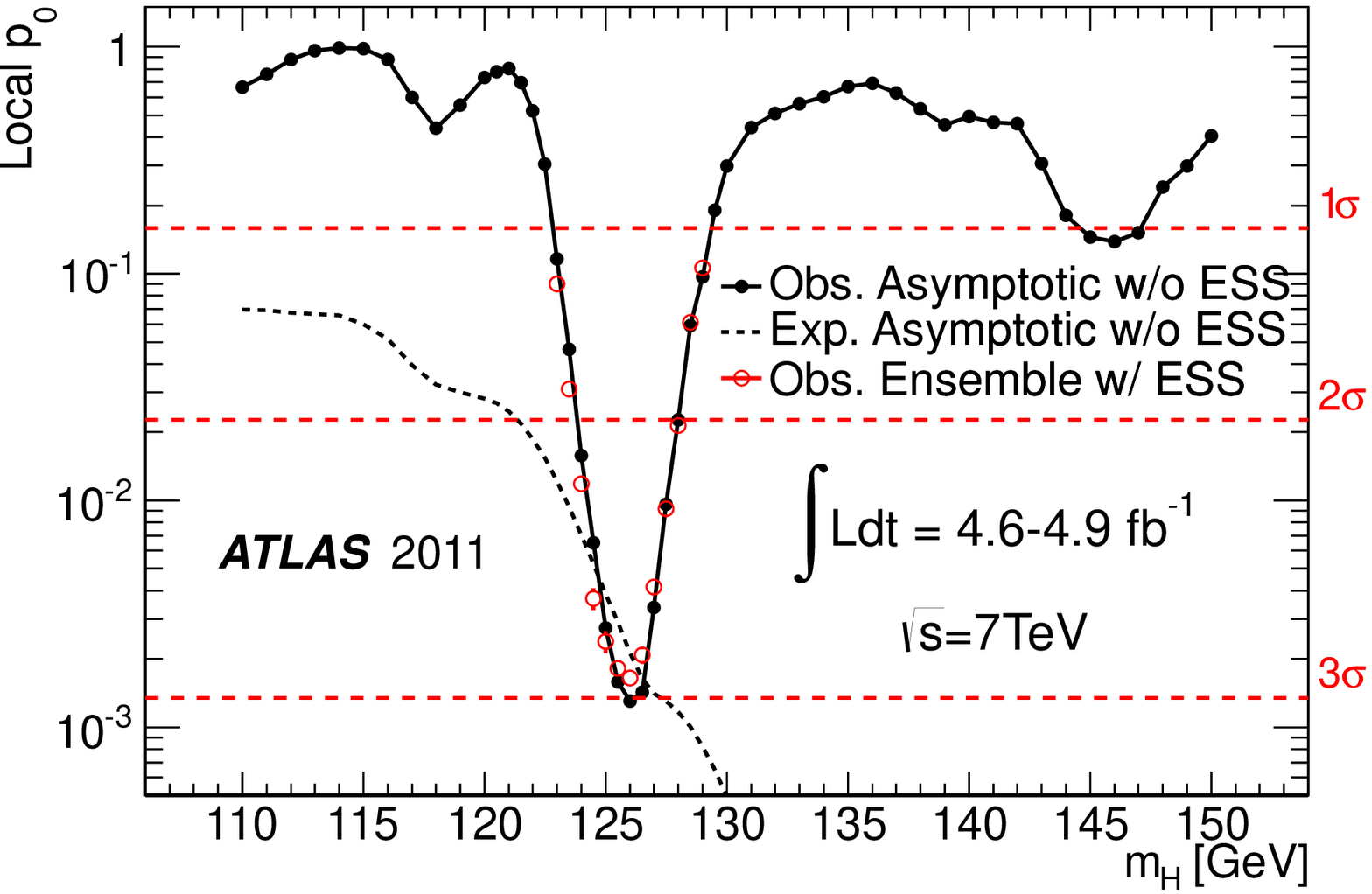}}
      \end{center}
      \caption{The local probability $p_0$ for a background-only
        experiment to be more signal-like than the observation, for
        individual channels and the combination. (a) In the full mass
        range of 110--600~\GeV\ and (b) in the low mass range of
        110--150~\GeV.  The full curves give the observed individual
        and combined $p_0$.  The dashed curves show the median
        expected value under the hypothesis of a SM Higgs boson signal
        at that mass. The combined observed local $p_0$ estimated
        using ensemble tests and taking into account energy scale
        systematic (ESS) uncertainties is illustrated in (c), the
        observed and expected combined results using asymptotic
        formulae is also shown.  The horizontal dashed lines
        indicate the $p_0$ corresponding to significances of
        1$\sigma$, 2$\sigma$, and 3$\sigma$ for (a), (b) and (c).}
  \label{fig:p0}
\end{figure}

\section{Significance of the Excess}
\label{S:p0}

The observed local $p$-values, calculated using the asymptotic
approximation, as a function of \mh\ and the expected value in the
presence of a SM Higgs boson signal at that mass are shown in
Fig.~\ref{fig:p0} in the entire search mass
range and in the low mass range. The asymptotic approximation has been
verified using ensemble tests which yield numerically consistent
results.

The largest significance for the combination is observed for
\mH=126~\GeV, where it reaches \significance\ with an expected value
in the presence of a signal at that mass of 2.9$\sigma$.  The observed
(expected) local significances for \mh$=$126~\GeV\ are 2.8$\sigma$
(1.4$\sigma$) in the \hgg\ channel and 2.1$\sigma$ (1.4$\sigma$) in
the \hZZllll\ channel. In the \hWWlnln\ channel, which has been
updated and includes additional data, the observed (expected) local
significance for \mh$=$126~\GeV\ is 0.8$\sigma$ (1.9$\sigma$); the
observed significance was previously
1.4$\sigma$~\cite{PreMoriondCombPaper}.

The significance of the excess is not very sensitive to energy scale
and resolution systematic uncertainties for photons and electrons;
however, the presence of these uncertainties leads to a small deviation
from the asymptotic approximation.  The observed $p_0$ including these
effects is therefore estimated using ensemble tests.  The results are
displayed in Fig.~\ref{fig:p0} as a function of \mH.  The effect of
the energy scale systematic uncertainties is an increase of
approximately 30\%\ of the corresponding local $p_0$.  The maximum 
local significance decreases slightly to \significanceESS.  The muon momentum scale
systematic uncertainties are smaller and therefore neglected.

The global $p$-values for the largest excess depends on the range of \mH\ and
the channels considered. The global $p_0$ associated with a
2.8$\sigma$ excess anywhere in the \hgg\ search domain of 110--150~\GeV\ is
approximately 7\%.  A 2.1$\sigma$ excess anywhere in the \hZZllll\
search range of 110--600~\GeV\ corresponds to a global $p_0$ of
approximately 30\%.

\begin{figure}[!h]
  \begin{center}
	\subfigure[]{ \includegraphics[width=.45\textwidth]{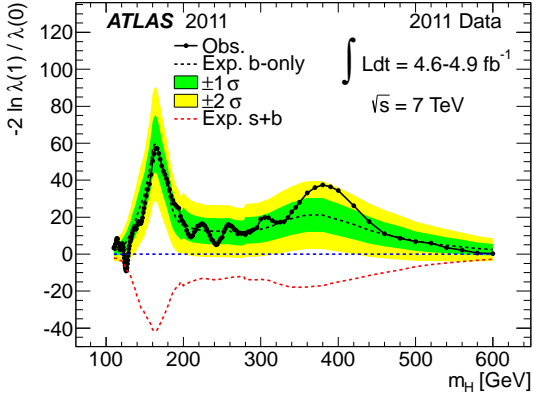} }
	\subfigure[]{    \includegraphics[width=.45\textwidth]{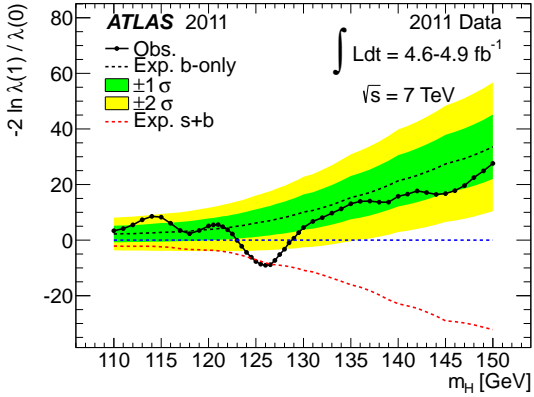} }
  \end{center}
  \caption{The ratio of profile likelihoods for $\mu=0$ and $\mu=1$
    as a function of the Higgs boson mass hypothesis.  The full line
    shows the observed ratio, the lower dashed line shows the median
    value expected under the signal-plus-background hypothesis, and
    the upper dashed line shows the median expected under the
    background-only hypothesis (a) for the full mass range and (b) the
    low mass range. The $\pm$1$\sigma$ and $\pm$2$\sigma$ intervals
    around the median background-only expectation are given by the
    green and yellow bands, respectively.}
  \label{fig:LR}
\end{figure}

The global probability for a \significanceESS\ excess in the combined
search to occur anywhere in the mass range 110--600~\GeV\ is estimated
to be approximately 15\%, decreasing to 5--7\%\ in the range
110--146~\GeV, which is not excluded at the 99\%~CL by
the LHC combined SM Higgs boson search~\cite{lhcCombination}.  
The data are observed to be consistent with the background-only hypothesis 
except for the region around $\mH=126$ \GeV.  The observed and expected ratio $-2 \ln(\lambda(1)/\lambda(0))$ is shown in Fig.~\ref{fig:LR}, 
which indicates a departure from the background-only hypothesis
similar to the signal-plus-background expectation.  

\FloatBarrier

\begin{figure}[h]
  \begin{center}
	\subfigure[]{  \includegraphics[width=.45\textwidth]{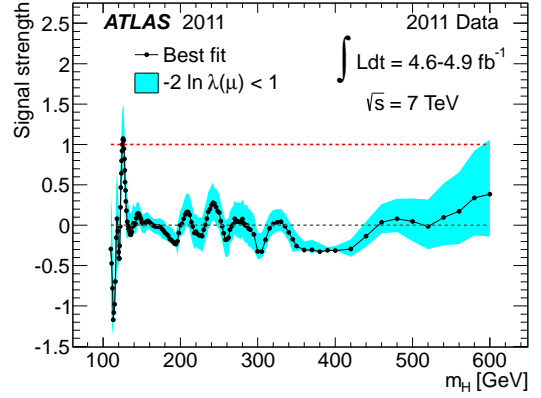} }
	\subfigure[]{    \includegraphics[width=.45\textwidth]{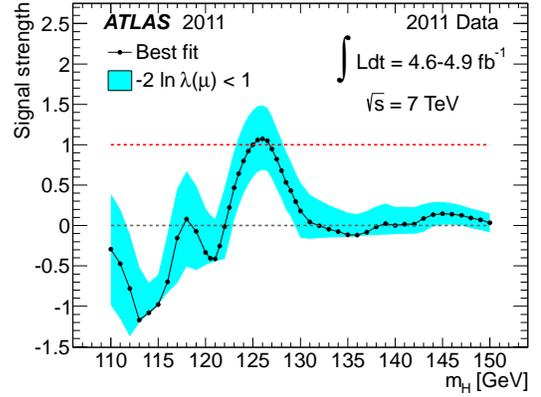} }
  \end{center}
  \caption{The combined best-fit signal strength $\mu$ as a function
    of the Higgs boson mass hypothesis (a) in the full mass range of
    this analysis and (b) in the low mass range.  The interval around
    $\hat{\mu}$ corresponds to a variation of $-2\ln \lambda(\mu)<1$.}
  \label{fig:muhat}
\end{figure}

\begin{figure*}[htb]
  \begin{center}
    \subfigure[]{ \includegraphics[width=.3\textwidth]{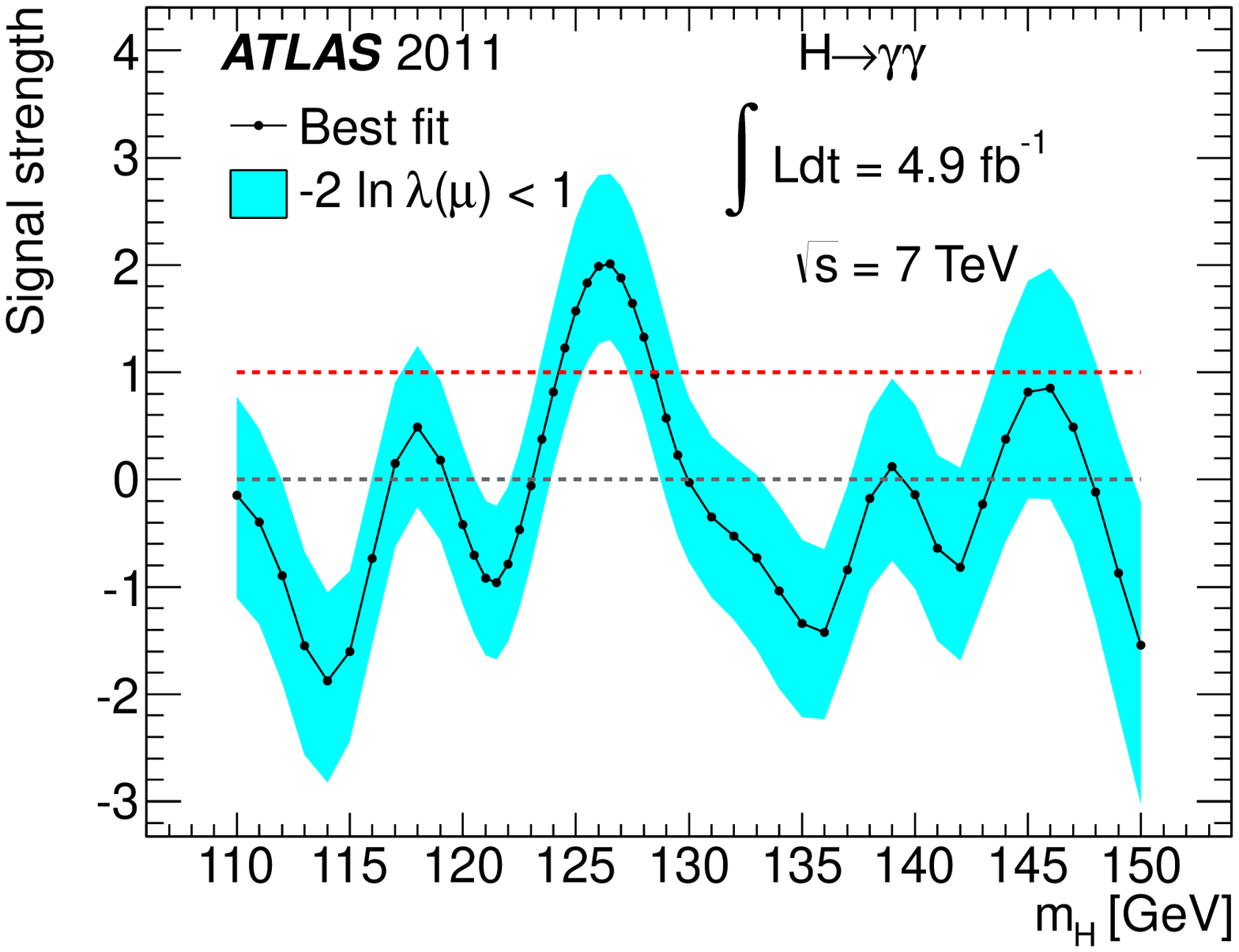}}
    \subfigure[]{ \includegraphics[width=.3\textwidth]{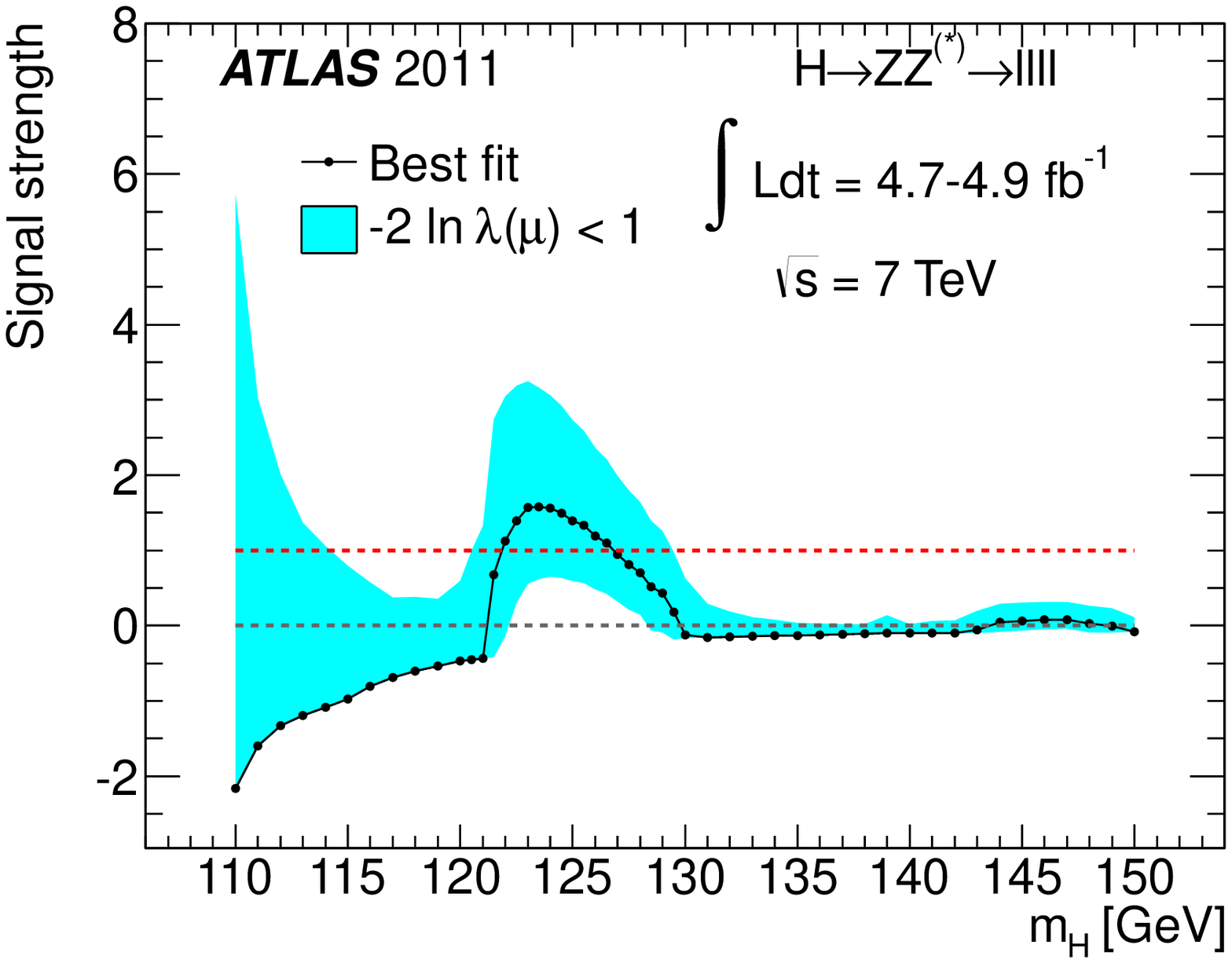}}
    \subfigure[]{ \includegraphics[width=.3\textwidth]{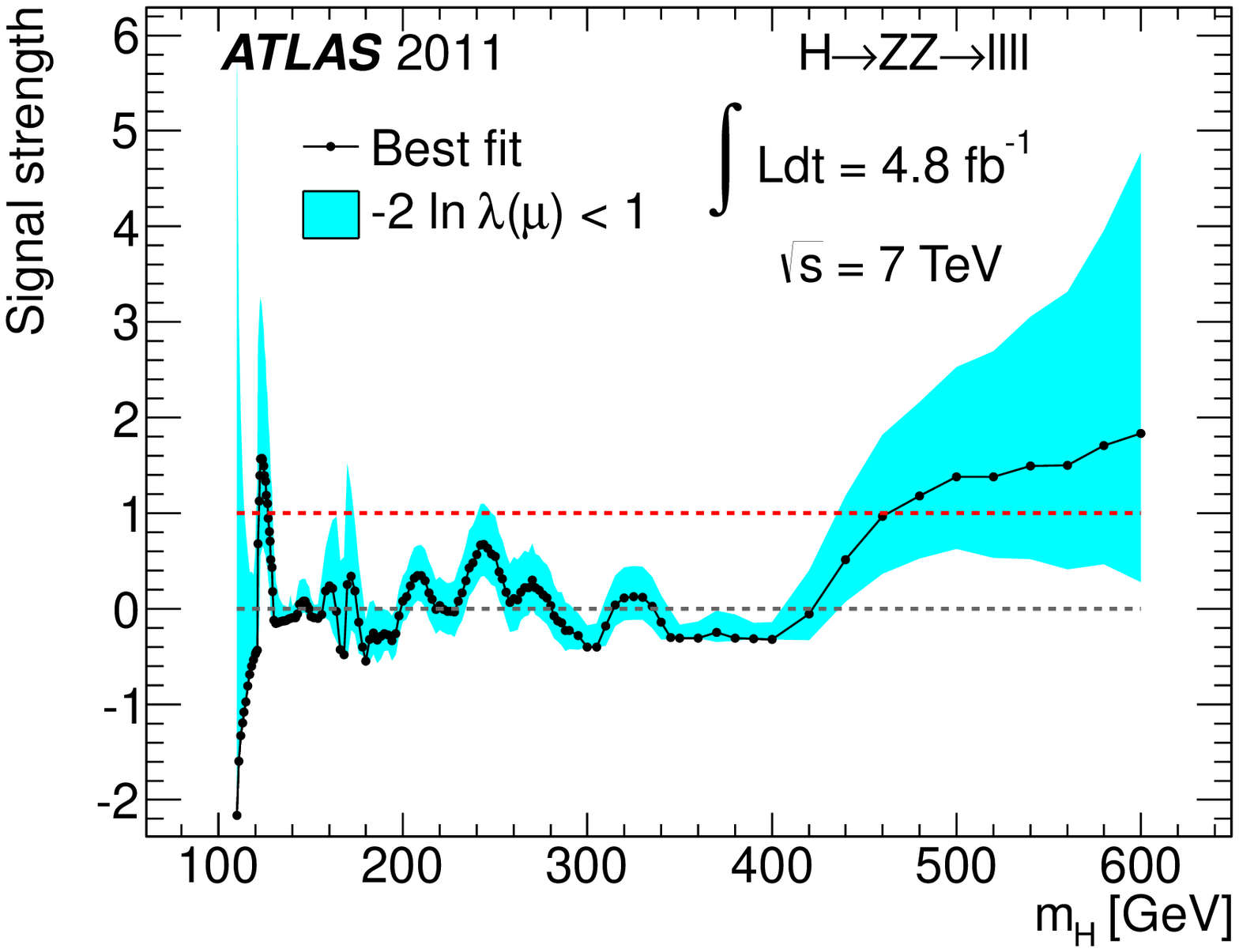}}
    \subfigure[]{ \includegraphics[width=.3\textwidth]{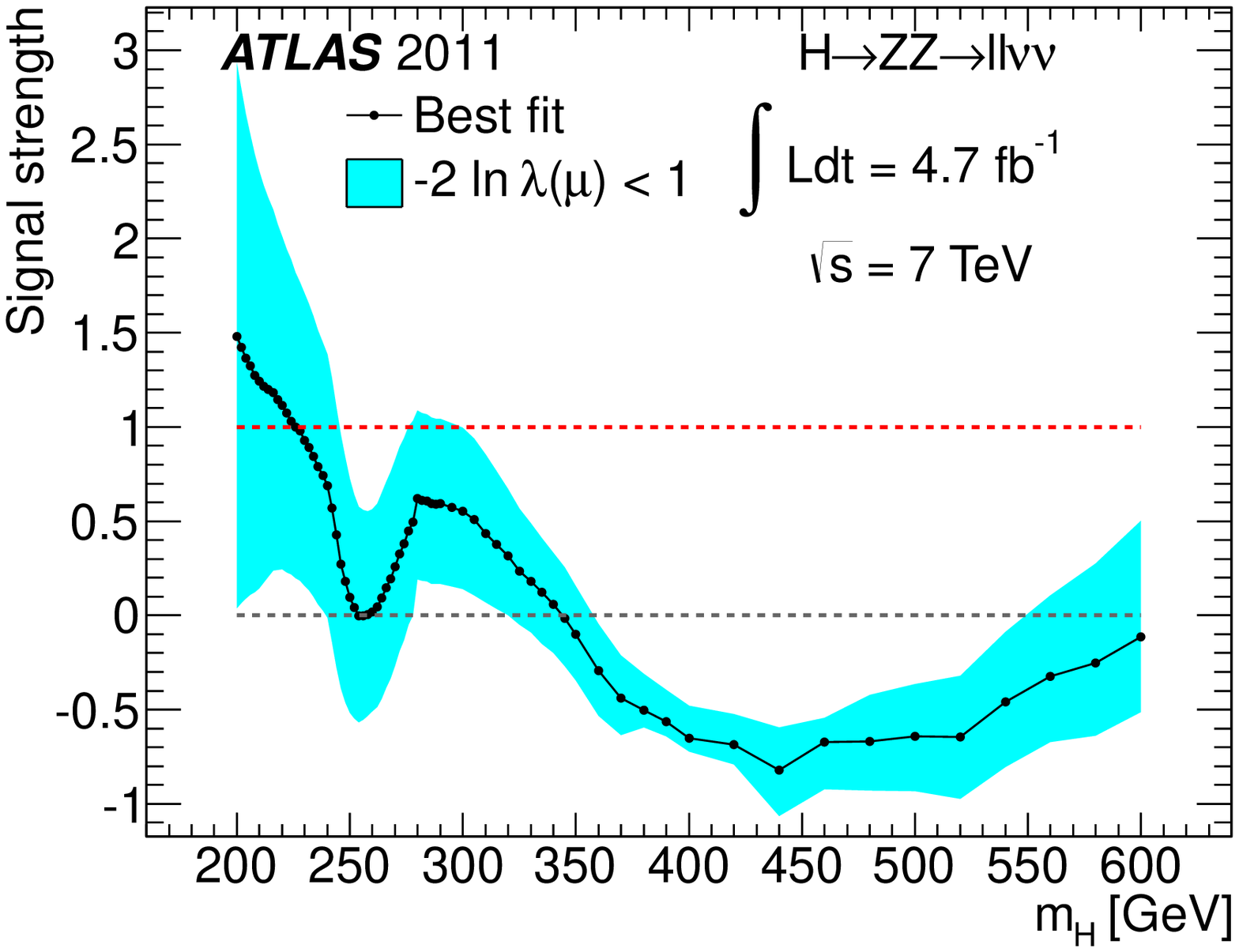}}
    \subfigure[]{ \includegraphics[width=.3\textwidth]{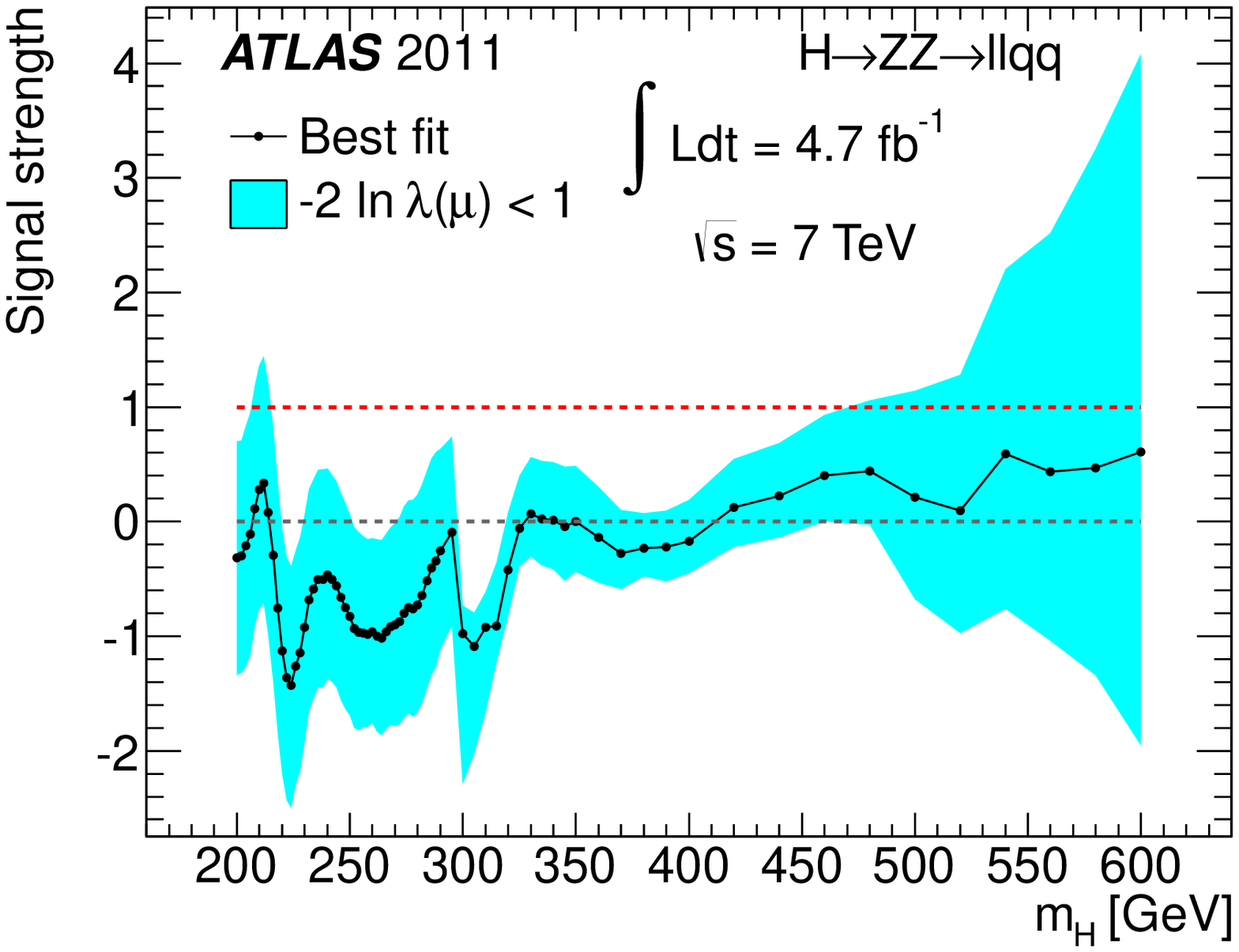}}
    \subfigure[]{ \includegraphics[width=.3\textwidth]{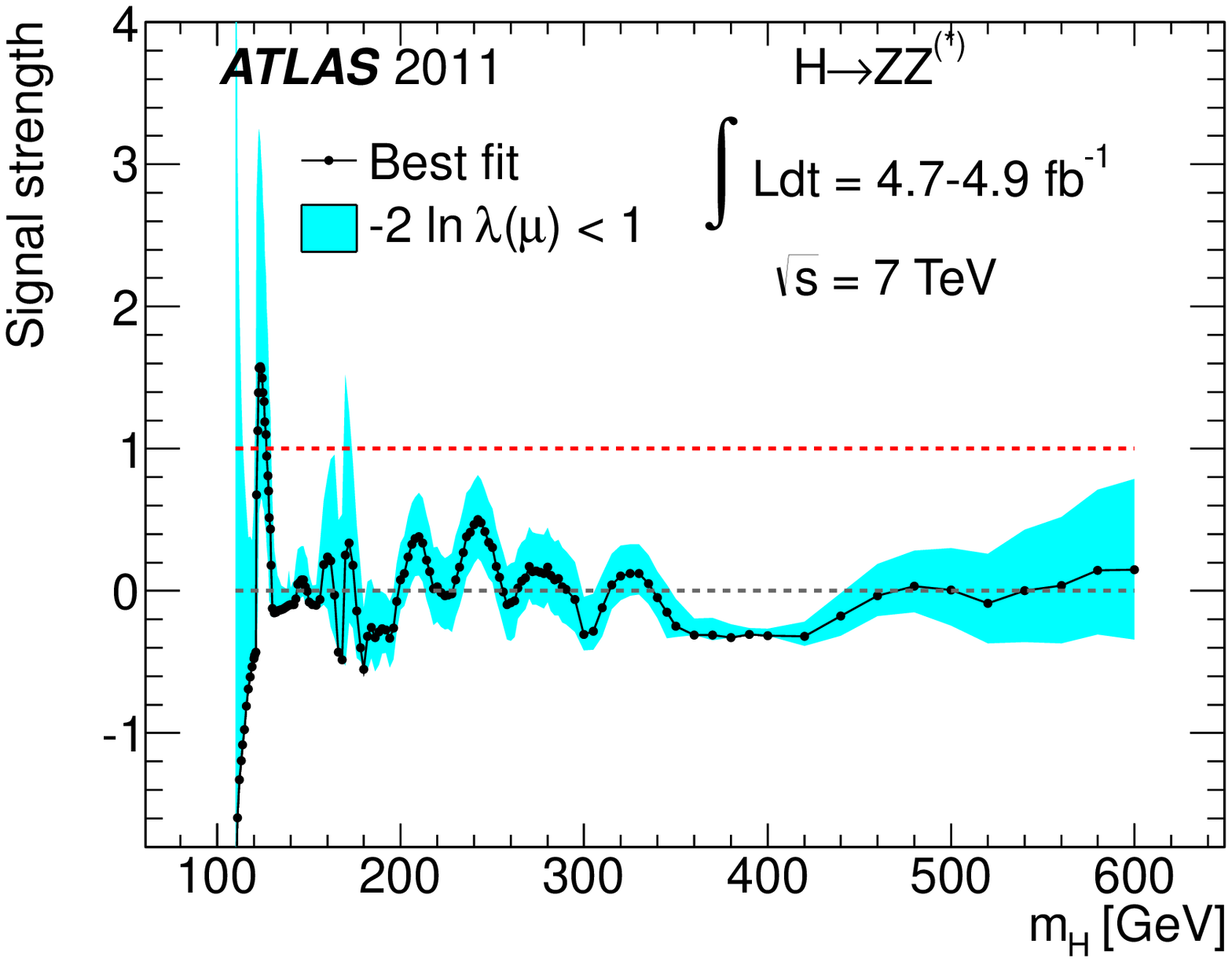}}
    \subfigure[]{ \includegraphics[width=.3\textwidth]{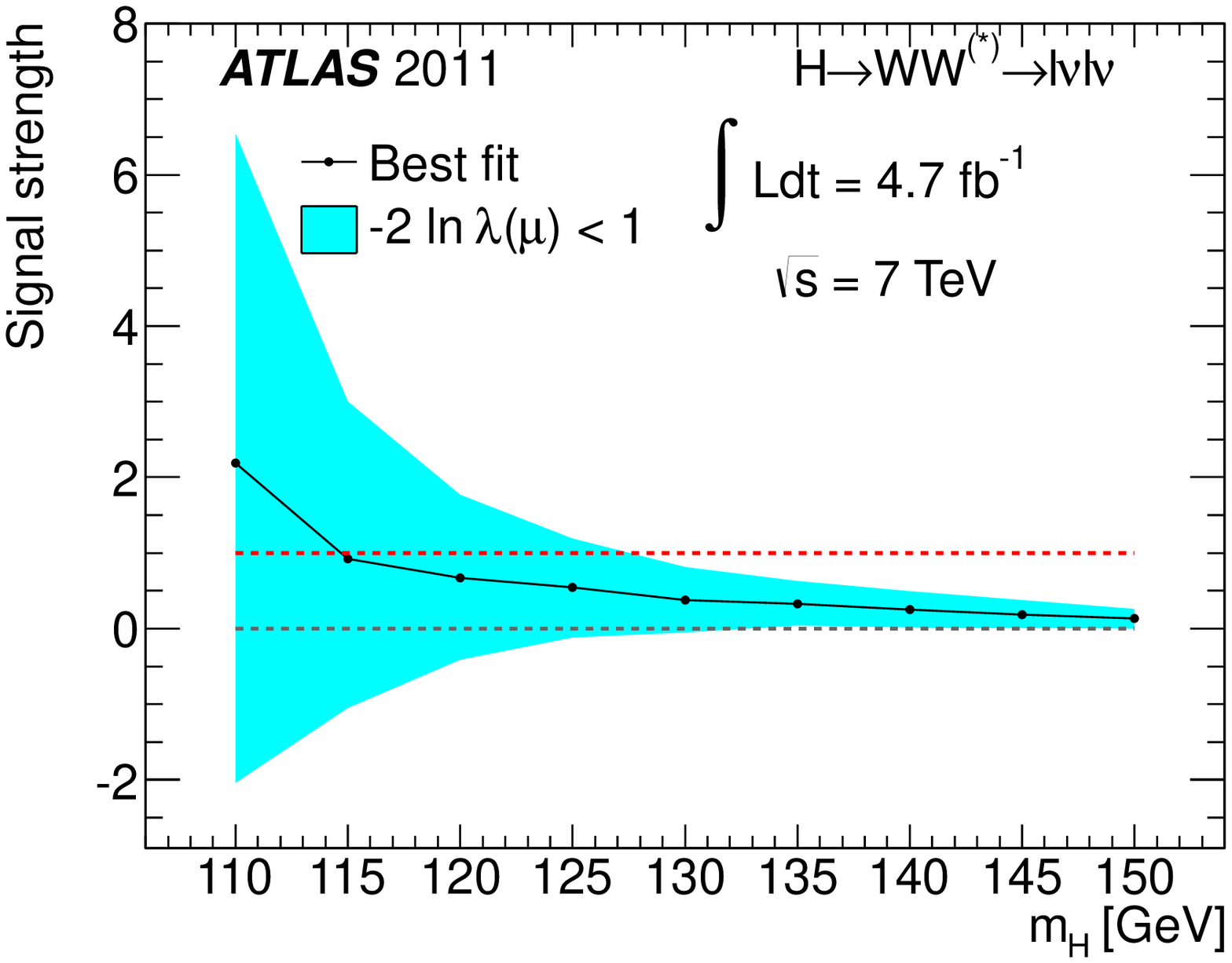}}
    \subfigure[]{ \includegraphics[width=.3\textwidth]{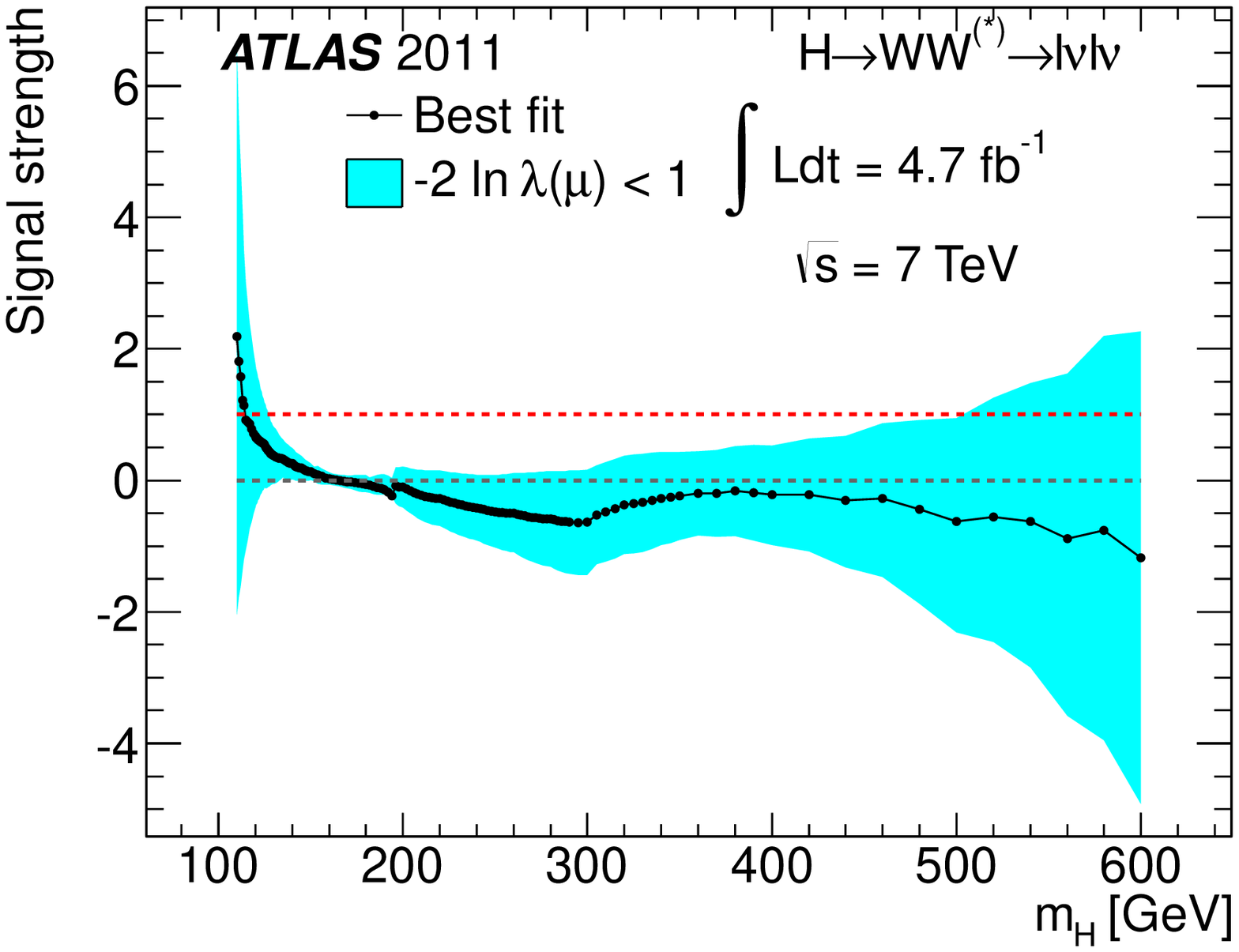}}
    \subfigure[]{ \includegraphics[width=.3\textwidth]{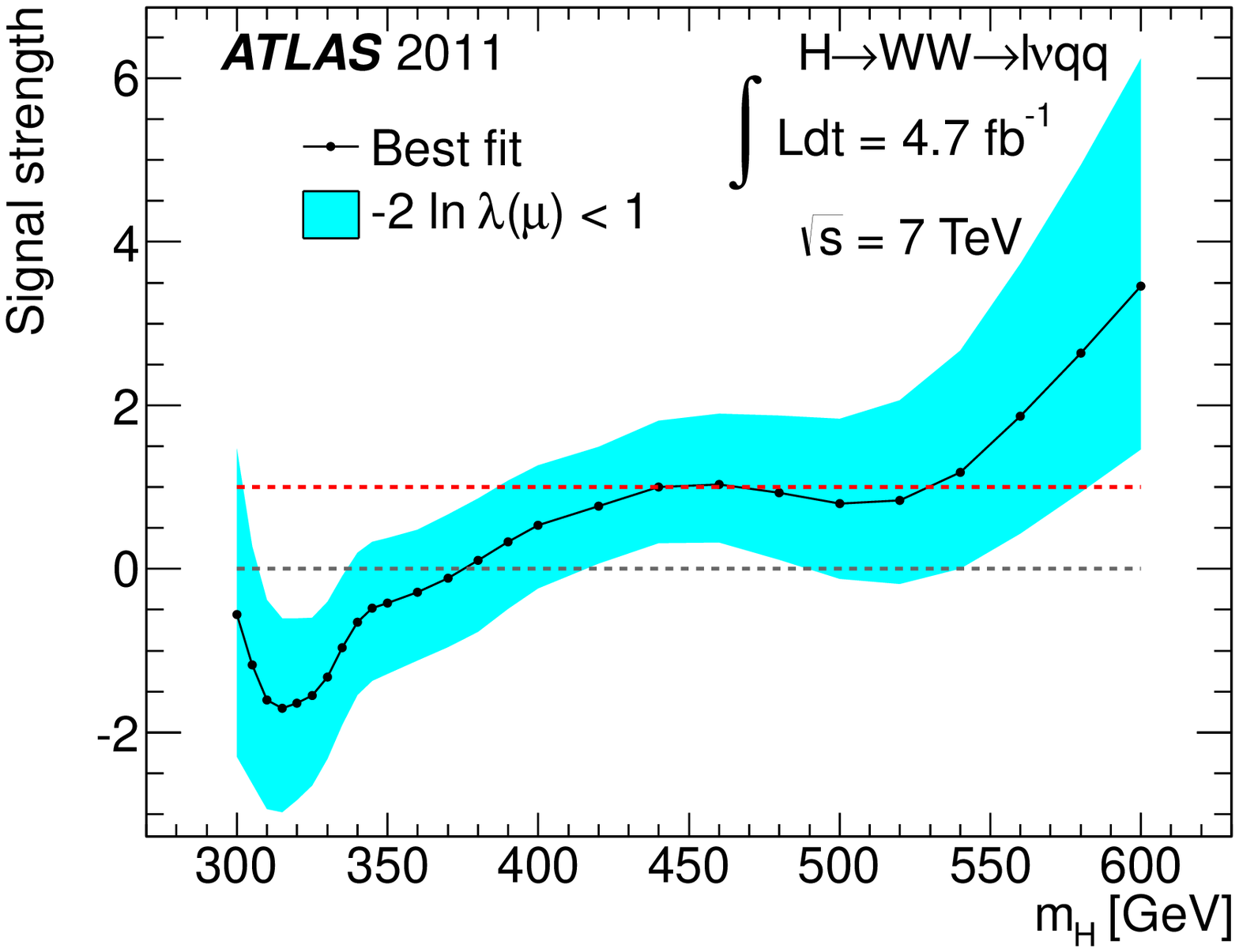}}
    \subfigure[]{ \includegraphics[width=.3\textwidth]{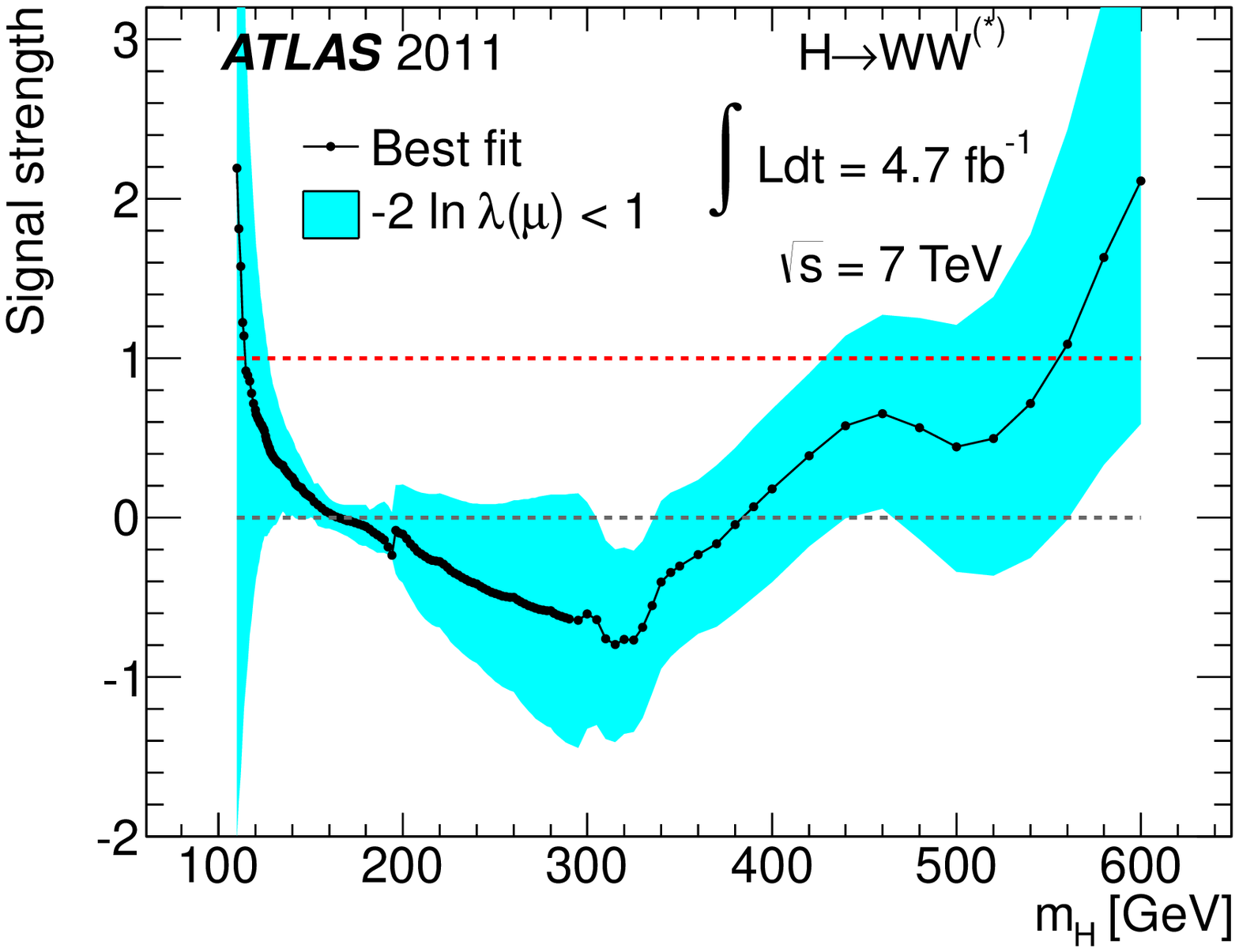}}
    \subfigure[]{ \includegraphics[width=.3\textwidth]{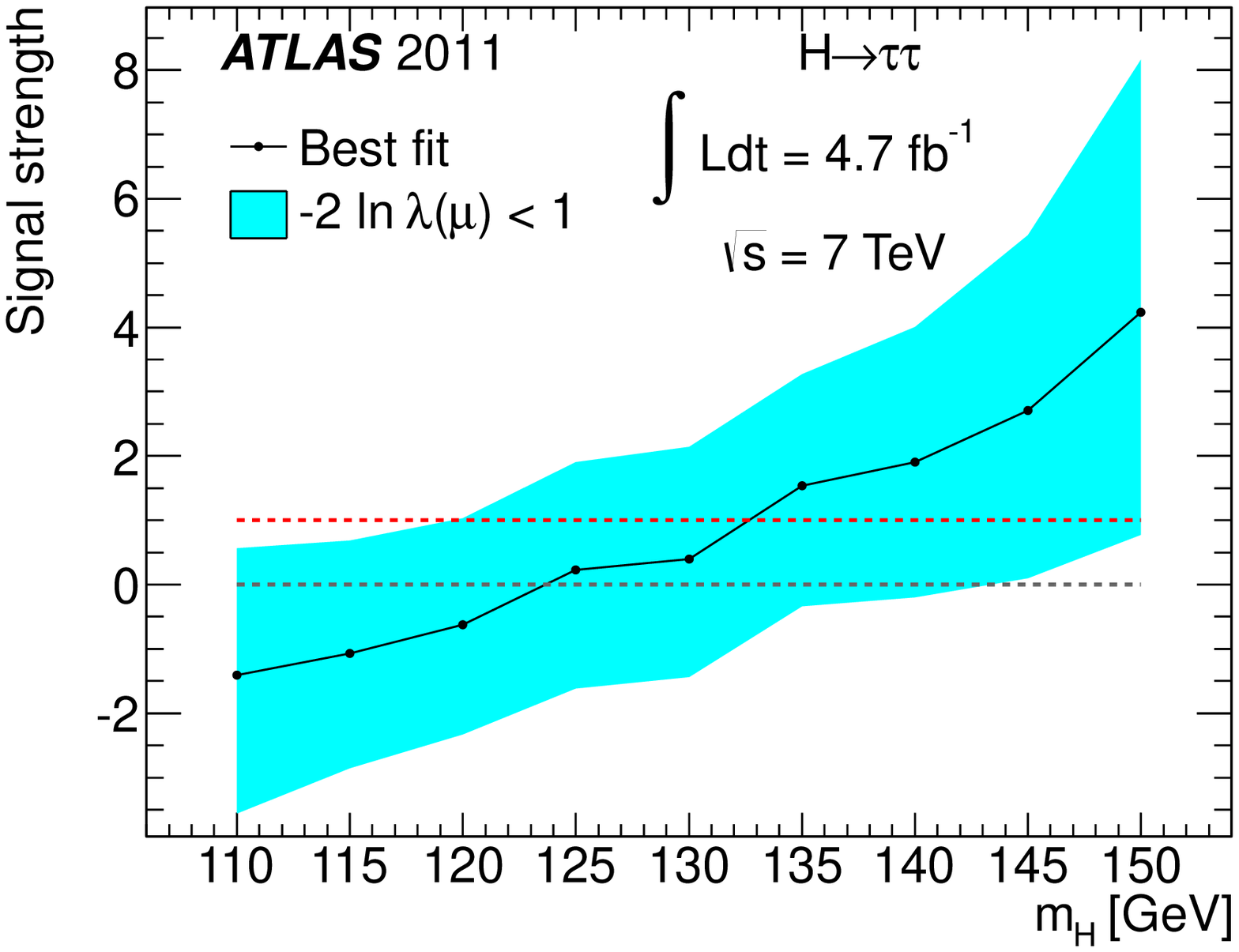}}
    \subfigure[]{ \includegraphics[width=.3\textwidth]{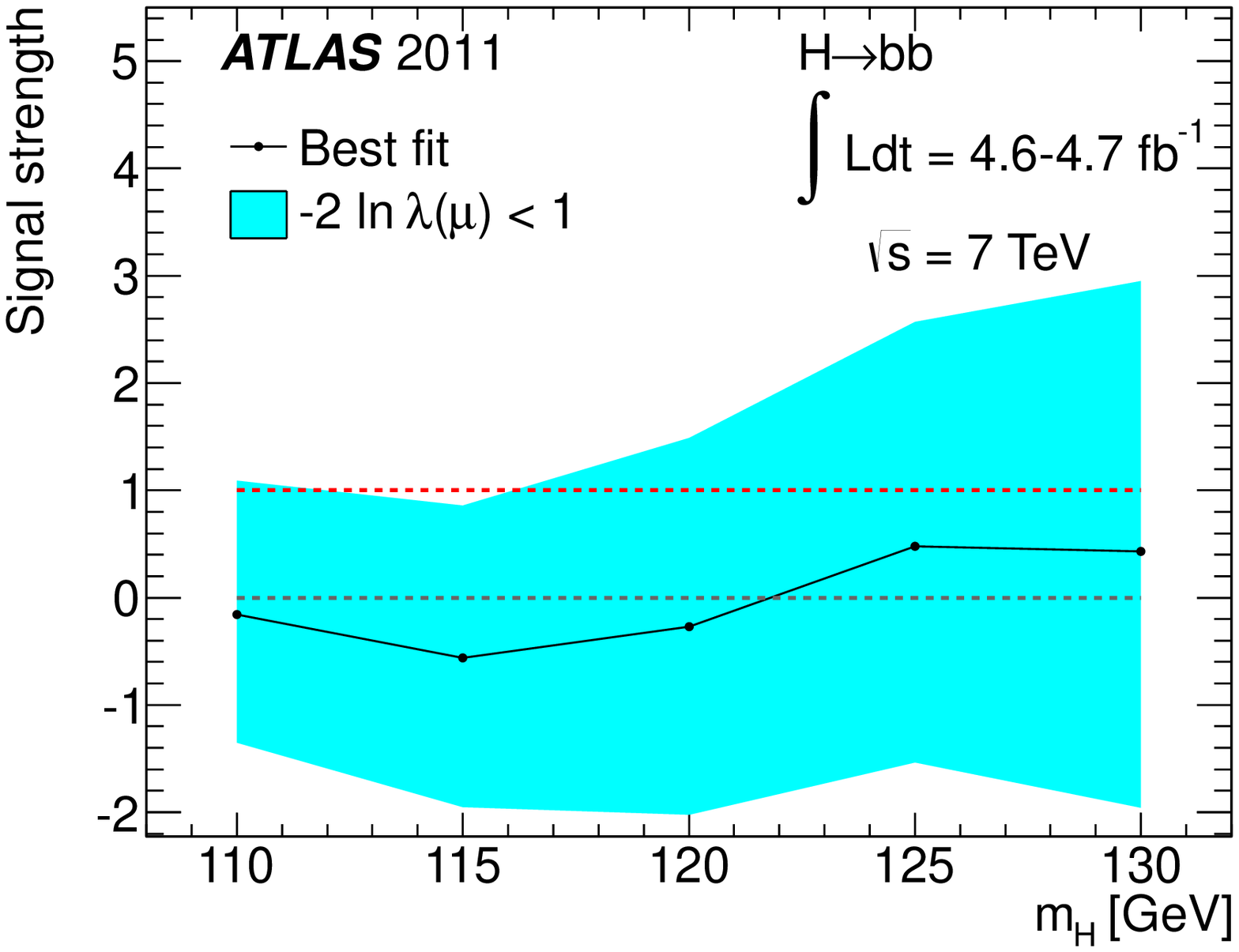}}
  \end{center}
  \caption{The best-fit signal strength $\mu$ as
    a function of the Higgs boson mass hypothesis for the channels: (a)  \hgg,
    (b)  \hZZllll\  in the low mass region, (c) 
    \hZZllll\ across the full search range, (d)  \hZZllnn, (e) 
    \hZZllqq, (f)  $H\to ZZ^{(*)}$ for all sub-channels across the full search range,
    (g)  \hWWlnln\  in the low mass region, (h) 
    \hWWlnln\ across the full search range, (i)  \hWWlnqq\ ,
    (j)  $H\to WW^{(*)}$ for all sub-channels in the full mass range, (k) 
    $H\to\tau\tau$, and (l) $H\to b\overline{b}$. 
     The band shows the interval around $\hat{\mu}$ corresponding
    to a variation of $-2\ln \lambda(\mu)<1$.}
  \label{fig:muhat_individual}
\end{figure*}

\begin{figure}[htb]
  \begin{center}
\mbox{\hspace{-0.4cm} 	\includegraphics[width=.52\textwidth]{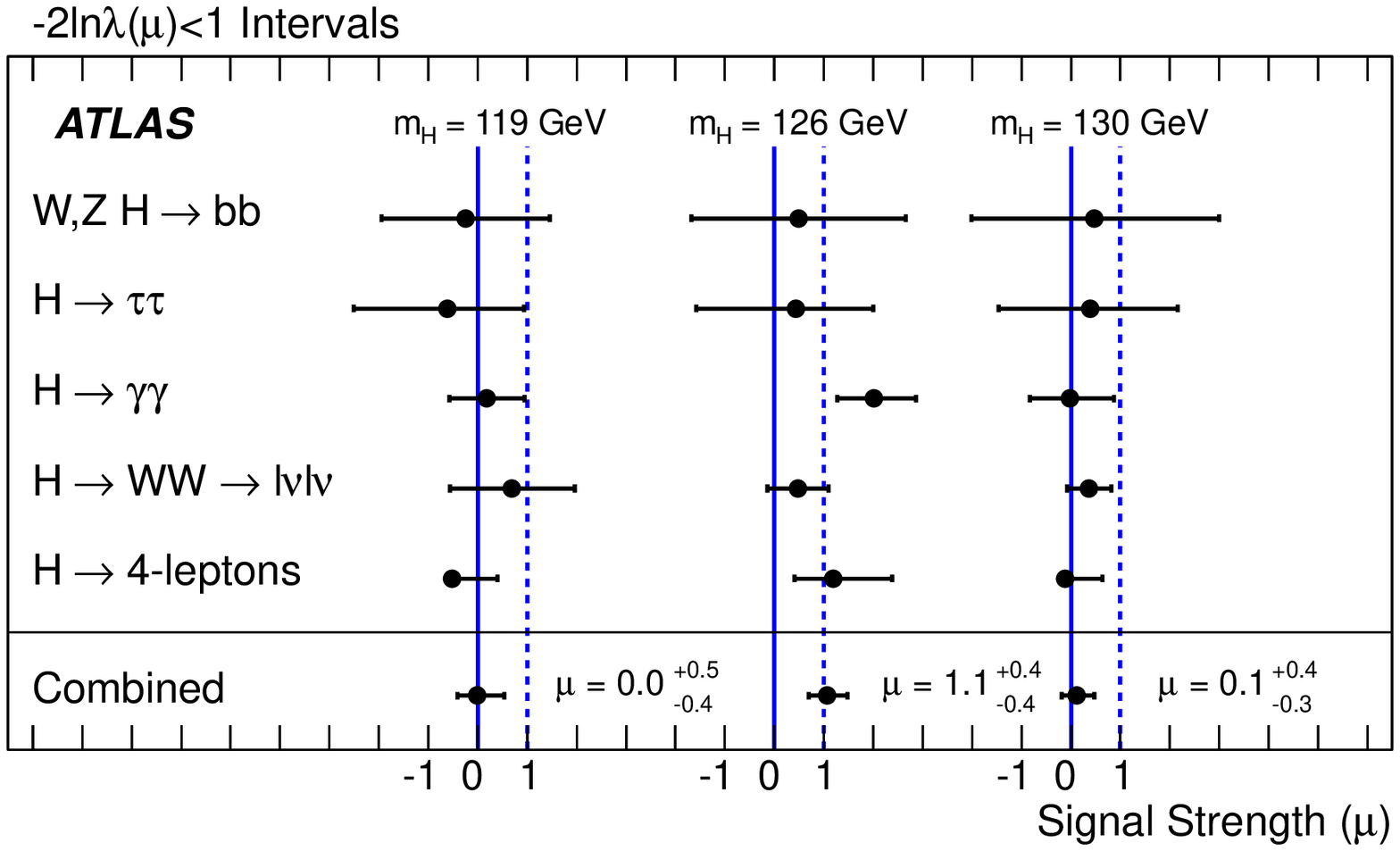}}
  \end{center}
  \caption{Summary of the individual and combined best-fit values of
    the strength parameter for three sample Higgs boson mass
    hypotheses 119~GeV, 126~GeV (where the maximum observed
    significance is reached) and~130 GeV.}
  \label{fig:muhatall}
\end{figure}

The best-fit value of $\mu$, denoted $\hat{\mu}$, is displayed for the
combination of all channels in Fig.~\ref{fig:muhat} and for individual
channels in Fig.~\ref{fig:muhat_individual} as a function of the \mH\
hypothesis. A summary of $-2\ln \lambda(\mu)<1$ intervals at three
specific Higgs boson mass hypotheses (\mH=119~\GeV, 126~\GeV\ and
130~\GeV) for each Higgs decay mode and the combination is given in
Fig.~\ref{fig:muhatall}. The bands around $\hat{\mu}$ illustrate the
$\mu$ interval corresponding to $-2\ln \lambda(\mu)<1$ and represent
an approximate $\pm 1\sigma$ variation. 
While the estimator $\hat{\mu}$ is allowed to be negative in
Figs.~\ref{fig:muhat} and \ref{fig:muhat_individual} in order to
illustrate the presence and extent of downward fluctuations, 
 the $\mu$ parameter is bounded to ensure non-negative values of the probability density functions in the individual
channels.  Hence, for negative $\hat{\mu}$ values close to the boundary,
the $-2\ln \lambda(\mu)<1$ region does not reflect a calibrated 68\%\ confidence
interval.  It should be noted that the $\hat{\mu}$ does not 
directly provide information on the relative strength of the production modes. The excess observed for $\mH=126\GeV$ corresponds to a
$\hat{\mu}$ of $1.1 \pm 0.4$, which is compatible with the
signal strength expected from a SM Higgs boson at that mass ($\mu=1$).

\section{Conclusion}
\label{S:Conclusions}

A combined search for the Standard Model Higgs boson has been
performed with the ATLAS detector in the $\sqrt{s}=7$ \TeV\ $pp$ collision data
collected in 2011, corresponding to an integrated luminosity of
4.6-4.9~\infb. The channels used in this combination are \hgg, \hbb,
\htt, \hZZllll, \hZZllqq, \hZZllnn, \hWWlnln, and \hWWlnqq.  The
observed exclusion ranges at the 95\%~CL are \lowerlowerObs\ to
\upperlowerObs, \lowerIsland\ to \upperIsland, and \lowerObs\ to
\upperObs, while Higgs boson masses between \lowerExp\ and \upperExp\
are expected to be excluded at the 95\%~CL. The mass region between
130.7~\GeV\ and 506~\GeV\ is excluded at the 99\%~CL.

The local significance of the observed excess when all channels are
combined is \significanceESS, in good agreement with the expected
significance in the presence of a SM Higgs boson with \mh=126~\GeV\ of
2.9$\sigma$. An estimate of the global probability for such
an excess to occur anywhere in the full explored Higgs boson mass
region (from 110~\GeV\ to 600~\GeV) is approximately 15\%.  The
global $p_0$ in the range not excluded at the 99\%\ CL
 by the LHC combined Higgs boson search results~\cite{lhcCombination} (from 110~\GeV\ to 146~\GeV) is approximately 5-7\%.

\section{Acknowledgements}

We thank CERN for the very successful operation of the LHC, as well as the
support staff from our institutions without whom ATLAS could not be
operated efficiently.

We acknowledge the support of ANPCyT, Argentina; YerPhI, Armenia; ARC,
Australia; BMWF, Austria; ANAS, Azerbaijan; SSTC, Belarus; CNPq and FAPESP,
Brazil; NSERC, NRC and CFI, Canada; CERN; CONICYT, Chile; CAS, MOST and NSFC,
China; COLCIENCIAS, Colombia; MSMT CR, MPO CR and VSC CR, Czech Republic;
DNRF, DNSRC and Lundbeck Foundation, Denmark; EPLANET and ERC, European Union;
IN2P3-CNRS, CEA-DSM/IRFU, France; GNAS, Georgia; BMBF, DFG, HGF, MPG and AvH
Foundation, Germany; GSRT, Greece; ISF, MINERVA, GIF, DIP and Benoziyo Center,
Israel; INFN, Italy; MEXT and JSPS, Japan; CNRST, Morocco; FOM and NWO,
Netherlands; RCN, Norway; MNiSW, Poland; GRICES and FCT, Portugal; MERYS
(MECTS), Romania; MES of Russia and ROSATOM, Russian Federation; JINR; MSTD,
Serbia; MSSR, Slovakia; ARRS and MVZT, Slovenia; DST/NRF, South Africa;
MICINN, Spain; SRC and Wallenberg Foundation, Sweden; SER, SNSF and Cantons of
Bern and Geneva, Switzerland; NSC, Taiwan; TAEK, Turkey; STFC, the Royal
Society and Leverhulme Trust, United Kingdom; DOE and NSF, United States of
America.

The crucial computing support from all WLCG partners is acknowledged
gratefully, in particular from CERN and the ATLAS Tier-1 facilities at
TRIUMF (Canada), NDGF (Denmark, Norway, Sweden), CC-IN2P3 (France),
KIT/GridKA (Germany), INFN-CNAF (Italy), NL-T1 (Netherlands), PIC (Spain),
ASGC (Taiwan), RAL (UK) and BNL (USA) and in the Tier-2 facilities
worldwide.


\bibliography{paper}



\onecolumngrid
\clearpage 
\begin{flushleft}
{\Large The ATLAS Collaboration}

\bigskip

G.~Aad$^{\rm 47}$,
T.~Abajyan$^{\rm 20}$,
B.~Abbott$^{\rm 110}$,
J.~Abdallah$^{\rm 11}$,
S.~Abdel~Khalek$^{\rm 114}$,
A.A.~Abdelalim$^{\rm 48}$,
O.~Abdinov$^{\rm 10}$,
R.~Aben$^{\rm 104}$,
B.~Abi$^{\rm 111}$,
M.~Abolins$^{\rm 87}$
% ...
\bigskip

% ...
$^{10}$ Institute of Physics, Azerbaijan Academy of Sciences, Baku, Azerbaijan\\
$^{11}$ Institut de F\'{i}sica d'Altes Energies and Departament de F\'{i}sica de la Universitat Aut\`{o}noma de Barcelona and ICREA, Barcelona, Spain\\
% ...
$^{20}$ Physikalisches Institut, University of Bonn, Bonn, Germany\\
% ...
$^{47}$ Fakult\"{a}t f\"{u}r Mathematik und Physik, Albert-Ludwigs-Universit\"{a}t, Freiburg, Germany\\
$^{48}$ Section de Physique, Universit\'{e} de Gen\`{e}ve, Geneva, Switzerland\\
% ...
$^{87}$ Department of Physics and Astronomy, Michigan State University, East Lansing MI, United States of America\\
% ...
$^{104}$ Nikhef National Institute for Subatomic Physics and University of Amsterdam, Amsterdam, Netherlands\\
% ...
$^{110}$ Homer L. Dodge Department of Physics and Astronomy, University of Oklahoma, Norman OK, United States of America\\
$^{111}$ Department of Physics, Oklahoma State University, Stillwater OK, United States of America\\
% ...
$^{114}$ LAL, Universit\'{e} Paris-Sud and CNRS/IN2P3, Orsay, France\\
% ...
\end{flushleft}


\end{document}